\documentclass[prc,preprintnumbers,unsortedaddress,amsmath,amssymb,floatfix]{revtex4}
\usepackage{comment}
\usepackage{amssymb}
\usepackage{color}
\usepackage{rotating}
\usepackage{amsmath}
\usepackage{ulem}
\usepackage{overpic}
\usepackage{graphicx}
\usepackage{dcolumn}
\usepackage{graphicx}
\usepackage{bm}
\usepackage{chngpage}
\usepackage{multirow}
\usepackage{slashed}
\usepackage{indentfirst}
\usepackage{booktabs}
\usepackage{amssymb,amsfonts}
\usepackage{amsmath}
\usepackage{xcolor}
\usepackage[colorlinks,citecolor=blue,linkcolor=blue,hypertex,breaklinks=true]{hyperref}

\newcommand{\be}{\begin{equation}}
\newcommand{\ee}{\end{equation}}
\newcommand{\bea}{\begin{eqnarray}}
\newcommand{\eea}{\end{eqnarray}}

\begin{document}
\title{Microscopic study of $M$1 resonances in Sn isotopes}

\author {Shuai Sun$\, ^{1,2}$, Li-Gang Cao$\, ^{1,2}$\footnote{Corresponding author: caolg@bnu.edu.cn}, Feng-Shou Zhang$\, ^{1,2,3}$\footnote{Corresponding author: fszhang@bnu.edu.cn}, Hiroyuki Sagawa${}^{4,5}$, and Gianluca Col\`{o}${}^{6}$ }

\affiliation{
$^{1}$Key Laboratory of Beam Technology of Ministry of Education, College of Nuclear Science and Technology,
Beijing Normal University, Beijing 100875, China\\
$^{2}$Institute of Radiation Technology, Beijing Academy of Science and Technology, Beijing 100875, China\\
$^{3}$Center of Theoretical Nuclear Physics, National Laboratory of Heavy Ion Accelerator of Lanzhou, Lanzhou 730000, China\\
$^{4}$RIKEN Nishina Center, Wako 351-0198, Japan\\
$^{5}$Center for Mathematics and Physics, University of Aizu,  Aizu-Wakamatsu 965-8560, Japan\\
$^{6}$Dipartimento di Fisica, Universit\`{a} degli Studi di Milano, and INFN, Sezione di Milano, Via Celoria 16, 20133 Milano, Italy
 }

\date{\today}

\begin{abstract}
The magnetic dipole ($M$1) resonances of even-even $^{112-120, 124}$Sn isotopes are investigated in the framework of the self-consistent Skyrme Hartree-Fock (HF) + BCS and Quasiparticle Random Phase Approximation (QRPA). The Skyrme energy density functionals SLy5 and T11 with and without tensor terms are adopted in our calculations. The mixed type pairing interaction is used to take care of the pairing effect for open-shell nuclei both in the ground and excited states calculations. The calculated magnetic dipole strengths are compared with available experimental data. The QRPA results calculated by SLy5 and T11 with tensor force show a better agreement with the experimental data than those without the tensor force. By analyzing the HF and QRPA strength distributions of $^{112}$Sn and $^{124}$Sn, we discuss the effect of tensor force on the $M$1 resonances in detail. It is found that the $M$1 resonance is sensitive to the tensor interaction, and favors especially a negative triplet-odd tensor one. Depending on the nucleus, a quenching factor of the $M$1 operator of about 0.71-0.95 is needed to reproduce the total observed transition strength. In our calculations, we also find some low-lying, pygmy-type magnetic dipole states distributed below 6.0 MeV, and they are formed mainly from the neutron configuration $\nu$2$d_{5/2}$$\rightarrow$$\nu$2$d_{3/2}$.
\end{abstract}

\maketitle

\section{Introduction}\label{part1}
The magnetic dipole ($M$1) resonance is one of the fundamental excitations of spin-flip type in finite nuclei \cite{harakeh2001giant,Fujita11,Heyde-M1,Richter95,Franz92}.
It has been studied experimentally and theoretically for several decades. The $M$1 resonance is known experimentally to include two major components. One is an orbital component at low excitation energy, it is found in deformed nuclei and called scissors mode. In spherical nuclei, the scissors mode is much suppressed. The other spin-flip component is found at an energy of around 8 MeV, contributing to most of the M1 strength. The scissors mode in deformed nuclei is interpreted as neutrons and protons vibrating with a small angle with respect to each other in a scissors-like motion, while the higher energy component describes a resonance-like structure made of proton and neutron spin-flip excitations. The study of $M$1 resonance is of great interest not only for the nuclear structure but also for nuclear astrophysics. It provides,
in addition to charge-exchange modes, an alternative chance to explore the nuclear interactions in spin and spin-isospin channels and can offer crucial information on nuclear structure \cite{Pietralla15}. The properties of the $M$1 resonance may impact the description of neutral current neutrino interactions in supernova \cite{Langanke04, Langanke08}, or the estimate the reaction cross sections in large-scale nucleosynthesis network calculations \cite{Loens12, Goriely16, Goriely19}.

In the past years, great efforts have been devoted to the study of $M$1 resonance in the framework of non-relativistic random phase approximation (RPA) with Skyrme or Gogny interactions, relativistic RPA, and shell model calculations. In the non-relativistic approaches, it is well known that the distribution of $M$1 resonance is very sensitive to the spin-dependent interactions. So, many studies have focused on the effect of spin-orbit and tensor interactions on the $M$1 strength distribution \cite{Goriely16, Goriely19,Vesely09,Nesterenko2010,Tselyaev19,Co09,Co12,Wenpw13,Wenpw14}. Recently, the self-consistent description of magnetic dipole resonance with relativistic energy density functionals has become available \cite{Oishi20,PaarM120,PaarM121,Changshi22}. The density-dependent point-coupling or density-dependent meson exchange interactions are adopted in the calculations. To properly describe the unnatural-parity  $M$1 resonance, the isovector-pseudovector interaction should be included in the residual interaction. In the case of the shell model calculations, the studies of the magnetic dipole resonance pay attention to the strengths at low energy for some selected nuclei \cite{Brown14,Schwengner17,Sieja18}.

Experimentally, the magnetic dipole resonance can be excited by inelastic scattering of protons, electrons and photons, it has been investigated for many years, and a rich amount of database has been built~\cite{harakeh2001giant,Fujita11,Heyde-M1,Richter95,Franz92,Pietralla08,M1exp1,M1exp2,M1exp3,M1exp4}.
Recently, electric and magnetic dipole responses along the even-even tin isotopes have been measured in an inelastic proton scattering experiment at RCNP~\cite{M1exp-Sn120,M1exp-Sn}. Total photoabsorption cross sections have been derived from the $E$1 and $M$1 strength distributions and show significant
differences compared to those from previous experiments. The magnetic dipole strengths in $^{112- 120,124}$Sn exhibit a broad distribution
between 6 and 12 MeV in all studied nuclei. The new magnetic dipole data in Sn isotopes provide a good opportunity to check the ability of existing nuclear
energy density functionals (EDFs) to reproduce the data. In Ref. \cite{PaarM121}, the authors have investigated $M$1 transitions in even-even $^{100-140}$Sn isotopes based on the relativistic EDFs, by raising many points for discussion. Up to now, a systematic investigation of the new database by non-relativistic models is still missing.

As we know, the tensor force plays a significant role in nuclear structure studies \cite{SAGAWA201476}. The shell evolution of the single-particle energies in some exotic nuclei can be well explained by the inclusion of tensor force \cite{Otsuka05,Browntensor06,COLO2007227,Brinktens07,Grasso07,Long07,Lesinski07,Dong11,Wangyz11}. Extensive efforts have also been undertaken to study the influence of the tensor force on the excited state properties of finite nuclei, like the spin and spin$-$isospin excitation modes \cite{Baitens09,Baitens10,Caotens09,Caotens11,Minatotens13,Cotens16,Xutens}. The tensor force also plays a role in the calculations of the response functions of infinite nuclear matter \cite{Pastoretens1,Pastoretens2,Pastoretens3}. The effect of the tensor force on heavy-ion collisions has been discussed within the
time-dependent Hartree-Fock (TDHF) method, it plays a non-negligible role in dynamic processes in nuclei \cite{Guotens1,Guotens2,Steventens}.  In this work, we will investigate the $M$1 resonances in even-even $^{112- 120,124}$Sn isotopes within the framework of Skyrme HF + BCS plus Quasiparticle RPA (QRPA).  The calculated results are compared to the experimental data from Refs. \cite{M1exp-Sn120,M1exp-Sn}. The effect of the tensor force on $M$1 resonances in even-even $^{112- 120,124}$Sn isotopes is discussed in detail in the present work. We will also pay attention to the quenching associated with the magnetic dipole
operator, which is a long-standing problem in nuclear structure \cite{Quench1,Tselyaev20,Caoquench}.

This article is organized as follows. The theoretical model is briefly reviewed in Sec.~\ref{part2}. In Sec.~\ref{part3}, the calculated results are compared with experimental data, the discussions on the effect of tensor force and quenching problem are also given in Sec.~\ref{part3}. The summary and some perspectives for future work are given in Sec.~\ref{part4}.

\section{Theoretical framework}\label{part2}

In this work, a  HF+BCS plus QRPA approach is  employed in the calculations. Since the theoretical framework of HF+BCS method is well known in the literature
(cf. Refs. \cite{Rowebook,Ringbook}), we briefly review only the QRPA main equations. The matrix equations of QRPA can be written as
\bea\label{RPA} \left( \begin{array}{cc}
A & B \\
-B^* &- A^* \end{array}  \right) \left( \begin{array}{c}
X^{\nu} \\
Y^{\nu}  \end{array} \right) =E_{\nu} \left( \begin{array}{c}
X^{\nu} \\
Y^{\nu}  \end{array} \right),
\eea
where $E_{\nu}$ is the eigenvalue of the ${\nu}$-th QRPA state and X$^{\nu}$, Y$^{\nu}$ are the corresponding forward and backward quasiparticle amplitudes, respectively. The details about the matrix elements A and B can be found in Ref. \cite{Skyrmerpa,Colorpa}.

The magnetic dipole operator is given by
\begin{equation}
\hat{F}(M1)=\mu_N \sum_{i=1}^A(g_l\vec{l}_i+g_s\vec{s}_i)=\mu_N \sum_{i=1}^A[g_l\vec{j}_i+(g_s-g_l)\vec{s}_i],
\end{equation}
where $\mu_N=e\hbar/2mc$ is the nuclear magneton. Since $j$ is a good quantum number for the
single-particle states in spherical nuclei, the first term
$g_l\vec j_i$ does not contribute to the transition matrix  for the $p$-$h$ type excitation since $j_p\neq j_h$ .  On the other hand, in open shell nuclei, the two quasi-particle  excitation with the same $j$ quantum number,
$j_1$=$j_2$,  contributes to the matrix element.  We should notice that even  in the $p$-$h$ type excitation,
the orbital contribution exists and it is  absorbed
in the $g$-factor of spin operator $g_s \rightarrow g_s-g_l$.

For the magnetic dipole operator, the reduced transition strength from the ground state to the excited state $\nu$ is written as
\begin{widetext}
\begin{eqnarray}
B^\nu(M1)&=&\frac{1}{2J+1}|\langle\nu||\hat{F}||g.s.\rangle|^2=\frac{1}{2J+1}\left|\sum_{c \geq d}(X^{\nu}_{cd}+Y^{\nu}_{cd})(v_{c}u_{d}+u_{c}v_{d})\langle c\|\hat{F}\|d\rangle\right|^{2}.
\end{eqnarray}
\end{widetext}

In the figures, the $M$1 discrete spectra are convoluted with Lorentzian distributions
\begin{eqnarray}
S_{M1}(E)=\sum_{\nu} B^\nu(M1)\frac{1}{\pi}\frac{\Gamma/2}{(E-E_\nu)^2+\Gamma^2/4},
\end{eqnarray}
where $\Gamma$ is the width and is taken equal to 2 MeV in present calculations.

The triplet-even and triplet-odd zero-range tensor terms of the Skyrme force are expressed as
\begin{align}
v_{T}= & \frac{T}{2}\left\{\left[\left(\bm{\sigma}_{1} \cdot \mathbf{k}^{\prime}\right)\left(\bm{\sigma}_{2} \cdot \mathbf{k}^{\prime}\right)-\frac{1}{3}\left(\bm{\sigma}_{1} \cdot \bm{\sigma}_{2}\right) \mathbf{k}^{\prime 2}\right] \delta\left(\mathbf{r}_{1}-\mathbf{r}_{2}\right)\right. \notag \\
& \left.+\delta\left(\mathbf{r}_{1}-\mathbf{r}_{2}\right)\left[\left(\bm{\sigma}_{1} \cdot \mathbf{k}\right)\left(\bm{\sigma}_{2} \cdot \mathbf{k}\right)-\frac{1}{3}\left(\bm{\sigma}_{1} \cdot \bm{\sigma}_{2}\right) \mathbf{k}^{2}\right]\right\} \notag \\
& +U\left\{(\bm{\sigma}_{1}\cdot\mathbf{k}')\delta(\bf{r}_{1}-\bf{r}_{2})(\bm{\sigma}_{2}\cdot\mathbf{k})-\frac{1}{3}(\bm{\sigma}_{1}\cdot \bm{\sigma}_{2})\mathbf{k}^{\prime} \cdot \delta\left(\mathbf{r}_{1}-\mathbf{r}_{2}\right) \bf{k}\right\},
\end{align}
where the operator $\mathbf{k}=\left(\bm{\nabla}_{1}-\bm{\nabla}_{2}\right)/2i$ acts on the right and $\mathbf{k}^{\prime}=-\left(\bm{\nabla}_{1}-\bm{\nabla}_{2}\right)/2i$ acts on the left. The coupling constants $T$ and $U$ denote the strengths of the triplet-even and triplet-odd tensor interactions, respectively.

It is known that the tensor force affects the spin-orbit mean potential. The spin-orbit potential is expressed as
\be
V_{\text {s.o. }}^{(q)}=U_{\text {s.o. }}^{(q)}\bf{l}\cdot\bf{s}
\ee
and
\begin{align}
U_{\text {s.o. }}^{(q)}=\frac{W_{0}}{2 r}\left(2 \frac{d \rho_{q}}{d r}+\frac{d \rho_{1-q}}{d r}\right)+\left(\alpha \frac{J_{q}}{r}+\beta \frac{J_{1-q}}{r}\right),
\end{align}
where $q = 0(1)$ is the quantum number $(1-t_z)/2$ ($t_z$ being the third isospin component) that distinguishes neutrons and protons. The first term on the right comes from the Skyrme spin-orbit interaction, and the second term,
including contributions from $J^2$ terms, comes from some exchange terms of the central force as well as from the tensor force.
The spin-orbit density $\bf{J}$ in spherical nuclei has only a  radial component whose expression reads
\be
J_q=\frac{1}{4\pi r^3}\sum_i v^2_i(2j_i+1)\left[j_i(j_i+1)-l_i(l_i+1)-\frac{3}{4}\right]R_i^2(r),
\ee
where $i = n, l, j$ runs over all states. The quantity $v^2_i$  is the occupation probability of each orbit determined by the BCS approximation
and $R_i(r)$ is the radial part of the HF single-particle wave function. $\alpha$ and $\beta$ in Eq.~(7) include both the central exchange terms and the tensor terms, that is, $\alpha=\alpha_{C}+\alpha_{T}$ and $\beta=\beta_{C}+\beta_{T}$. The central exchange contributions are written in terms of the usual Skyrme parameters
\begin{align}
\alpha_{C} & =\frac{1}{8}\left(t_{1}-t_{2}\right)-\frac{1}{8}\left(t_{1} x_{1}+t_{2} x_{2}\right), \notag \\
\beta_{C} & =-\frac{1}{8}\left(t_{1} x_{1}+t_{2} x_{2}\right),
\end{align}
while the tensor contributions are expressed as
\begin{align}
\alpha_{T}=\frac{5}{12} U, \quad \beta_{T}=\frac{5}{24}(T+U).
\end{align}

In the HF+BCS plus QRPA calculations, we take an effective density-dependent zero-range pairing interaction,
\begin{equation}
V_{pair}(\textbf{r}_{1},\textbf{r}_{2})=V_{0}\left [1-\eta\left (\frac{\rho(\textbf{r})}{\rho_{0}}\right )\right ]\delta(\textbf{r}_{1}-\textbf{r}_{2}),
\end{equation}
where $\rho(\textbf{r})$ is the particle density, and $\rho_{0}=0.16\ fm^{-3}$ is the density at nuclear saturation. The parameter $\eta$ represents the pairing type, when $\eta$ is either 1.0, 0.5 or 0.0, it means we adopt surface, mixed or volume pairing interaction. The mixed pairing interaction is used in our calculations, as it is very effective in describing many properties of finite nuclei \cite{Dobacpair,Caoisgmr,Yamagamipair}.  The pairing strength $V_{0}$ is adjusted to reproduce the empirical neutron gap in $^{120}$Sn ($\Delta_n$=1.392 MeV). Then the same value is adopted for the calculations of other Sn isotopes.

\begin{figure}[htb]
 \includegraphics[width=0.4\textwidth]{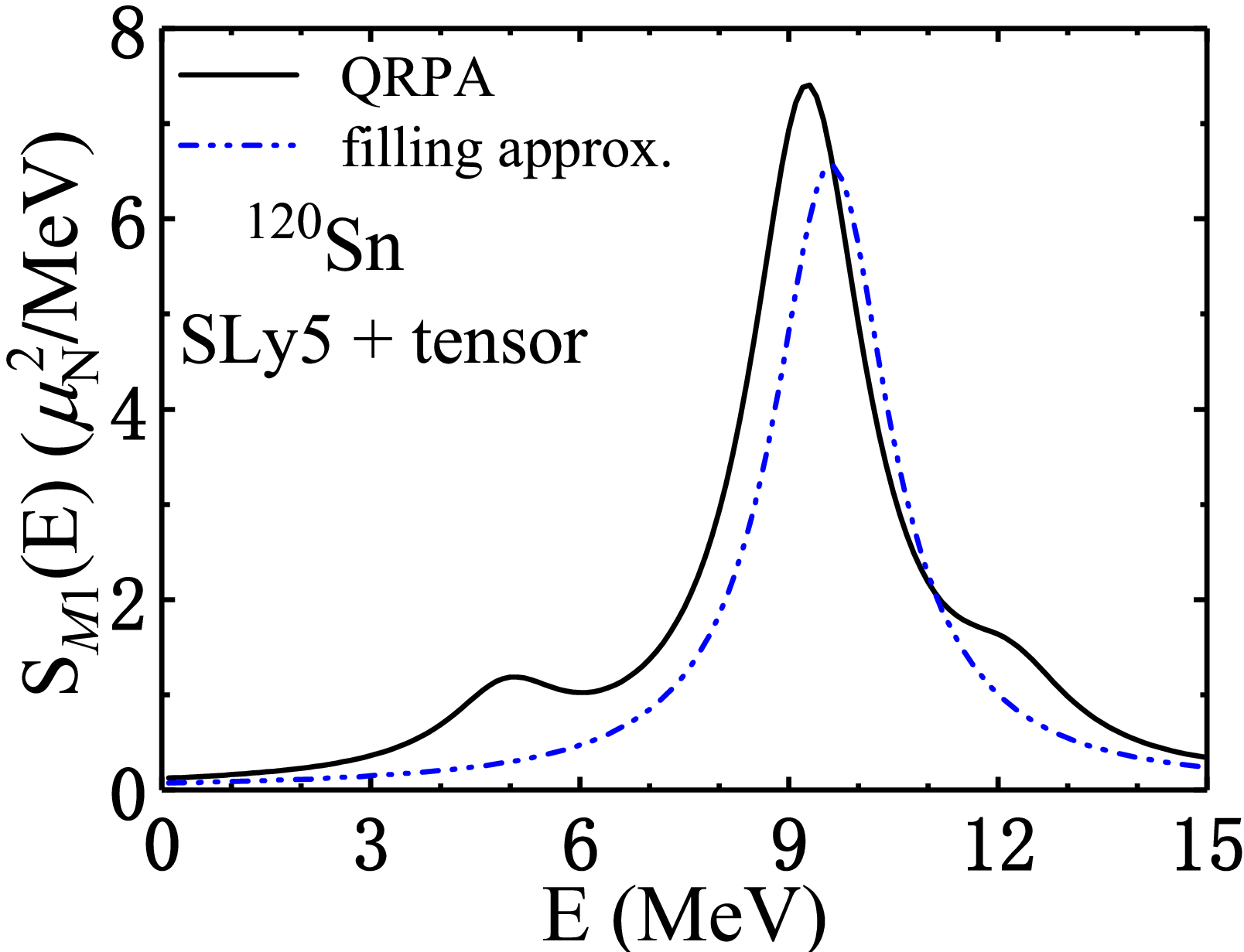}
\caption{(color online) The $M$1 strength distributions of $^{120}$Sn in filling approximation and QRPA, respectively. The calculated strengths are convoluted by a
Lorentzian shape with a width of  2.0 MeV. }  \label{f1}
\end{figure}

In order to investigate the effect of pairing on the $M$1 strength distribution, as an example, the filling approximation\cite{Bertulani09,Bertulani091} and QRPA
calculations are performed for $^{120}$Sn using SLy5 with tensor force, and the results are shown in Fig.~\ref{f1}. In the filling approximation, pairing is neglected completely,
that is, the p-p interaction is also dropped in the QRPA
matrix. The $M$1 strength distribution given by the filling approximation shows a unimodal structure with a peak at energy around 9.6 MeV, which is mainly coming from the proton configuration $\pi$1$g_{9/2}$$\rightarrow$$\pi$1$g_{7/2}$. In the QRPA result, the main peak is shifted downward to 9.2 MeV. Besides the $M$1 main peak, two additional $M$1 pygmy resonance states emerge at energies around 5.0 and 12.3 MeV, respectively. The word {\it pygmy state} is often used for a low-lying $E$1 state with smaller strength than giant dipole resonances. On the other hand, many authors employ this word without implying any special multipole, but just referring to the fact that the strength is smaller with respect to giant resonances\cite{Yoshida17}. Thus it can be used even for low-lying $M$1 or other multipoles having the smaller strength. We find that the low-lying state comes from the neutron quasiparticle configuration $\nu$2$d_{5/2}$$\rightarrow$$\nu$2$d_{3/2}$ while the high-energy state is due to the $\nu$1$h_{11/2}$$\rightarrow$$\nu$1$h_{9/2}$ neutron quasiparticle configuration. The discrepancy between the results of filling approximation and QRPA stems from the particles scattering around the Fermi surface, i.e., the neutron state 2$d_{3/2}$ changes from fully occupied to partially occupied, while the neutron 1$h_{11/2}$ state turns from being empty to being partially filled.  Notice that the states 2$d_{5/2}$ and 2$d_{3/2}$ are below the Fermi level while 1$h_{11/2}$ and 1$h_{9/2}$ are above that.   These changes by the pairing correlations allow the relatively strong transitions from $\nu$2$d_{5/2}$ to $\nu$2$d_{3/2}$ and $\nu$1$h_{11/2}$ to $\nu$1$h_{9/2}$. The above discussion shows that the effect of pairing on the $M$1 resonance is
substantial and make appreciable difference from the filling approximation.

\section{Results and discussions}\label{part3}

In present study, all the calculations assume a spherical shape for the even-even Sn isotopes. The quasiparticle states are obtained by solving HF+BCS in coordinate space with a box boundary condition and the size of the box is 24 fm. We have checked that the predicted ground state properties of Sn isotopes, such as binding energies, charge radii, agree well with the experimental data. After solving the HF+BCS equation in coordinate space, we build up a model space of two-quasiparticle configurations for $M$1 excitation, and then we solve the QRPA matrix equations in the model space. The major shell configurations up to $\Delta N=8$ are
 adopted to build up the QRPA model space, which is large enough to allow the convergence of the results.

\subsection{Skyrme interactions for $M$1}

\begin{figure}[htb]
\includegraphics[width=0.4\textwidth]{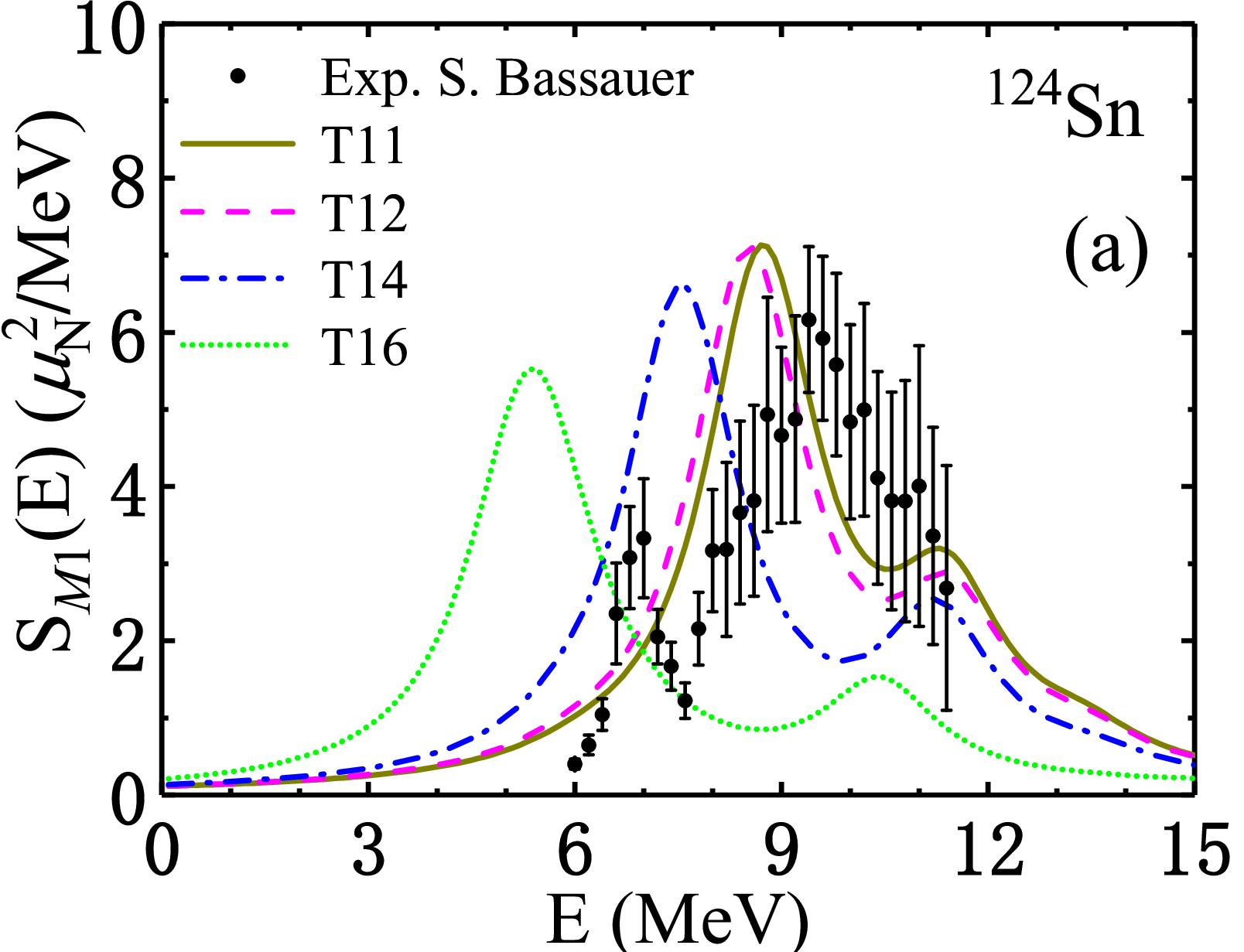}
\includegraphics[width=0.4\textwidth]{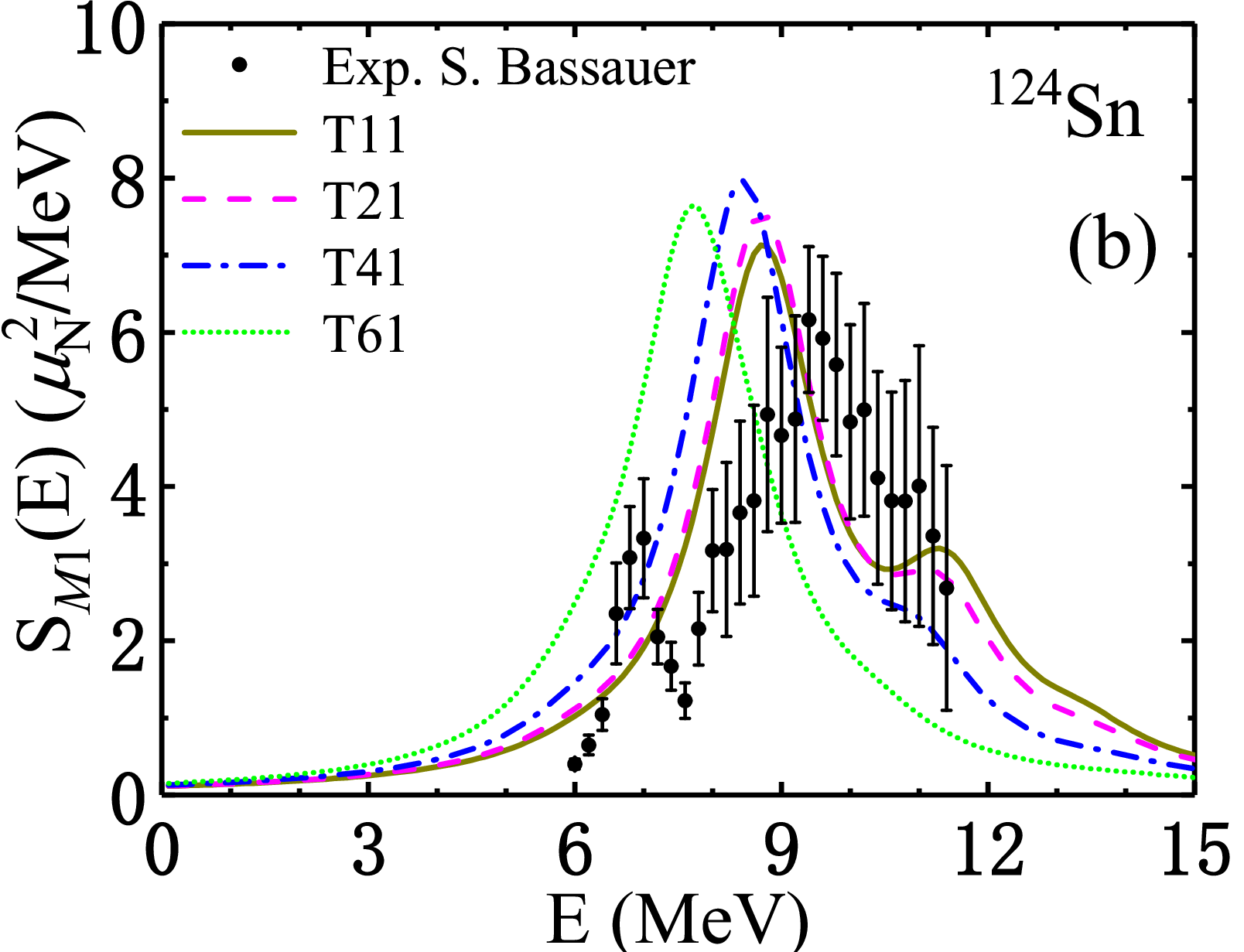}
\includegraphics[width=0.4\textwidth]{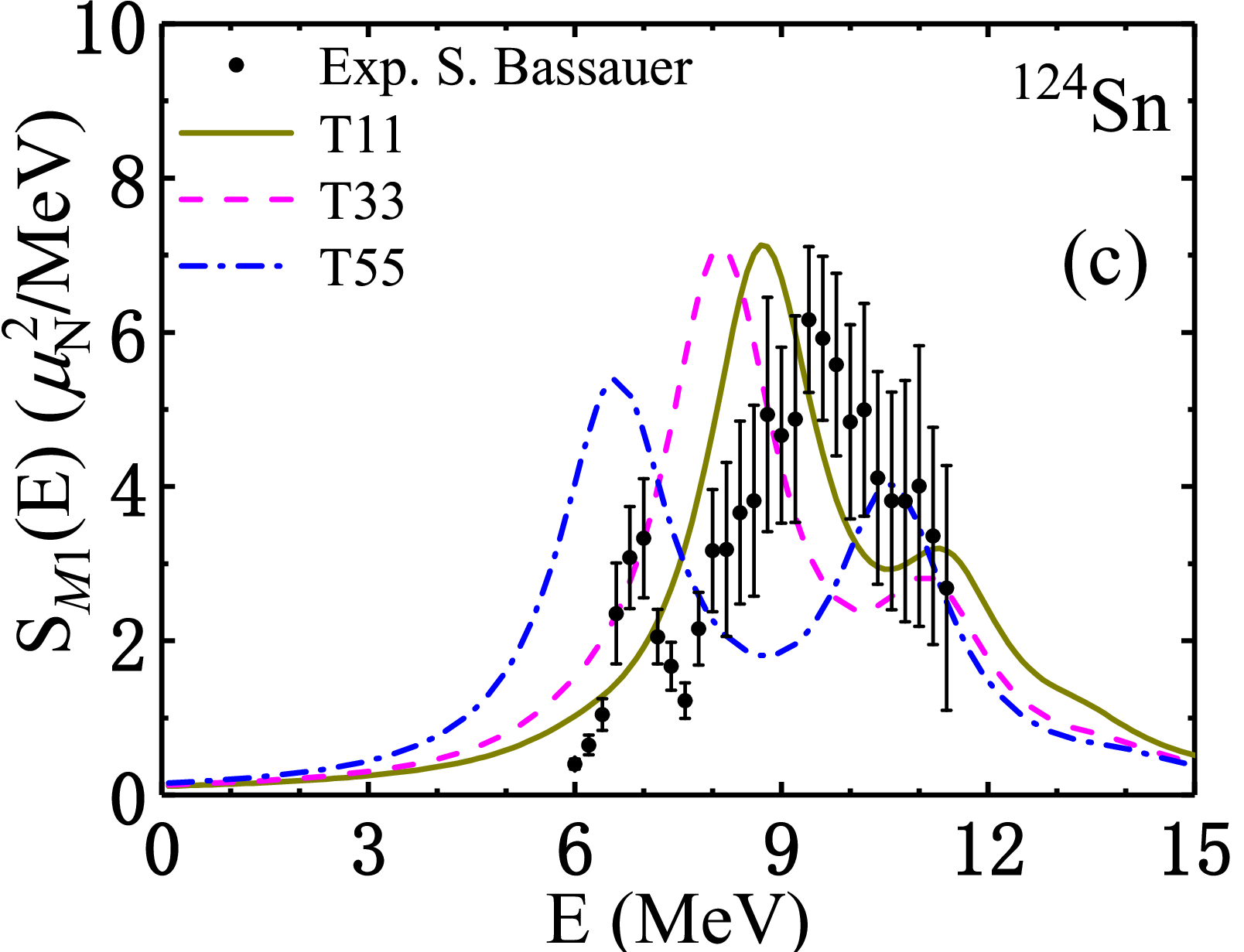}
\includegraphics[width=0.4\textwidth]{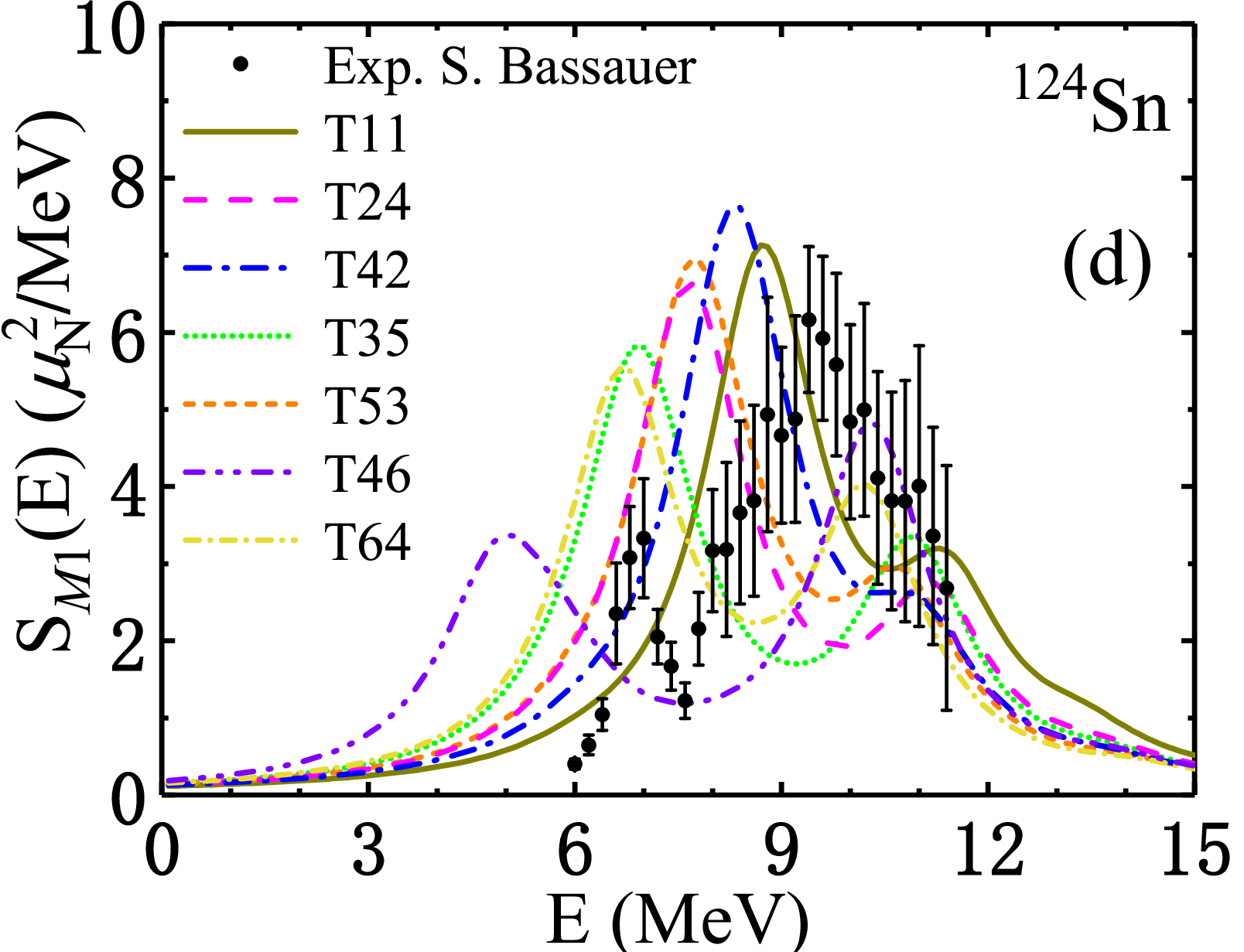}
\includegraphics[width=0.4\textwidth]{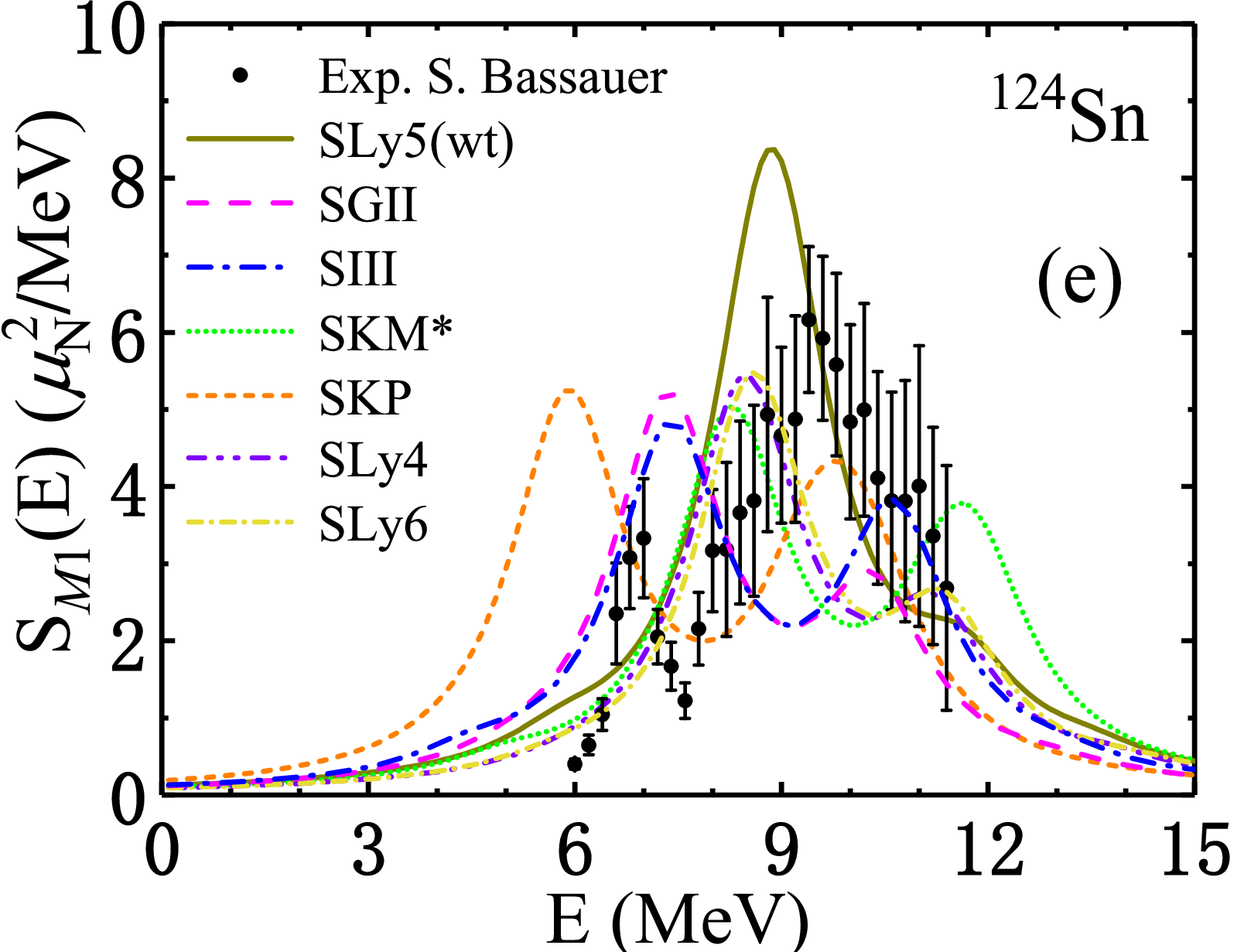}
\includegraphics[width=0.4\textwidth]{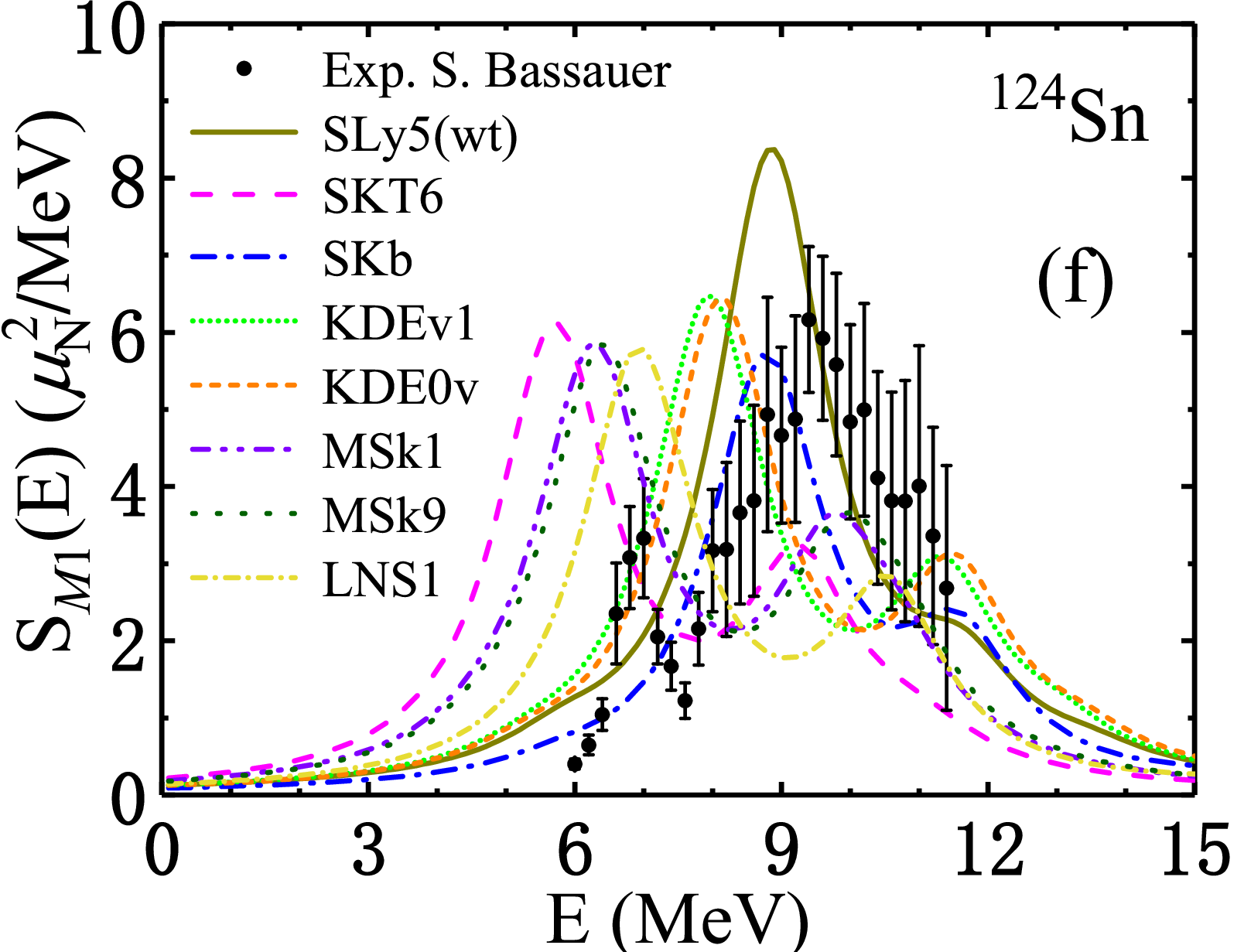}
\caption{(color online) The QRPA strength distributions of $^{124}$Sn calculated by using several  T$ij$ and other Skyrme EDFs.
In Fig.~\ref{f2} (a)-(d), the T$ij$ EDFs are employed: (a) by changing $\alpha$ for a fixed value
$\beta=-$60\ MeV$\,$fm$^5$; (b) by changing $\beta$ for a fixed value $\alpha=-$60\ MeV$\,$fm$^5$;
(c) by fixing $\alpha=\beta$=($-$60, 60, 180)\ MeV$\,$fm$^5$ that corresponds to $i=j$=(1,3,5);
 (d) corresponds to the cases $\alpha\neq\beta$ that are not shown in Figs. (a)-(c), including T$24$, T$42$, T$35$, T$53$, T$46$~and~T$64$, except T$11$.
(e) and (f) show  commonly used Skyrme EDFs without the tensor terms, except SLy5 which includes the tensor terms. Experimental data  are taken from Bassauer {\it et al.} \cite{M1exp-Sn}.}  \label{f2}
\end{figure}

We adopt various commonly used Skyrme EDFs for the $M$1 calculations of $^{124}$Sn to examine both the model dependence and the role of tensor interaction.  The adopted Skyrme EDFs with tensor terms are SLy5 with the tensor force \cite{COLO2007227} and some of the T$ij$ interactions \cite{Lesinski07}.  As representatives without the tensor
 terms,  we employ SLy4 and SLy6 \cite{Chabanat98}, SIII \cite{SIII}, SGII \cite{SGII}, SkM* \cite{SKMs}, SkP \cite{SKP},  KDE0v and KDEv1 \cite{KDE0v}, SkT6 \cite{SKT6}, MSK1 \cite{MSK13}, MSK9 \cite{MSK9},  SKb \cite{SKb}, and LNS1 \cite{LNS1} interactions.  The calculated and experimental strength distributions are shown in Fig.~\ref{f2}. Fig.~\ref{f2} (a)~(Fig.~\ref{f2} (b)) shows the results of T$ij$ family specified  by  T$1j$~(T$i1$) sets,  in which  the indices~$i$~and~$j$~refer to the coefficients of 
 the proton-neutron ($\beta$) and like-particle ($\alpha$) spin-orbit densities  in Eq.~(7),
 \begin{align}
&\alpha = \alpha_C+\alpha_T=60\ (j - 2)\ \textrm{MeV}\,\textrm{fm}^{5},\notag\\
&\beta = \beta_C+\beta_T=60\ (i - 2)\ \textrm{MeV}\,\textrm{fm}^{5}.
\end{align}
These T$ij$ family members are chosen to investigate the evolution of the strength distributions
by increasing the $\alpha$~($\beta$) value at a given  $\beta$~($\alpha$) value.
Fig.~\ref{f2} (c) shows results of the T$11$, T$33$~and~T$55$~sets, which are chosen to investigate the strength evolution
given by the parameter sets of  the $i$ = $j$ family members. Fig.~\ref{f2} (d) shows the results of other T$ij$ parameter sets, including T$24$, T$42$, T$35$, T$53$, T$46$~and~T$64$. The results of other Skyrme EDFs without the tensor terms  (except SLy5 with the tensor) are shown in Fig.~\ref{f2} (e) and (f).

The protocols for the determination of these Skyrme EDFs are as follows.
SLy5  is a  Skyrme EDF given by Lyon group, the detailed information can be found in Ref. \cite{Chabanat98}. The list of constraints used to construct the cost function $\chi^{2}$ for the minimization reads: the binding energies and the charge radii of~$^{16}$O, $^{40,48}$Ca, $^{56}$Ni, $^{132}$Sn and~$^{208}$Pb; the spin$-$orbit splitting of the neutron~$3p$~state in~$^{208}$Pb; the energy per particle in the nuclear matter ($E/A$~$\simeq$~$-$16~MeV) at the saturation density ($\rho_{0}$~$\simeq$~0.16~fm$^{-3}$), the incompressibility modulus~($K_{\infty}$~$\simeq$~230~MeV) and the symmetry energy coefficient~($a_{s}$~$\simeq$~32~MeV) at the saturation density
  of nuclear matter; the equation of state of neutron matter predicted by Wiringa $et~al.$~in Ref. \cite{Wiringa88}; the enhancement factor $\kappa$ of the Thomas-Reiche-Kuhn sum rule ($\kappa$ = 0.25); $x_{2}$ was fixed to be $-1.0$.
The SLy5 functional can be considered
as a ``standard'' Skyrme functional that performs well for many observables like masses, natural parity 
non charge-exchange
excitations, predictions of drip lines and the structure of neutron stars.  It is employed here as a benchmark of what can be obtained for $M$1 while the fit of the EDF has not been focused on spin properties.  Later, Col\`o~$et~al.$~included the tensor terms perturbatively in the SLy5 interaction in order to reproduce the evolutions of single-particle energies  of Z=50 isotopes and N=82 isotones \cite{COLO2007227}.

The T$ij$ parametrizations were proposed in Ref. \cite{Lesinski07}, where indices~$i$~and~$j$~refer to the proton-neutron ($\beta$) and like-particle ($\alpha$) coupling constants
given  in Eq.~(12).
 The fit protocol of T$ij$ sets is similar to that of SLy5 parametrization, but has  three differences: (a)~the values for~$\alpha$~and~$\beta$~were fixed beforehand  for each T$ij$ member and then other parameters were optimized for the protocol.  This means
that the tensor terms were excluded in the fit procedure but fixed {\it a priori}; (b)~the binding energies of $^{90}$Zr and $^{100}$Sn were added to the set of data; (c)~the constraint~$x_{2}=-1$~ imposed on the SLy5 parametrization
was released and the parameter $x_2$ had been included in the optimization process.
 By using these T$ij$ EDFs, we aim at pinpointing the specific effect of tensor terms on $M$1: since tensor terms affect the spin-orbit splitting, the
effect of the point (a) on $M$1 is clear, while (b) and (c) are to some extent details that do not matter too much in the present context.

\subsection{Correlations between $M$1 unperturbed energies and the spin-orbit strength $W_0$ as well as the tensor terms}
\begin{figure}[htb]
\includegraphics[width=0.5\textwidth]{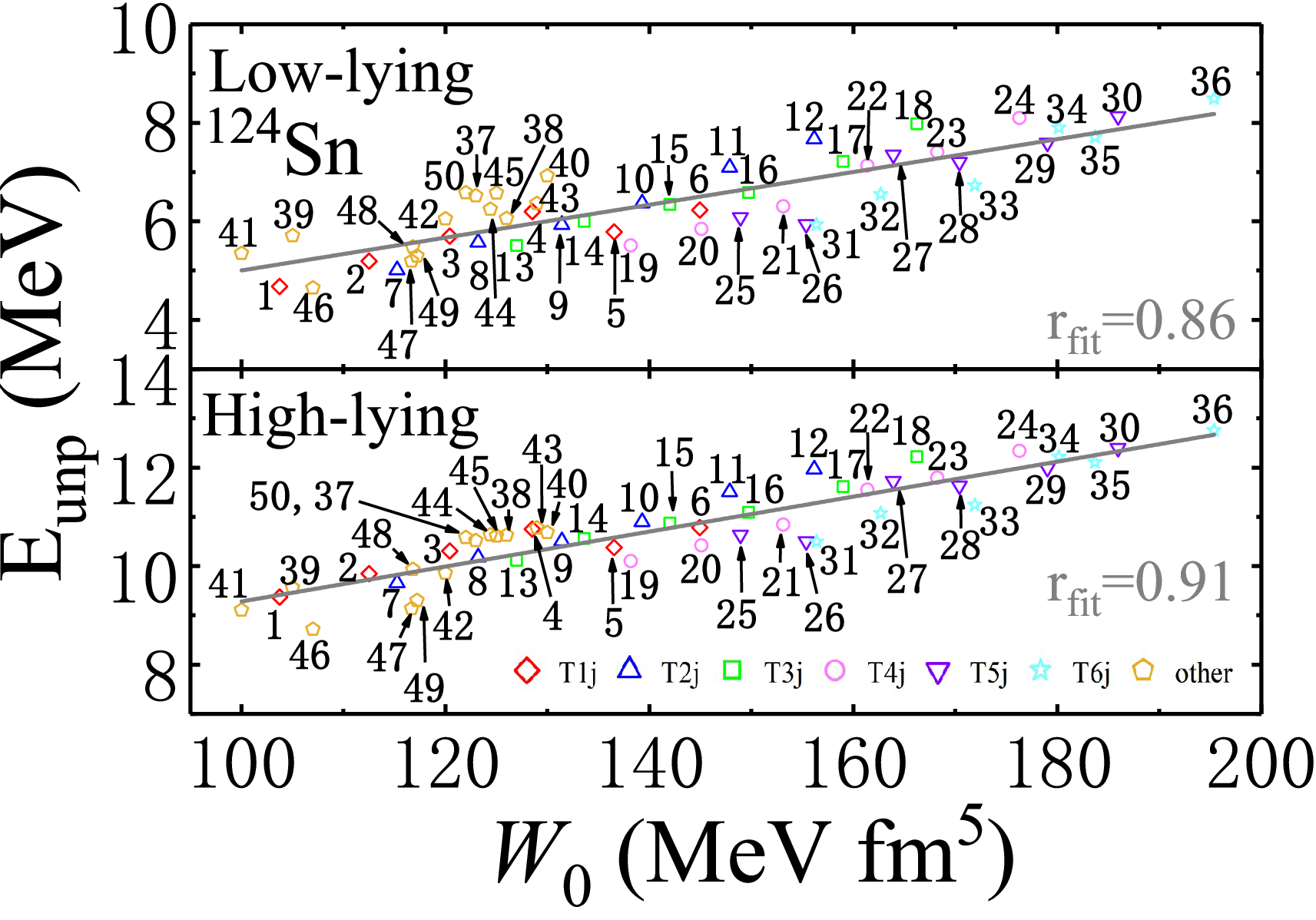}
\caption{(color online) The energies of unperturbed $M$1 peaks for $^{124}$Sn as a function of $W_{0}$. In the calculations, the Skyrme EDFs without tensor terms are employed.  The computed data points are labeled, here and in what follows, by numbers: 1~=~T11, 2~=~T12, 3~=~T13, 4~=~T14, 5~=~T15, 6~=~T16, 7~=~T21, 8~=~T22, 9~=~T23, 10~=~T24, 11~=~T25, 12~=~T26, 13~=~T31, 14~=~T32, 15~=~T33, 16~=~T34, 17~=~T35, 18~=~T36, 19~=~T41, 20~=~T42, 21~=~T43, 22~=~T44, 23~=~T45, 24~=~T46, 25~=~T51, 26~=~T52, 27~=~T53, 28~=~T54, 29~=~T55, 30~=~T56, 31~=~T61, 32~=~T62, 33~=~T63, 34~=~T64, 35~=~T65, 36~=~T66, 37~=~SLy4, 38~=~SLy5, 39~=~SGII, 40~=~SKM*, 41~=~SKP, 42~=~SIII, 43~=~KDE0v, 44~=~KDEv1, 45~=~SKb, 46~=~SKT6, 47~=~MSK1, 48~=~LNS1, 49~=~MSK9, 50~=~SLy6. The grey lines correspond to the results of the linear fits. }  \label{f3}
\end{figure}

In order to clarify the role of the tensor terms of the Skyrme EDFs,  we study first  the correlation
between the unperturbed energies of $M$1 states and the spin$-$orbit strength~$W_{0}$ in the case of
$^{124}$Sn.  When the tensor force is not involved, it is expected that the spin$-$orbit splitting is mainly governed by the spin$-$orbit strength $W_{0}$
together with some contributions from $\alpha_C$ and $\beta_C$ in Eq. (9). Since the $M$1 unperturbed excitation energy is mainly given by the
$p-h$ type excitation between the spin-orbit partners, the excitation energies of $M$1 peaks are sensitive to the spin$-$orbit strength $W_{0}$. In Fig. \ref{f3}, we show the unperturbed low-lying and high-lying $M$1 states for 50 different Skyrme EDFs.
There are two main unperturbed configurations in $^{124}$Sn: the proton configuration 1$g_{9/2}$$\rightarrow$1$g_{7/2}$ and the neutron one 1$h_{11/2}$$\rightarrow$1$h_{9/2}$.
The former corresponds to the low-lying $M$1 state, while the latter corresponds to the high-lying one.
As displayed in Fig.~\ref{f3}, it is found that there are clear linear correlations between the energies of unperturbed $M$1 peaks and $W_{0}$.  The correlation coefficients are
$r_{fit}$=0.86 and 0.91 for the low-lying and high-lying states, respectively.

\begin{figure}[htb]
\includegraphics[width=0.5\textwidth]{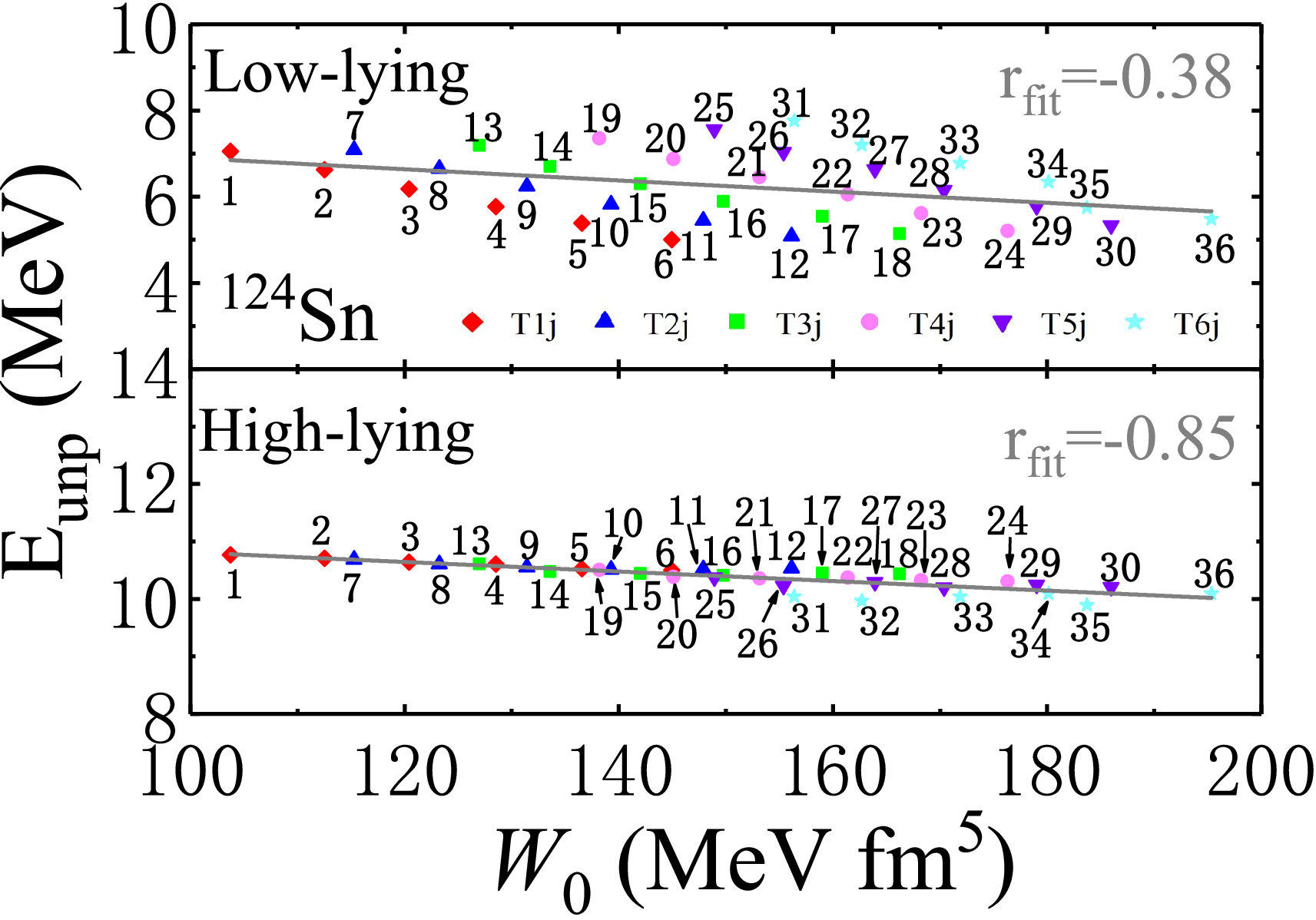}
\caption{(color online) The energies of low-lying and high-lying unperturbed $M$1 peaks for $^{124}$Sn as a function of $W_{0}$, calculated by using~T$ij$~sets with tensor force. }  \label{f4}
\end{figure}

Next we demonstrate what happens in the case of EDFs with tensor terms, i.e., for the T$ij$  family.
The $W_0$ dependence of the energies of low-lying and high-lying $M$1 states is  shown for all the parameter sets of
the T$ij$  family in Fig. \ref{f4}.   Here we do not see any clear  correlation between the energies and the spin-orbit coupling strength $W_0$, in contrast to Fig.~\ref{f3}.  Curiously, even a weak anti-correlation between the energies and $W_0$ appears in Fig. \ref{f4}.  This is because the spin-orbit splitting
of the like-particle has two contributions, from $W_0$ and the like-particle
spin-orbit current weighted by
$\alpha$ in Eq. (7).  In the optimization process, the value $W_0$ is optimized for a given  $\alpha$ value to reproduce the empirical spin-orbit splitting
in the protocol. Because of this cross-talk feature  of $W_0$ and $\alpha$, the results of the $M$1 energies do not
show any linear dependence with positive slope
 on  $W_0$.

\begin{figure}[htb]
\includegraphics[width=0.5\textwidth]{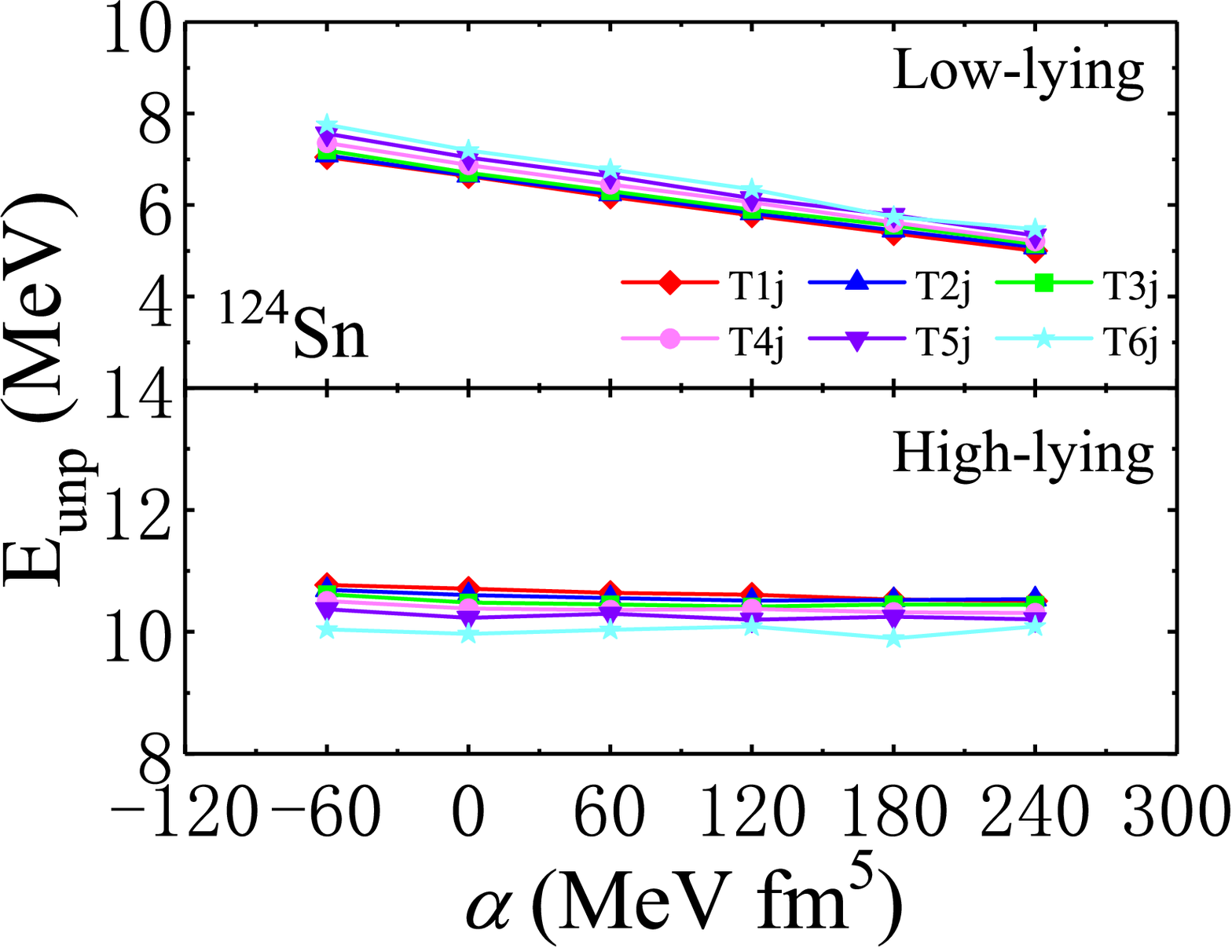}
\caption{(color online) The energies of low-lying and high-lying unperturbed $M$1 peaks for $^{124}$Sn are shown as a function of~$\alpha$~by fixing $\beta$.}
 \label{f5}
\end{figure}

We have also calculated how the excitation energies of the unperturbed $M$1 peaks depend on the
strength of the tensor terms, $\alpha$ or $\beta$.
We find an anti-correlation between $\alpha$ and
the energies of the low-lying $M$1 states, as shown in Fig. \ref{f5}, but a very weak correlation
between the $M$1 energies and $\beta$. The anti-correlation on $\alpha$ is due to the feature of this
like-particle term,
which has an opposite sign with respect to the spin-orbit strength $W_0$, i.e., the larger the value
of $\alpha$
the smaller the spin-orbit splitting is.

\subsection{QRPA correlations and tensor terms}

\begin{figure}[htb]
\includegraphics[width=0.4\textwidth]{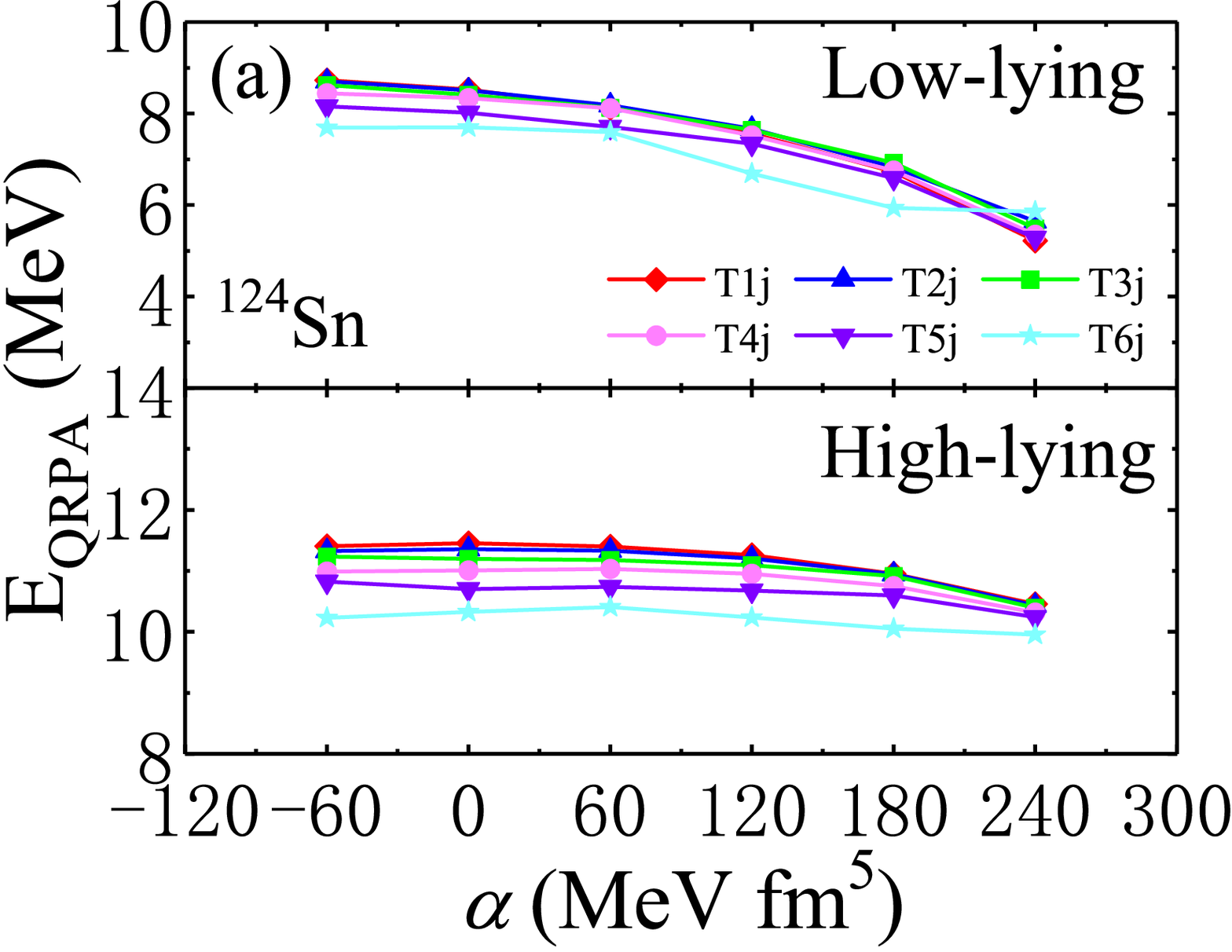}
\includegraphics[width=0.4\textwidth]{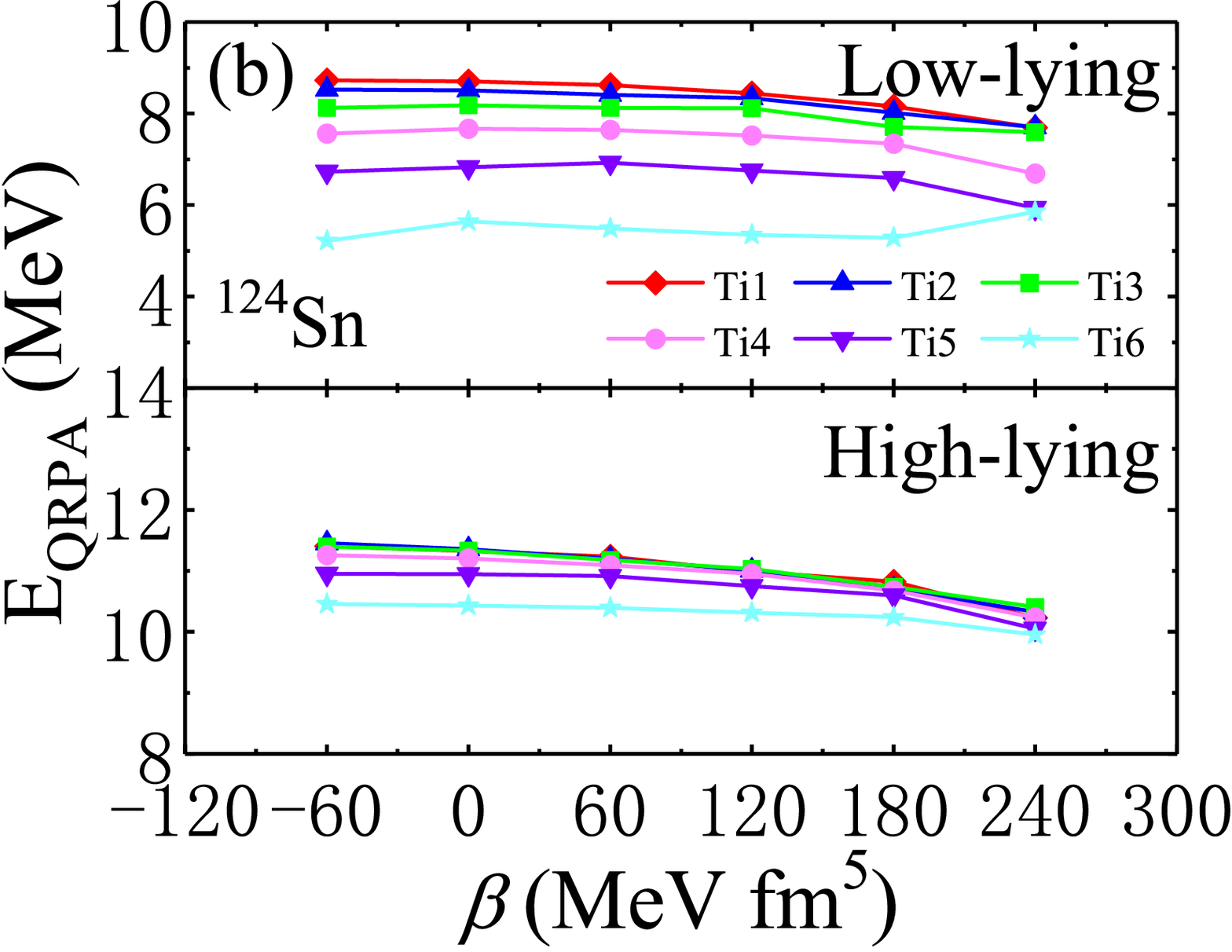}
\caption{(color online) The QRPA energies of low-lying and high-lying $M$1 peaks for $^{124}$Sn as a function of~$\alpha$~for each fixed~$\beta$~(figure~(a)), and as a function of~$\beta$~for each fixed~$\alpha$~(figure~(b)). The theoretical values are calculated by using the T$ij$ parameter sets with tensor force.}
 \label{f6}
\end{figure}

We now study  how the QRPA energies correlate with the tensor terms and also
with the main part of the QRPA residual interactions, which are associated with the Landau parameters
$G_0$ and $G_0'$. Figure \ref{f6} (a) and (b) show the correlations  between the low-lying and high-lying $M$1 states and the tensor terms $\alpha$  and $\beta$, respectively. In the panel (a), the low-lying $M$1 peaks show a clear
anti-correlation with the value of $\alpha$, which can be understood because of a smaller spin-orbit
splitting caused by a larger $\alpha$ value, as discussed above. On the other hand, the correlation
is rather weak for the high-lying states. It might be due to the strong QRPA correlations for the high-lying states
as will be discussed below. In the right panel, the correlations between the $M$1 energies and $\beta$
value is very modest,
showing a small anti-correlation effect.

\begin{figure}[htb]
\includegraphics[width=0.5\textwidth]{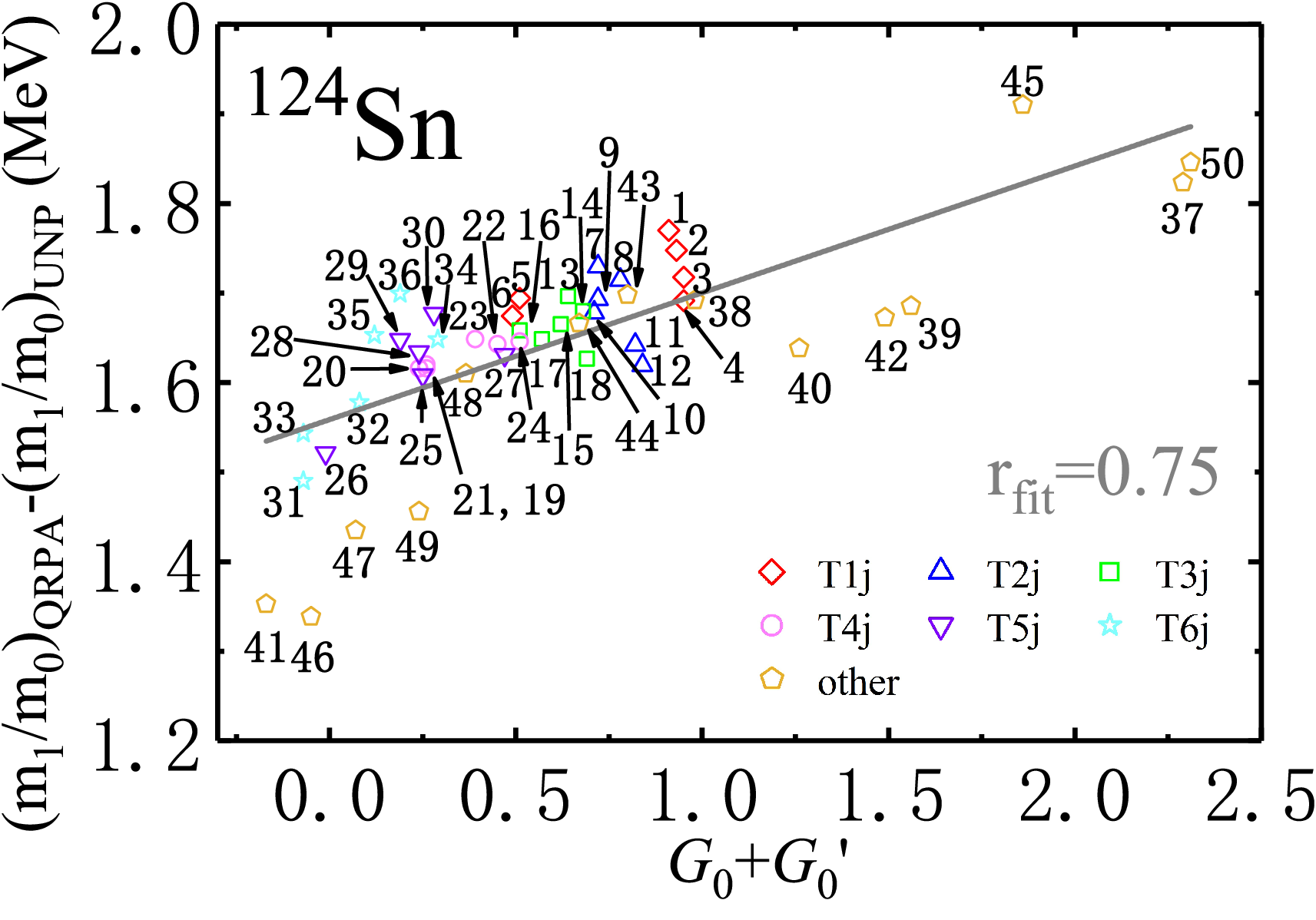}
\caption{(color online) The difference of the centroid energies of the QRPA and
unperturbed response for $^{124}$Sn as a function of $G_{0}+G_{0}'$.
All sets of $Tij$ family members and other Skyrme forces are adopted in the QRPA calculations. The grey line corresponds to the result of linear fit.}  \label{f7}
\end{figure}

In Fig. \ref{f7}, we study the correlation between the sum of the Landau parameters $G_{0}+G_{0}'$ and the difference of centroid energies of QRPA and unperturbed strengths, where the centroid energy is defined
by the ratio of the energy-weighted sum rule $m_1$ to the non-energy-weighted sum rule $m_0$, ($m_1$/$m_0$)$_{\text{QRPA}}$ and ($m_1$/$m_0$)$_{\text{UNP}}$ in the figure are the calculated centroid energies of QRPA and unperturbed strengths, respectively. We can see a clear linear correlation between the sum of Landau parameters and the centroid energies. This correlation can be understood within a simple two levels model in the following.

In the QRPA response, there are two main unperturbed configurations in $^{124}$Sn: the proton configuration 1$g_{9/2}$$\rightarrow$1$g_{7/2}$ and the neutron one 1$h_{11/2}$$\rightarrow$1$h_{9/2}$, as already mentioned
in the previous subsection. The energy of the proton (neutron) two-quasiparticle configuration can be represented as $\varepsilon_{p}\left(\varepsilon_{n}\right)$. The QRPA matrix is, then, schematically expressed as
\begin{equation}
\left(
  \begin{array}{cc}
\varepsilon_{p}+v_{0} & v_{1} \\
v_{1} & \varepsilon_{n}+v_{0}
  \end{array}
\right),\notag
\end{equation}
where  $v_{0} = G_{0} + G_{0}^{\prime}$  is the pp or nn interaction while  $v_{1}  = G_{0} - G_{0}^{\prime}$  is the pn interaction.
We remind that
$G_{0}$ and $G_{0}^{\prime}$~represent the Landau parameters in the spin channel~$\vec{\sigma}_{1}\cdot\vec{\sigma}_{2}$~and spin-isospin channel~$(\vec{\sigma}_{1}\cdot\vec{\sigma}_{2})(\vec{\tau}_{1}\cdot\vec{\tau}_{2})$, respectively.
If we diagonalise the matrix, the two eigenvalues can be written as
\begin{align}
\hbar \omega_{1} = \frac{\varepsilon_{p}+\varepsilon_{n}}{2}+v_{0}-\frac{\sqrt{\left(\varepsilon_{p}+\varepsilon_{n}\right)^{2}+4 v_{1}^{2}}}{2}, \notag\\
\hbar \omega_{2} = \frac{\varepsilon_{p}+\varepsilon_{n}}{2}+v_{0}+\frac{\sqrt{\left(\varepsilon_{p}+\varepsilon_{n}\right)^{2}+4 v_{1}^{2}}}{2}. \notag
\end{align}
On the other hand, according to the definition of the centroid energy
\begin{align}
&\left(\frac{m_{1}}{m_{0}}\right)_{\mathrm{QRPA}}=\frac{1}{2}\left(\hbar \omega_{1}+\hbar \omega_{2}\right), \notag\\
&\left(\frac{m_{1}}{m_{0}}\right)_{\mathrm{UNP}}=\frac{1}{2}\left(\varepsilon_{p}+\varepsilon_{n}\right). \notag
\end{align}
Therefore, there is a correlation:
$$
\left(\frac{m_{1}}{m_{0}}\right)_{\mathrm{QRPA}}-\left(\frac{m_{1}}{m_{0}}\right)_{\mathrm{UNP}}= G_{0} + G_{0}^{\prime}.
$$
This positive correlation is clearly demonstrated in Fig.~\ref{f7}.

We will now compare the calculated $M$1 strength distributions of
$^{124}$Sn with the experimental data obtained by $(p,p')$ scattering in Ref. \cite{M1exp-Sn} .
The QRPA strength distributions of $^{124}$Sn calculated by using several T$ij$ and other Skyrme EDFs are shown
in Fig. \ref{f2}.
In Fig.~\ref{f2} (a)-(d), the T$ij$ EDFs are employed: (a) changing $\alpha$ for a fixed value
$\beta=-$60 MeV$\,$fm$^5$; (b) changing $\beta$ for a fixed value $\alpha=-$60 MeV$\,$fm$^5$;
(c) in the case $\alpha=\beta=$ ($-$60, 60, 180) MeV$\,$fm$^5$, corresponding to $i=j$=(1,3,5);
(d) in the cases $\alpha\neq\beta$ that are not shown in Figs. (a)-(c), including T$24$, T$42$, T$35$, T$53$, T$46$~and~T$64$, except T$11$.
(e) and (f) in the case of commonly used Skyrme EDFs without the tensor terms, except SLy5 which has the tensor terms.

As expected from Fig. \ref{f6} (a), the peak position of the $M$1 strength becomes lower for larger $\alpha$.
The same trend can be seen also in Fig. \ref{f6} (b) when varying the value of $\beta$, while the change
of peak energy is rather modest. One can find the same trend also in Figs. \ref{f2} (a) and (b), i.e.,
the larger tensor terms give lower peak energies. In Figs. \ref{f2} (e) and (f), the results depend
on both the Landau parameters and the spin-orbit coupling $W_0$.

Eventually, from Fig~\ref{f2}, it is found that the sets T11 and SLy5 with tensor force
give better description of the strength distribution in $^{124}$Sn compared to the other parameter
sets, in terms of the peak height and the peak position. Because of this reason, in the following, the T11
and SLy5 Skyrme EDFs with and without tensor terms \cite{Chabanat98,COLO2007227,Lesinski07}
will be studied in more detail.
Table I displays the values of $T$, $U$, $\alpha$, $\beta$, $\alpha_C$, $\beta_C$, $\alpha_T$ and $\beta_T$
in Eqs. (7), (9) and (10) for the Skyrme parameter sets T11 and SLy5. It is found that the values of $\alpha$ of the two parameter sets are
negative. $\alpha$ can be positive in some T$ij$ sets, by definition, from Eq. (12).
As a counter example to T$11$ and SLy5, the T15 parameter set is also chosen in the following calculations for $M$1 states.

\begin{table}[htb]
\caption{Parameters of the tensor terms and $J^2$ terms in units of MeV$\,$fm$^{5}$. }\label{t1}
\centering
\setlength\tabcolsep{2mm}{
\begin{tabular}{ccccccccc}\\
\hline\hline
\specialrule{0em}{2pt}{2pt}
  &  $\textit{T}$~~~  & $\textit{U}$ & $\alpha$ & $\beta$  & $\alpha_C$ & $\beta_C$ & $\alpha_T$ & $\beta_T$ \\
\specialrule{0em}{2pt}{2pt}
\hline
\specialrule{0em}{2pt}{2pt}
SLy5 &  888.0 &  $-$408.0  & $-$89.8  & 51.1 & 80.2 & -48.9 & -170.0 & 100.0    \\
\specialrule{0em}{1pt}{1pt}
T11 &  258.9 &  $-$342.8  & $-$60.0  & $-$60.0  & 82.8 &-42.5 & -142.8 & -17.5  \\
\specialrule{0em}{1pt}{1pt}
T15 &  $-$500.9 & 173.3  & 180.0  & $-$60.0 & 107.8  & 8.3  &72.2  & -68.3    \\
\specialrule{0em}{2pt}{2pt}
\hline\hline
\end{tabular}
}
\end{table}

\begin{figure}[htb]
 \includegraphics[width=0.4\textwidth]{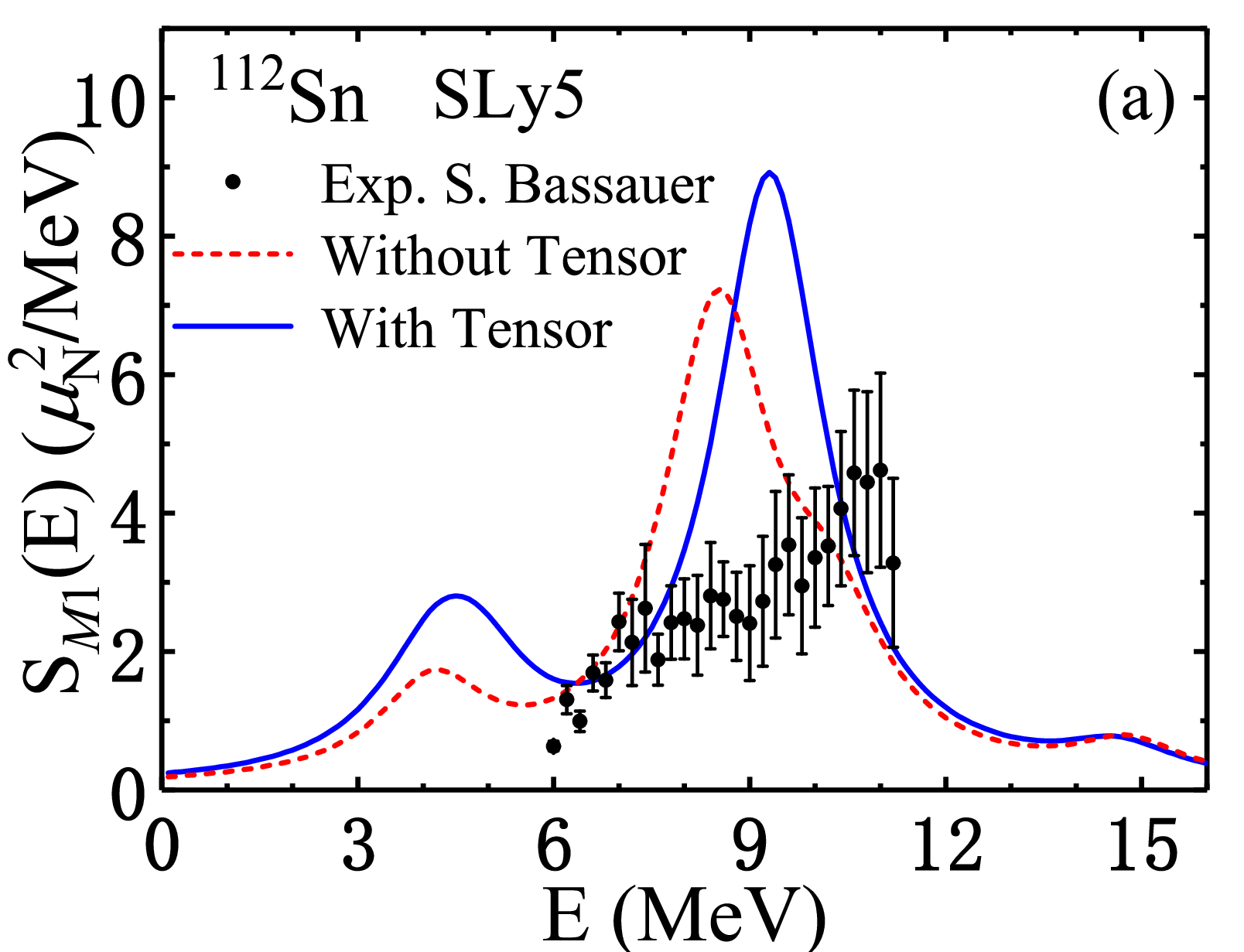}
 \includegraphics[width=0.4\textwidth]{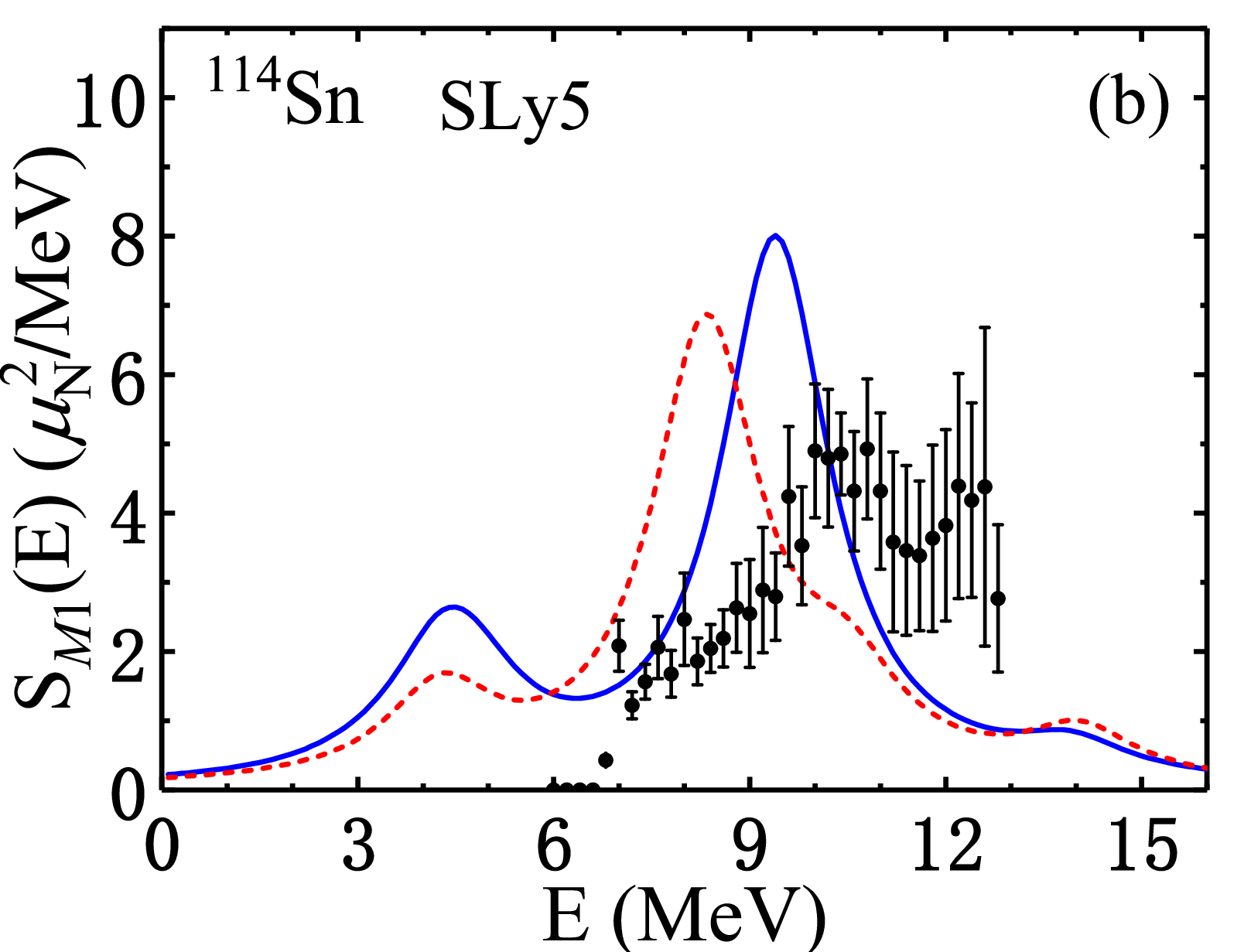}
 \includegraphics[width=0.4\textwidth]{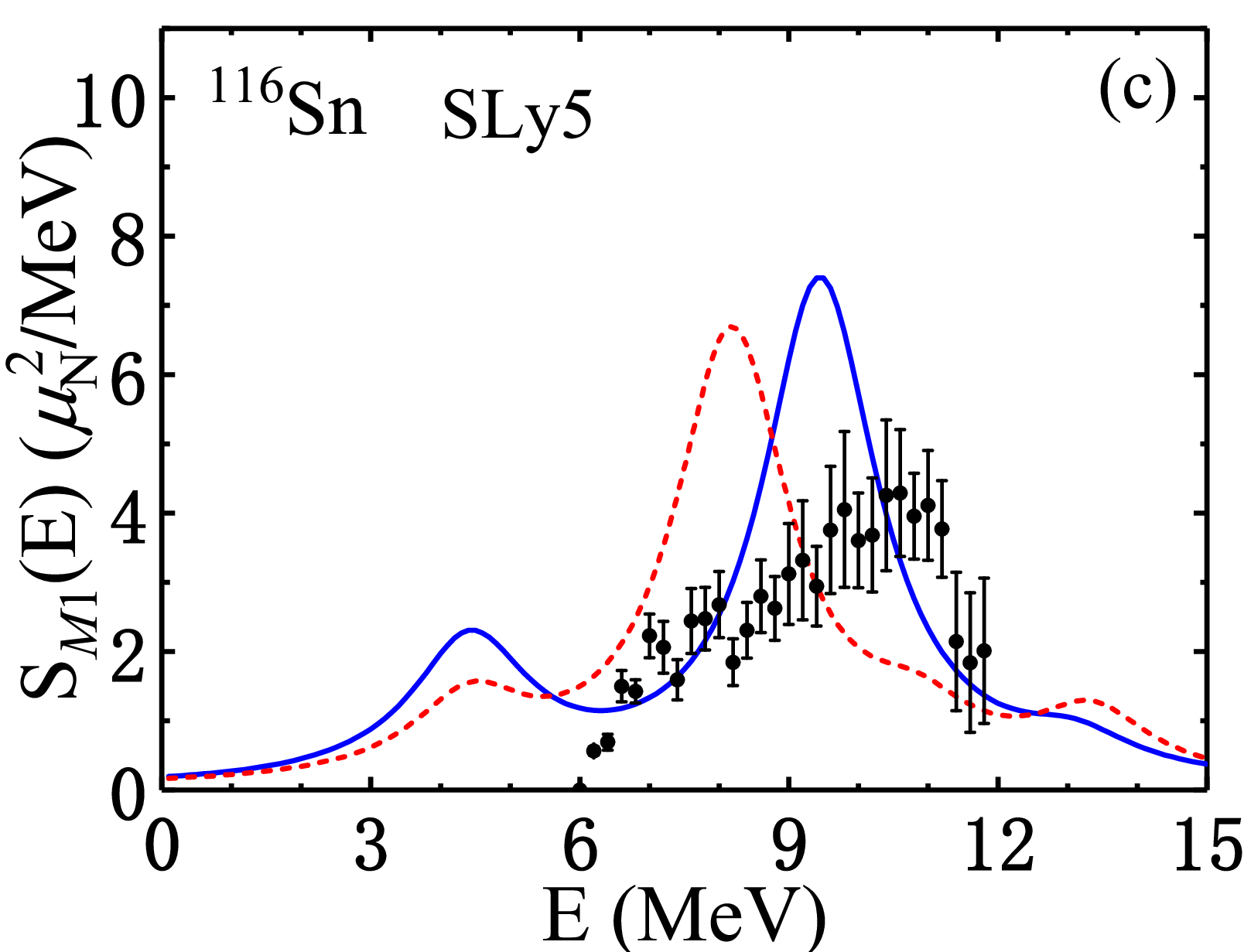}
 \includegraphics[width=0.4\textwidth]{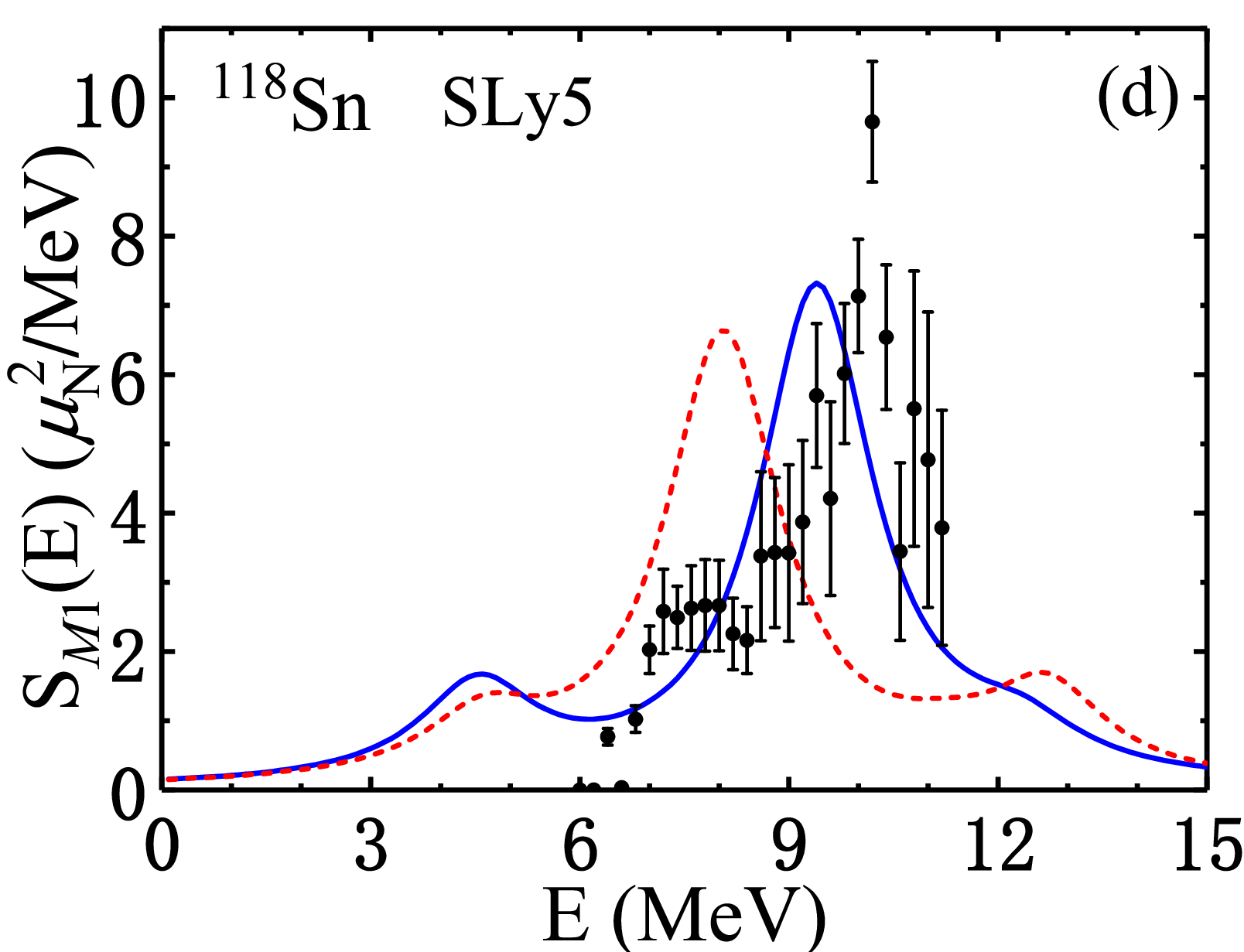}
 \includegraphics[width=0.4\textwidth]{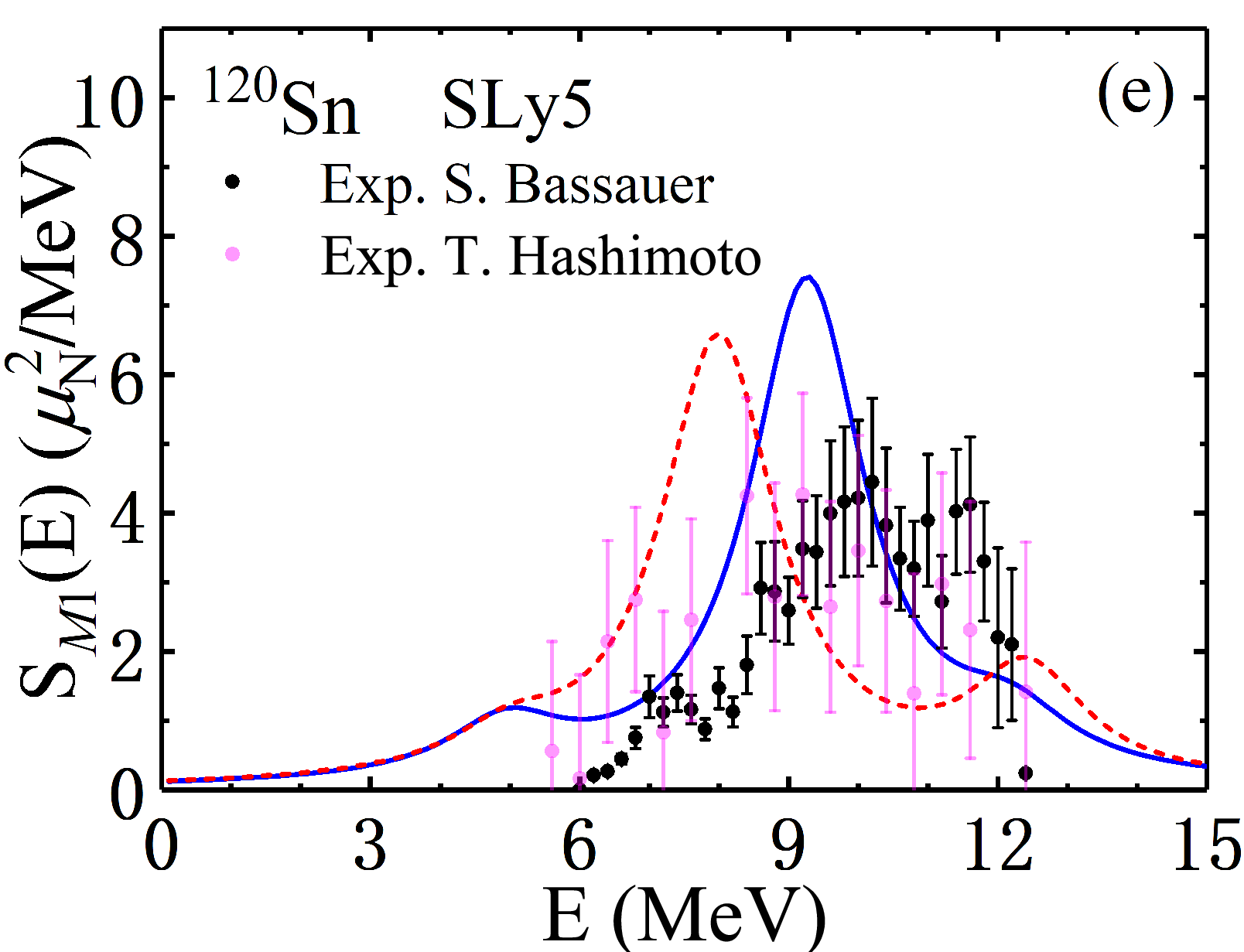}
 \includegraphics[width=0.4\textwidth]{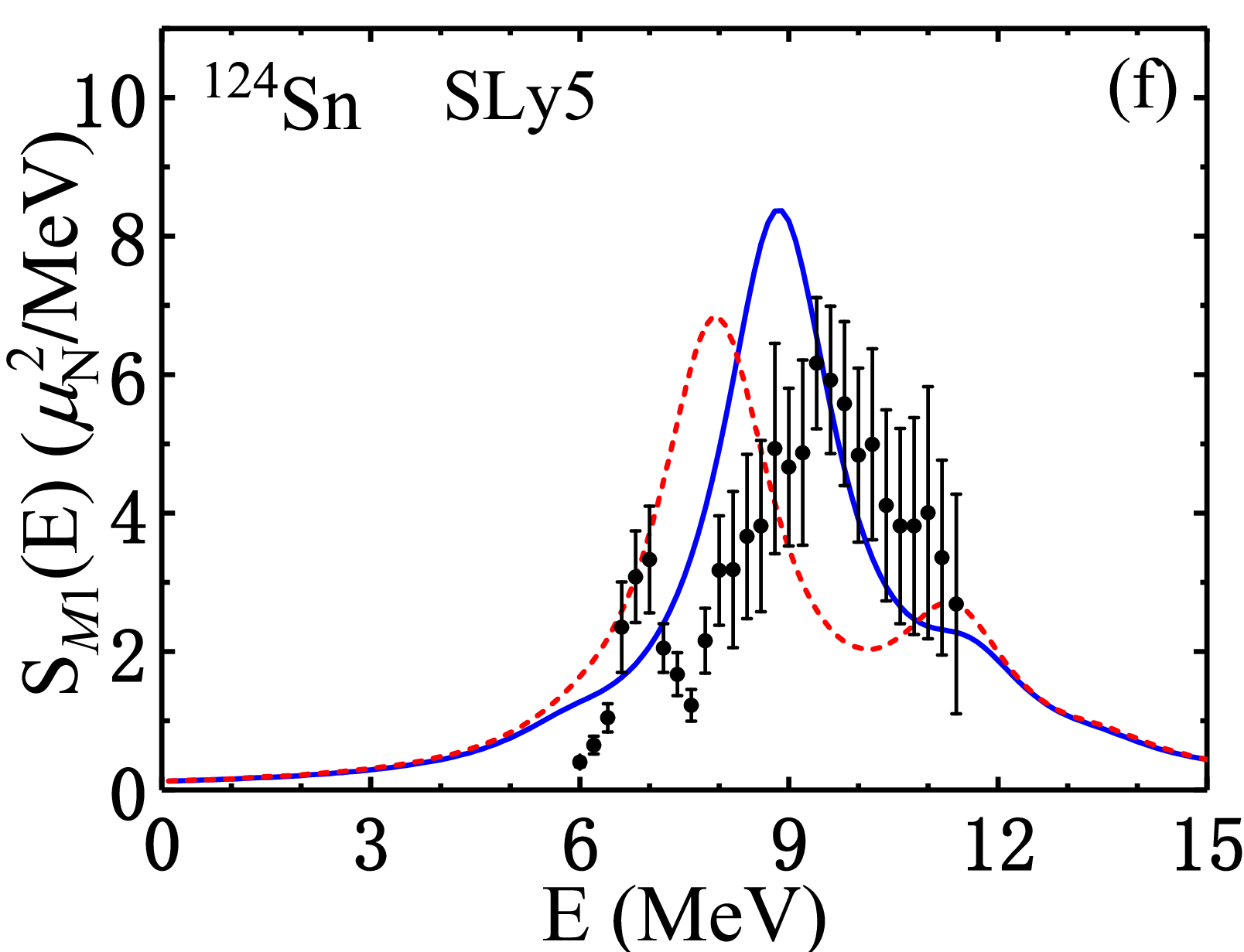}
\caption{(color online) The QRPA strength distributions of $^{112-120, 124}$Sn, calculated by using the SLy5 Skyrme interaction. The results with and without tensor terms are both shown, and compared with the experimental data \cite{M1exp-Sn120,M1exp-Sn}. Calculated strengths are convoluted by a Lorentzian shape with a width of  2.0 MeV.}\label{f8}
\end{figure}

\begin{figure}[htb]
 \includegraphics[width=0.4\textwidth]{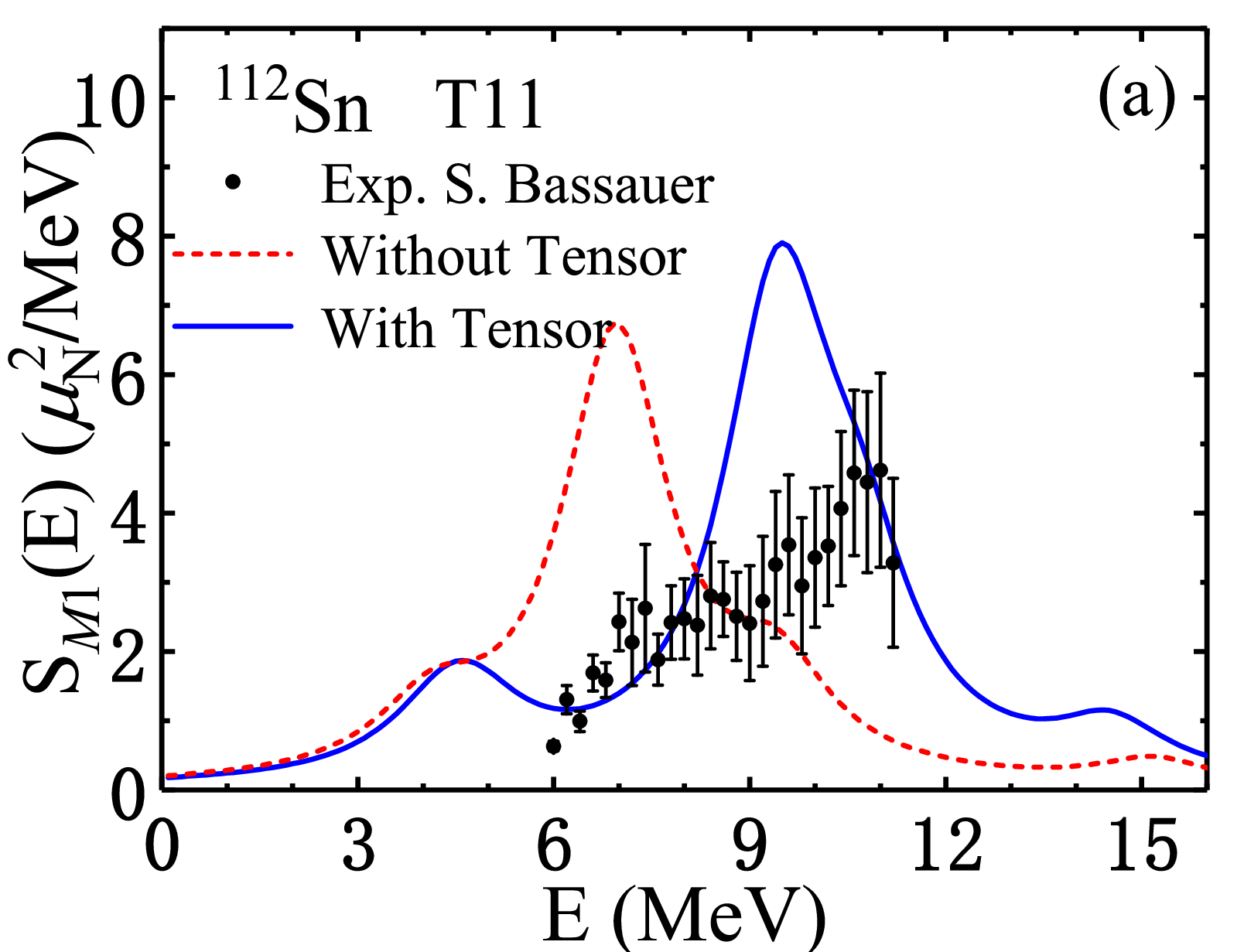}
 \includegraphics[width=0.4\textwidth]{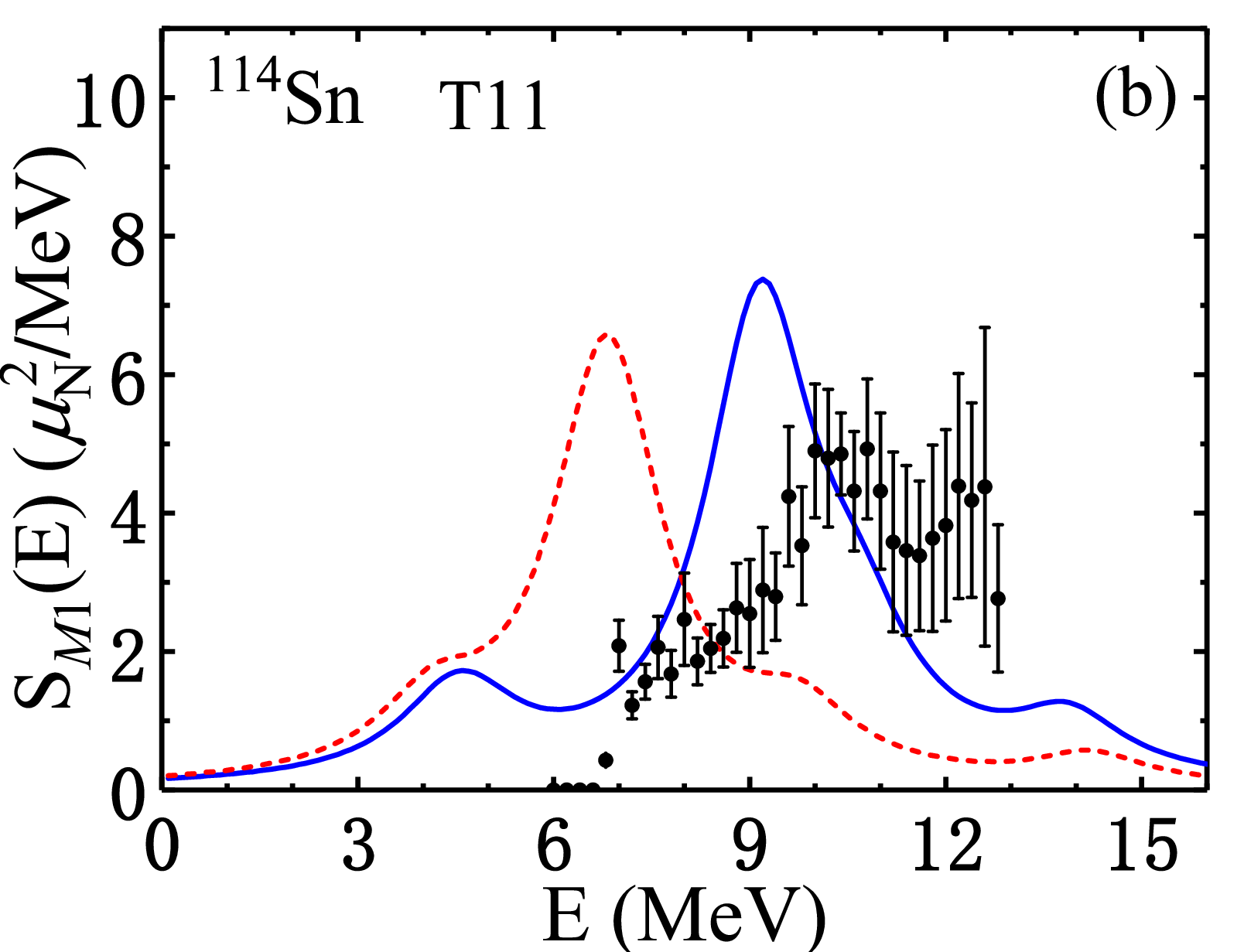}
 \includegraphics[width=0.4\textwidth]{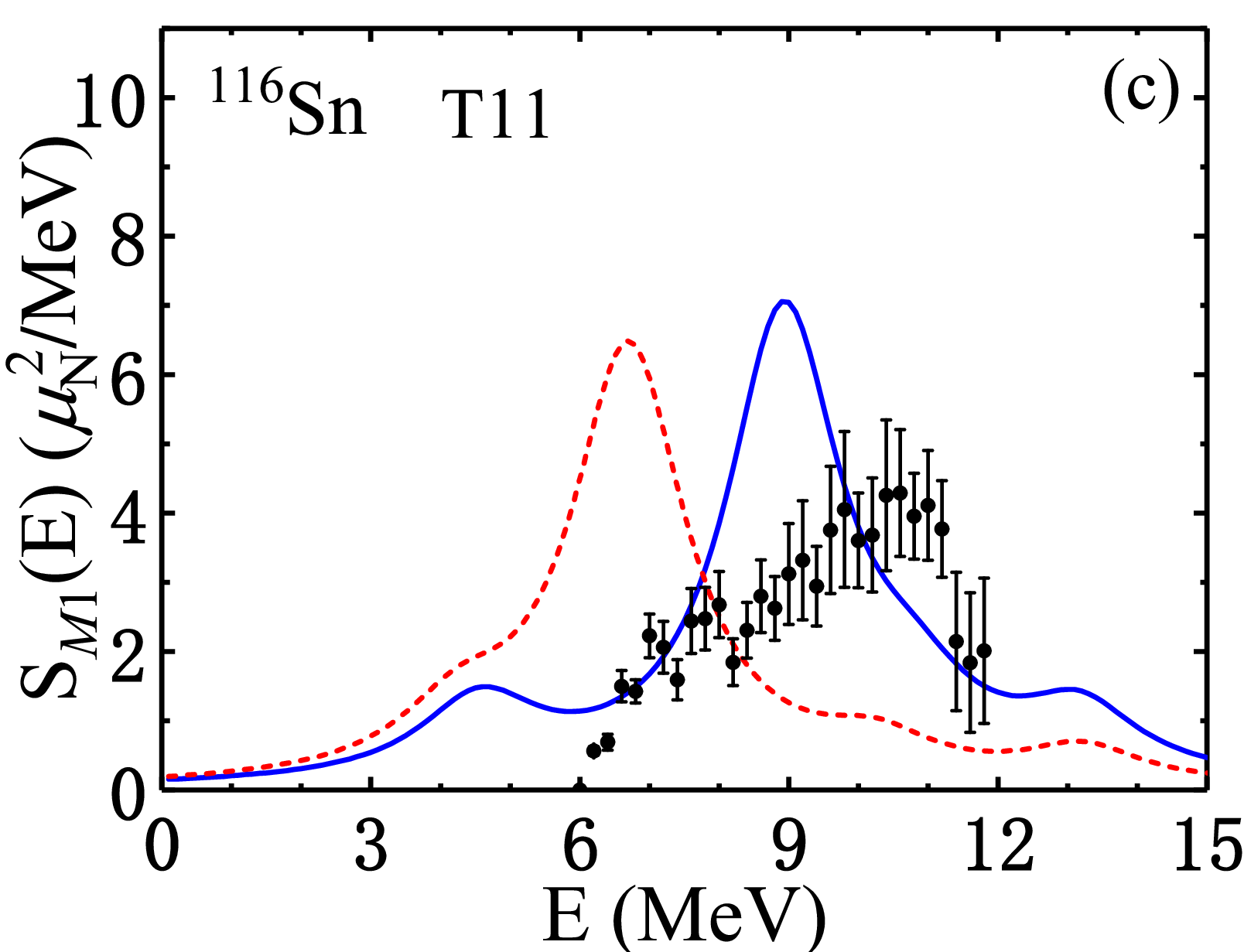}
 \includegraphics[width=0.4\textwidth]{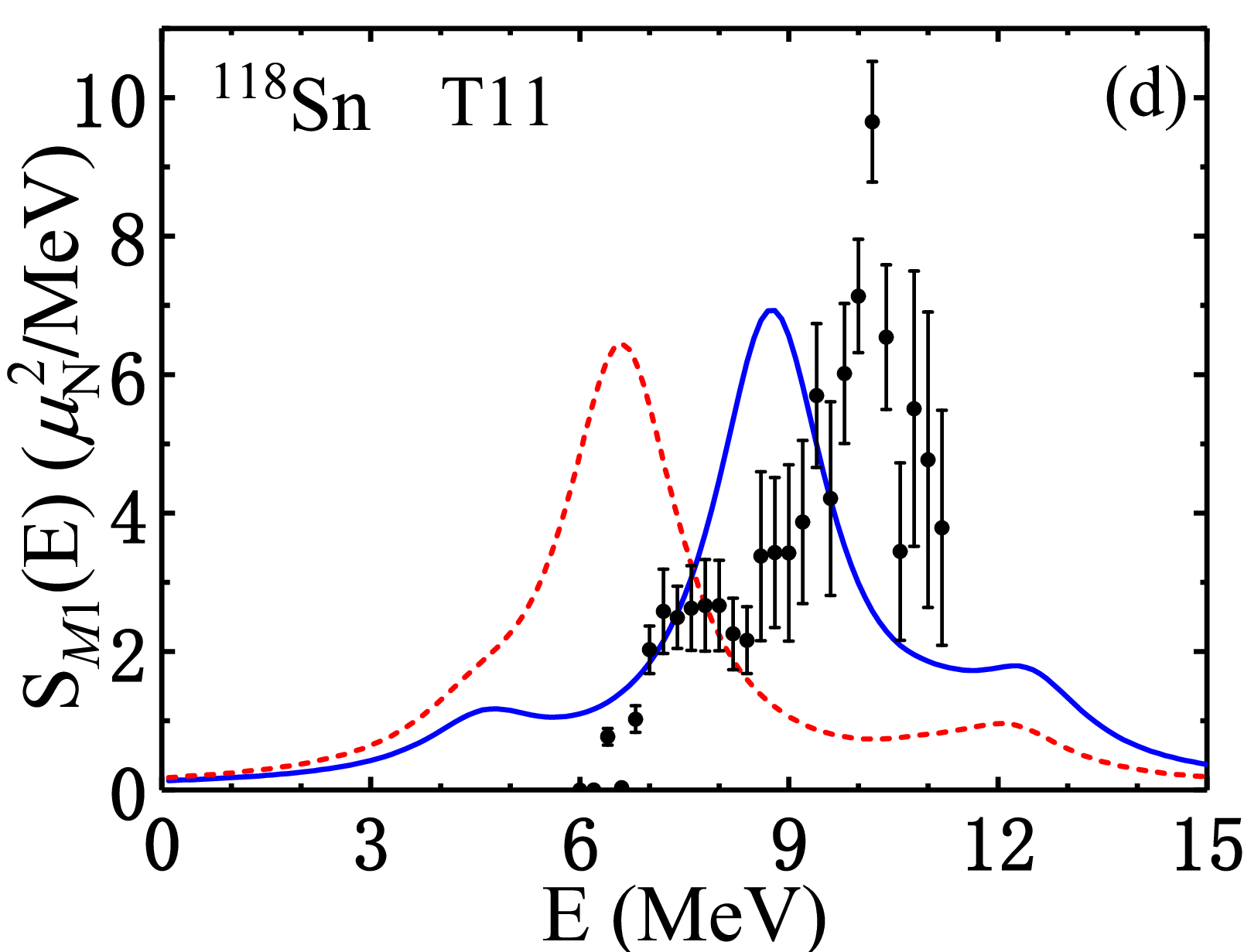}
 \includegraphics[width=0.4\textwidth]{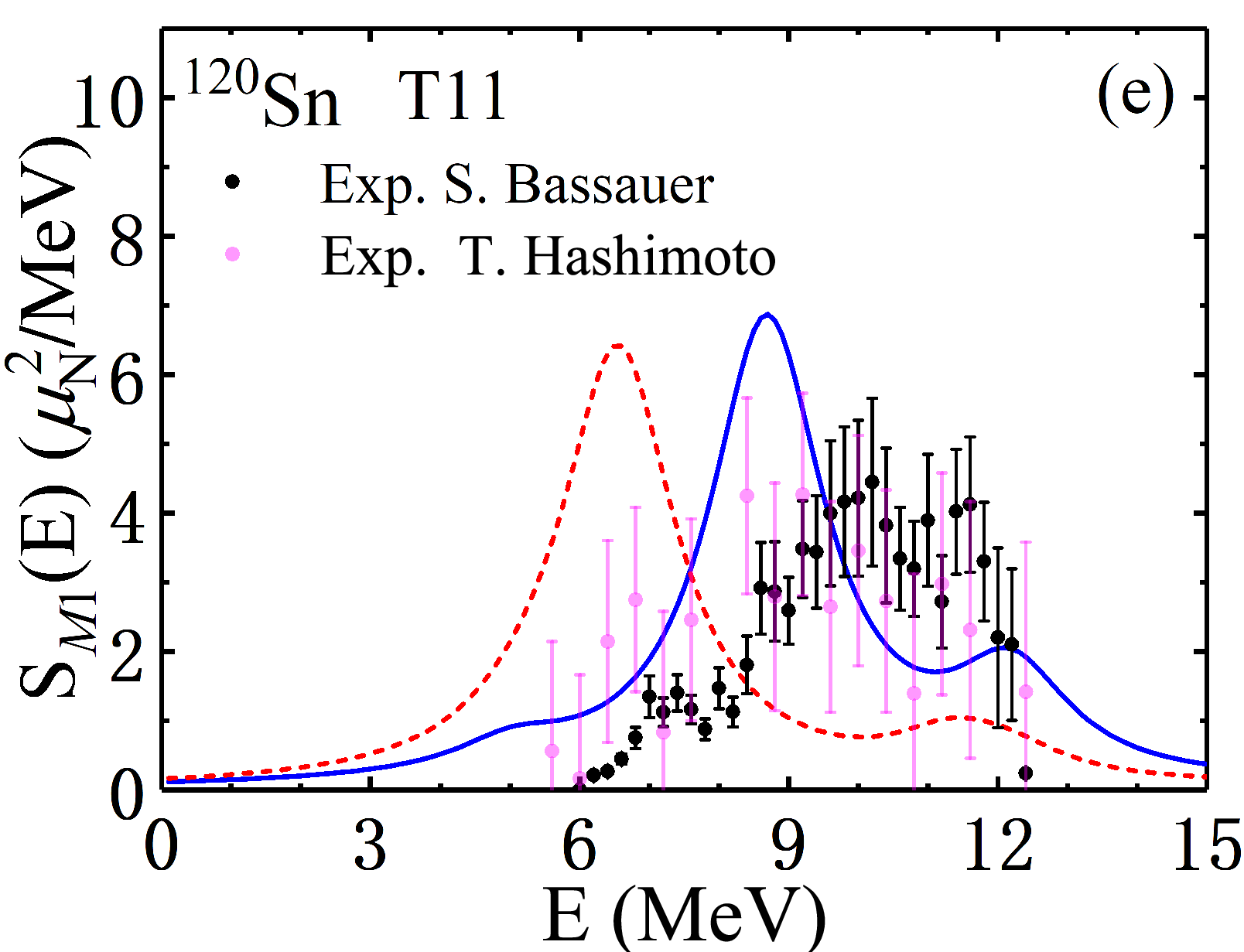}
 \includegraphics[width=0.4\textwidth]{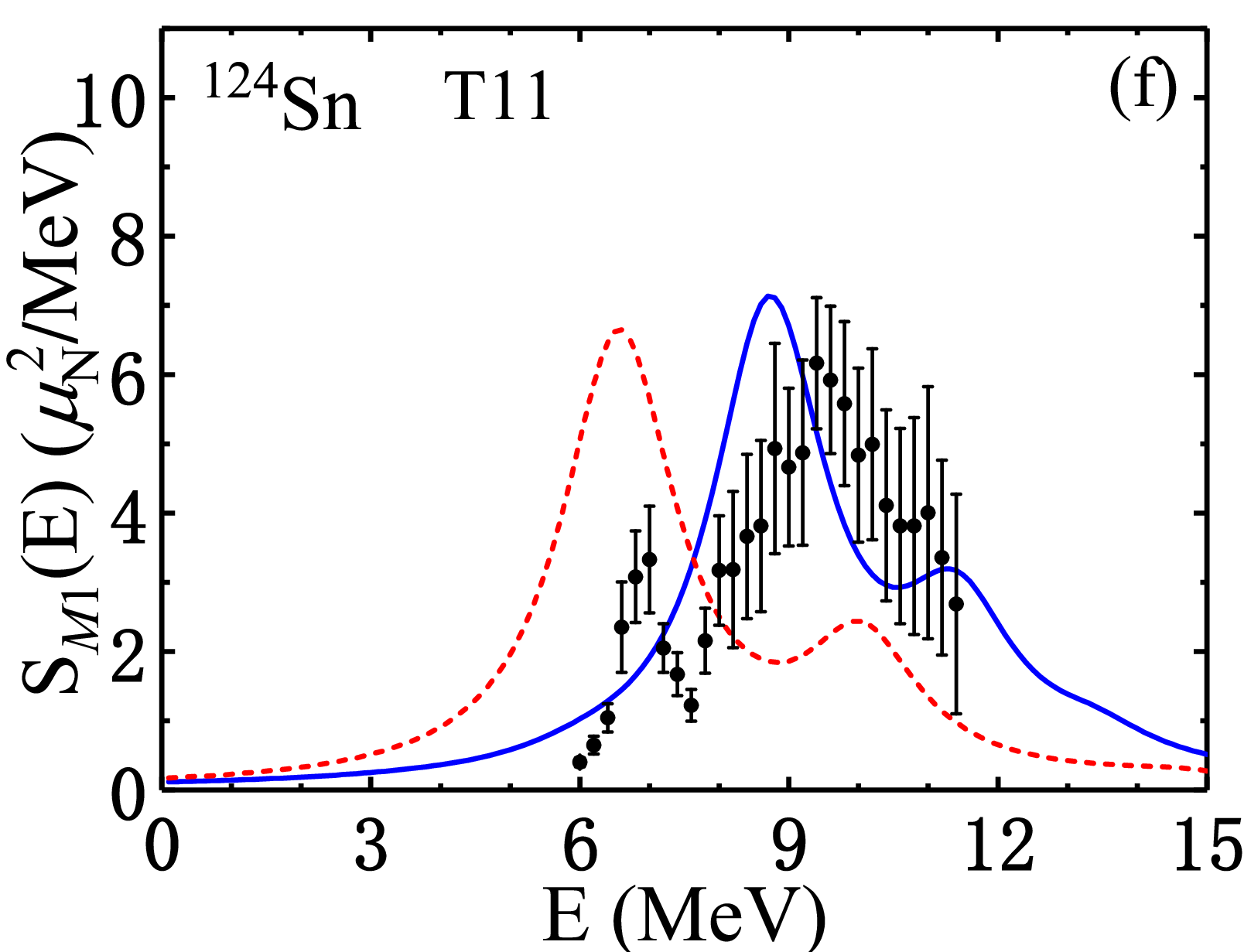}
\caption{(color online) The same as Fig. 8, but calculated with T11.}\label{f9}
\end{figure}

\begin{figure}[htp]
 \includegraphics[width=0.4\textwidth]{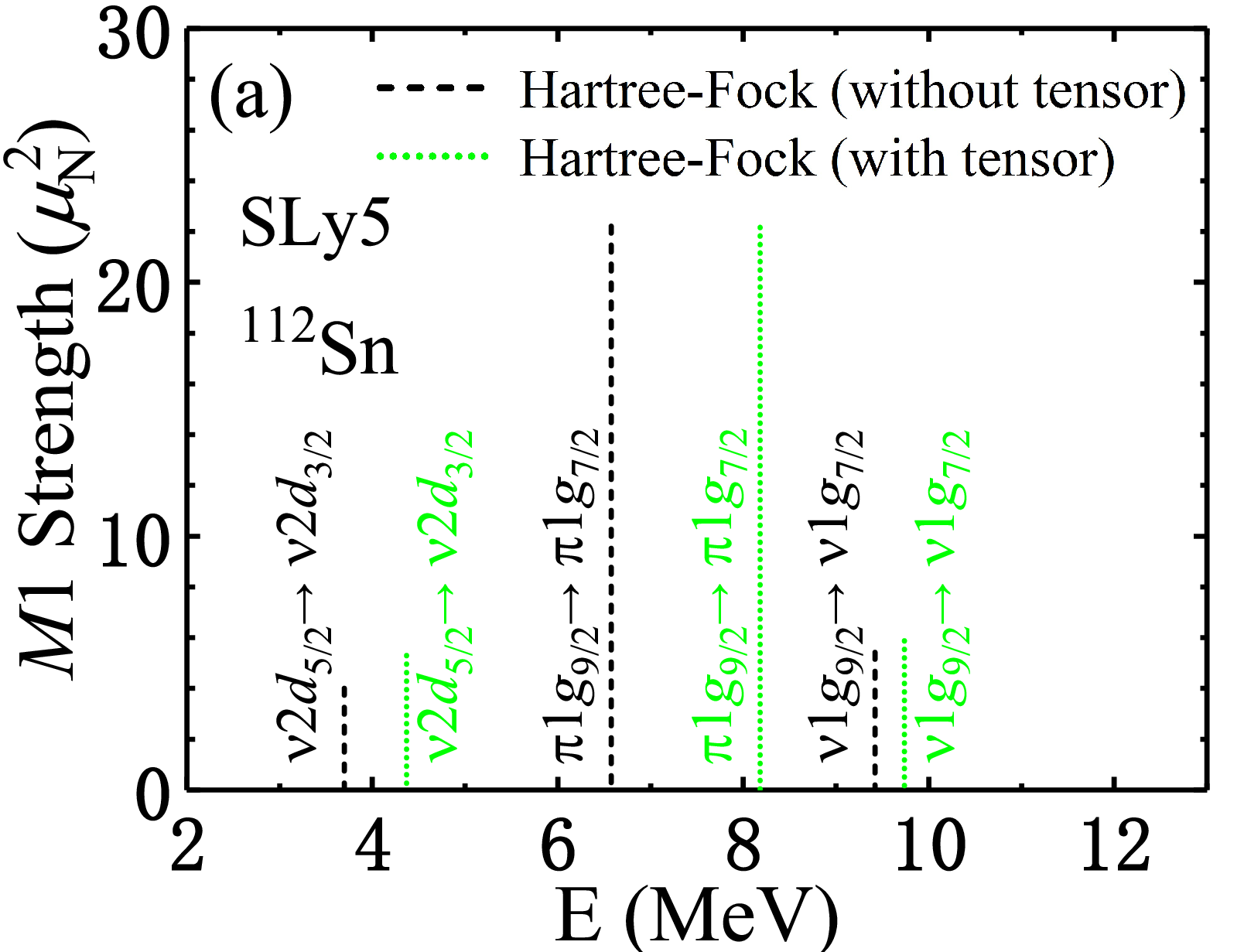}
 \includegraphics[width=0.4\textwidth]{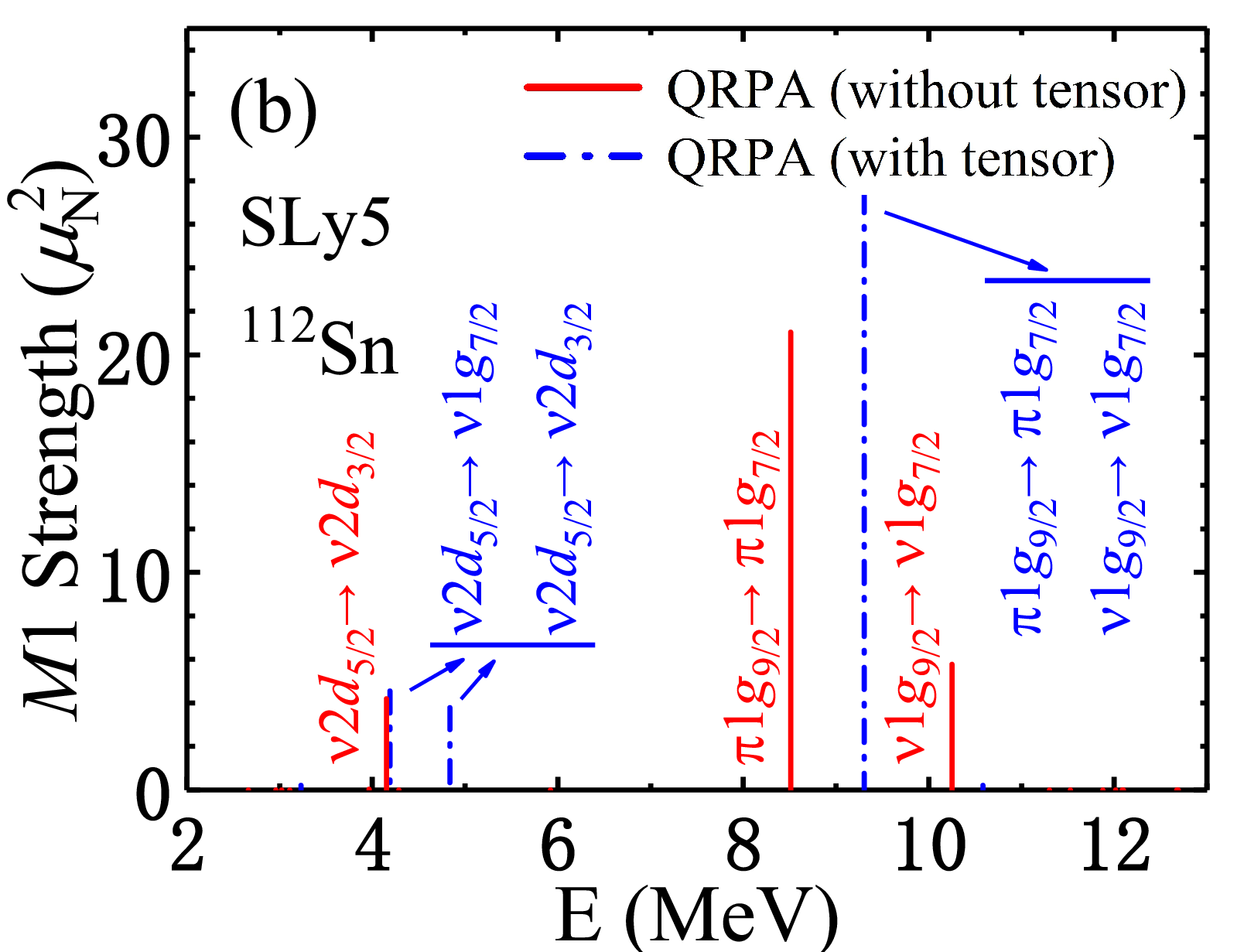}
 \includegraphics[width=0.4\textwidth]{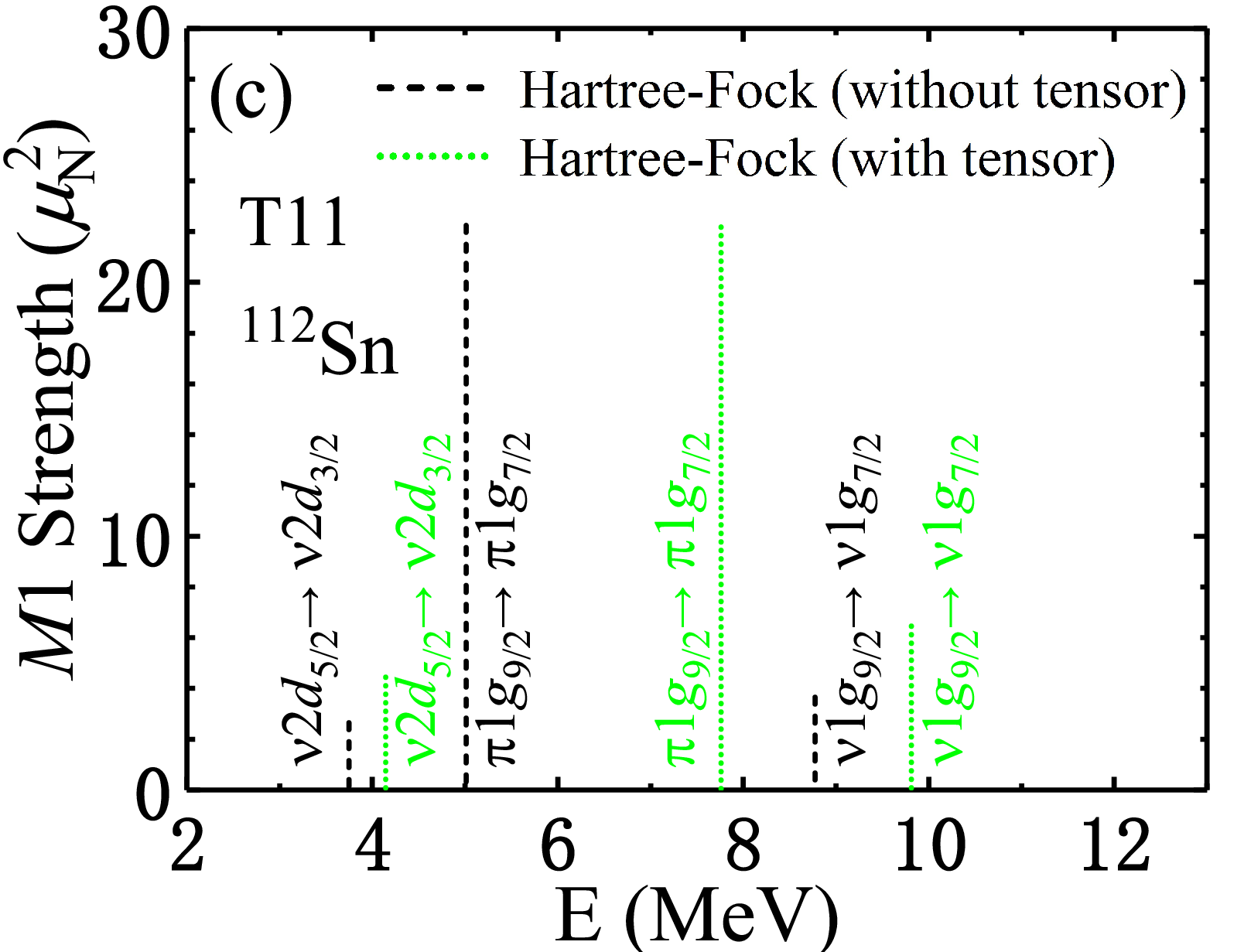}
 \includegraphics[width=0.4\textwidth]{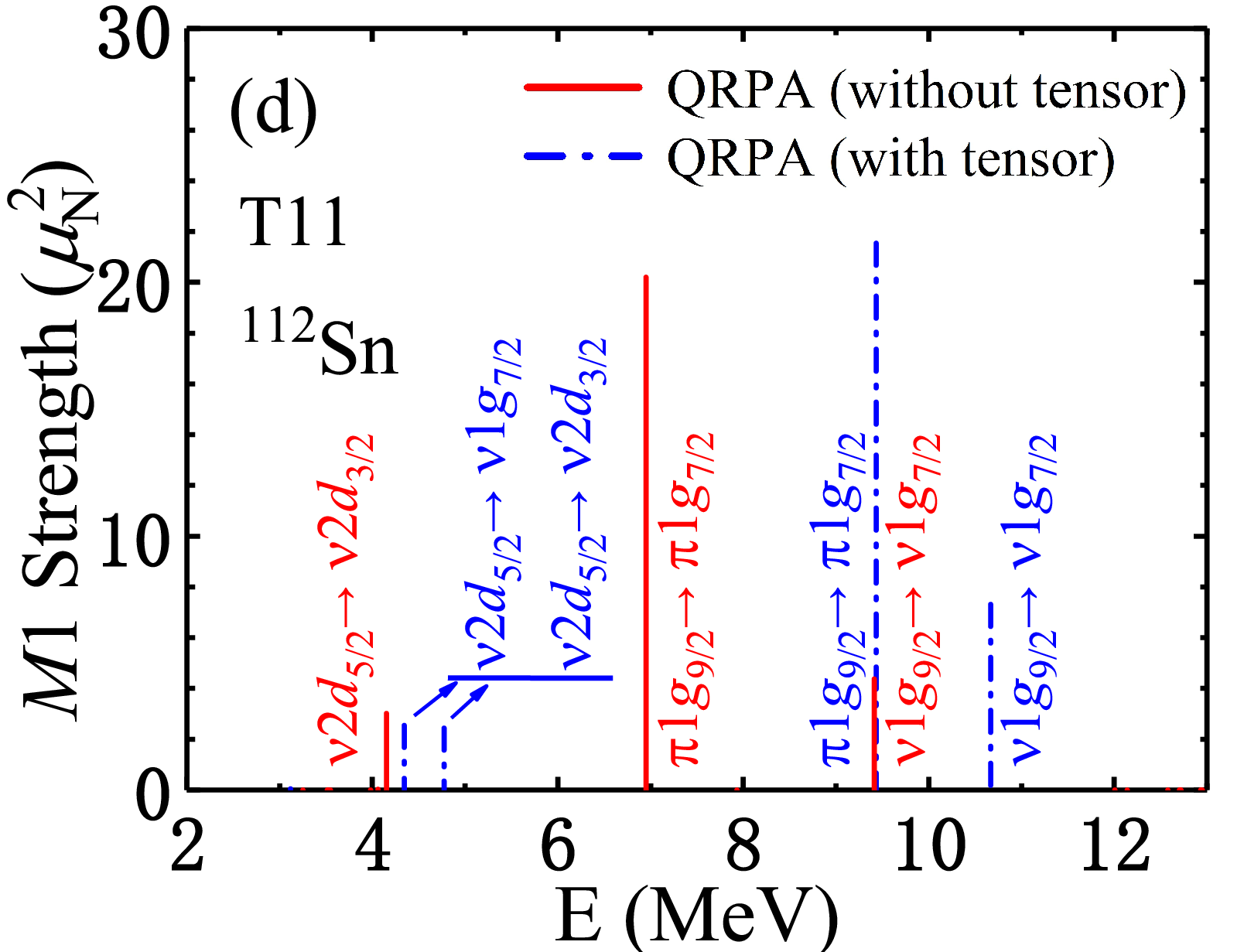}
\caption{(color online) The $M$1 Hartree-Fock and QRPA strength distributions of $^{112}$Sn obtained using the SLy5 and T11 parameter sets in the cases with and without tensor force. }
	\label{f10}
\end{figure}

 \begin{table}
\caption{The Hartree-Fock and QRPA energies, two-quasiparticle configurations which give the main contribution to the excited state, and $X_\nu^2-Y_\nu^2$ (only for QRPA state). The results are calculated for $^{112}$Sn with SLy5 and T11 interactions  with and without tensor force.}\label{t2}
\begin{ruledtabular}
\begin{tabular}{cccccccccccccc}
force &     \multicolumn{6}{c}{without tensor}    &   &  \multicolumn{6}{c}{with tensor}   \\
     &     \multicolumn{2}{c}{Hartree-Fock} & & \multicolumn{3}{c}{QRPA}   &   &  \multicolumn{2}{c}{Hartree-Fock}  & & \multicolumn{3}{c}{QRPA}  \\
\hline
   SLy5  & E$_\nu$   & config. &    &  E$_\nu$  &  config.  &  X$_\nu$$^2$-Y$_\nu$$^2$  &  & E$_\nu$   & config. &    &  E$_\nu$  &  config.  &  X$_\nu$$^2$-Y$_\nu$$^2$     \\
\hline
         & 3.70      & $(2d_{\frac{5}{2}}2d_{\frac{3}{2}}^{-1})^\nu$  &    &  4.15  & $(2d_{\frac{5}{2}}2d_{\frac{3}{2}}^{-1})^\nu$  &  99.1  &  & 4.37   & $(2d_{\frac{5}{2}}2d_{\frac{3}{2}}^{-1})^\nu$ &    &  4.19  &  $(2d_{\frac{5}{2}}1g_{\frac{7}{2}}^{-1})^\nu$  &  62.0     \\
         &        &    &    &     &     &  &     &      &   &  &  &  $(2d_{\frac{5}{2}}2d_{\frac{3}{2}}^{-1})^\nu$  &  36.2     \\
         & 6.57      & $(1g_{\frac{9}{2}}1g_{\frac{7}{2}}^{-1})^\pi$  &    &  8.51  & $(1g_{\frac{9}{2}}1g_{\frac{7}{2}}^{-1})^\pi$  &  99.5  &  & 8.18   & $(1g_{\frac{9}{2}}1g_{\frac{7}{2}}^{-1})^\pi$&    &  4.84  &  $(2d_{\frac{5}{2}}1g_{\frac{7}{2}}^{-1})^\nu$  &  37.2     \\
        &        &    &    &     &     &  &     &      &   &  &  &  $(2d_{\frac{5}{2}}2d_{\frac{3}{2}}^{-1})^\nu$  &  59.8     \\
         & 9.42      & $(1g_{\frac{9}{2}}1g_{\frac{7}{2}}^{-1})^\nu$  &    &  10.25  & $(1g_{\frac{9}{2}}1g_{\frac{7}{2}}^{-1})^\nu$  &  99.0  &  & 9.74   & $(1g_{\frac{9}{2}}1g_{\frac{7}{2}}^{-1})^\nu$&    &  9.31  &  $(1g_{\frac{9}{2}}1g_{\frac{7}{2}}^{-1})^\pi$  &  78.4     \\
        &        &    &    &     &     &  &     &      &   &  &  &  $(1g_{\frac{9}{2}}1g_{\frac{7}{2}}^{-1})^\nu$  &  15.7     \\
 \hline
   T11  & E$_\nu$   & config. &    &  E$_\nu$  &  config.  &  X$_\nu$$^2$-Y$_\nu$$^2$  &  & E$_\nu$   & config. &    &  E$_\nu$  &  config.  &  X$_\nu$$^2$-Y$_\nu$$^2$     \\
\hline
         & 3.75      & $(2d_{\frac{5}{2}}2d_{\frac{3}{2}}^{-1})^\nu$  &    &  4.15  & $(2d_{\frac{5}{2}}2d_{\frac{3}{2}}^{-1})^\nu$  &  98.8  &  & 4.14   & $(2d_{\frac{5}{2}}2d_{\frac{3}{2}}^{-1})^\nu$ &    &  4.34  &  $(2d_{\frac{5}{2}}1g_{\frac{7}{2}}^{-1})^\nu$  &  53.4     \\
         &        &    &    &     &     &  &     &      &   &  &  &  $(2d_{\frac{5}{2}}2d_{\frac{3}{2}}^{-1})^\nu$  &  46.2     \\
         & 5.01      & $(1g_{\frac{9}{2}}1g_{\frac{7}{2}}^{-1})^\pi$  &    &  6.95  & $(1g_{\frac{9}{2}}1g_{\frac{7}{2}}^{-1})^\pi$  &  99.6  &  & 7.76   & $(1g_{\frac{9}{2}}1g_{\frac{7}{2}}^{-1})^\pi$&    &  4.78  &  $(2d_{\frac{5}{2}}1g_{\frac{7}{2}}^{-1})^\nu$  &  46.3     \\
        &        &    &    &     &     &  &     &      &   &  &  &  $(2d_{\frac{5}{2}}2d_{\frac{3}{2}}^{-1})^\nu$  &  52.0     \\
         & 8.78      & $(1g_{\frac{9}{2}}1g_{\frac{7}{2}}^{-1})^\nu$  &    &  9.41  & $(1g_{\frac{9}{2}}1g_{\frac{7}{2}}^{-1})^\nu$  &  99.5  &  & 9.81   & $(1g_{\frac{9}{2}}1g_{\frac{7}{2}}^{-1})^\nu$&    &  9.43  &  $(1g_{\frac{9}{2}}1g_{\frac{7}{2}}^{-1})^\pi$  &  96.6     \\
        &        &   &    &     &    &     &  &    &   &    &  10.67  &  $(1g_{\frac{9}{2}}1g_{\frac{7}{2}}^{-1})^\nu$  &  97.2     \\
  \end{tabular}
\end{ruledtabular}
\end{table}

\begin{figure}[htp]
 \includegraphics[width=0.4\textwidth]{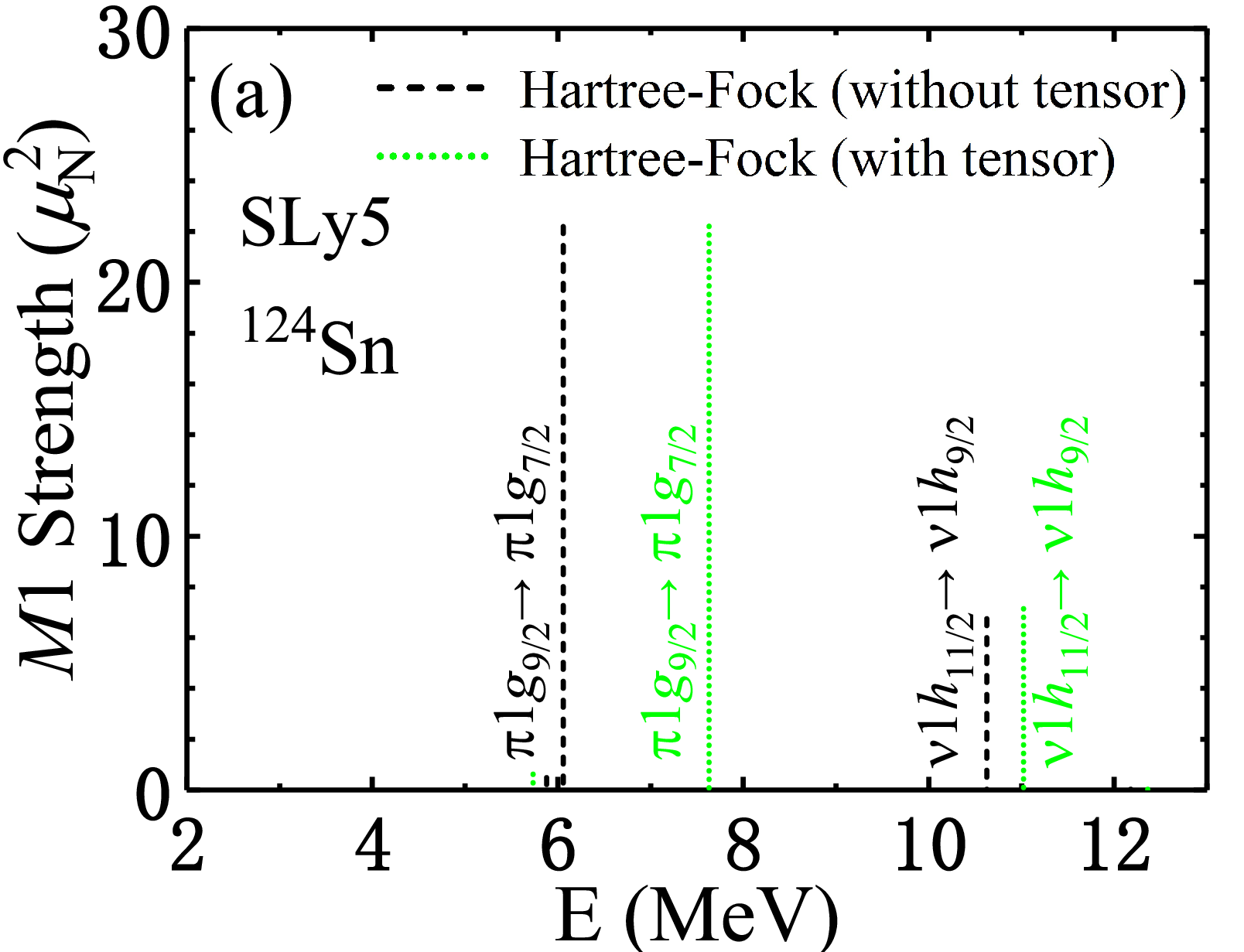}
 \includegraphics[width=0.4\textwidth]{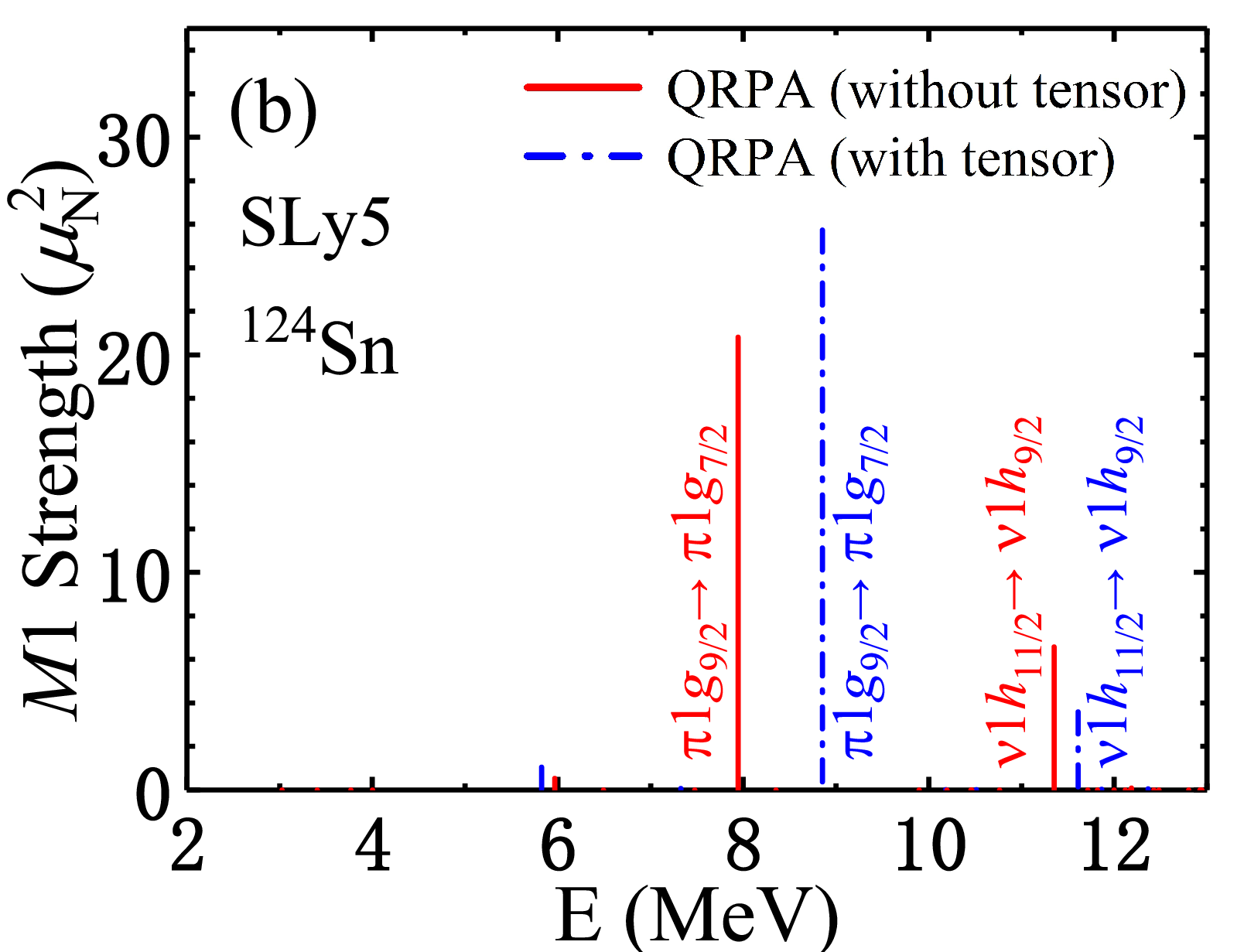}
 \includegraphics[width=0.4\textwidth]{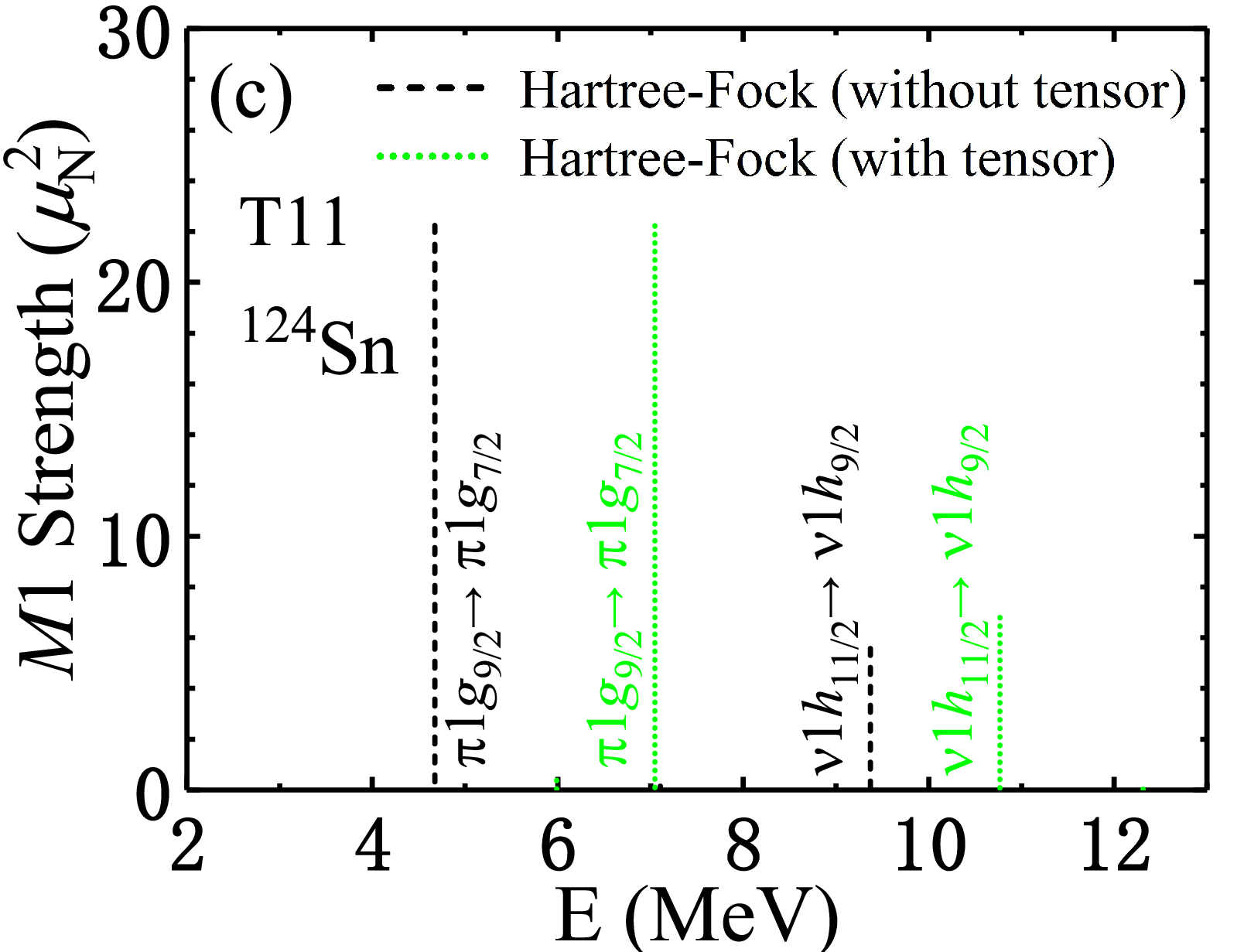}
 \includegraphics[width=0.4\textwidth]{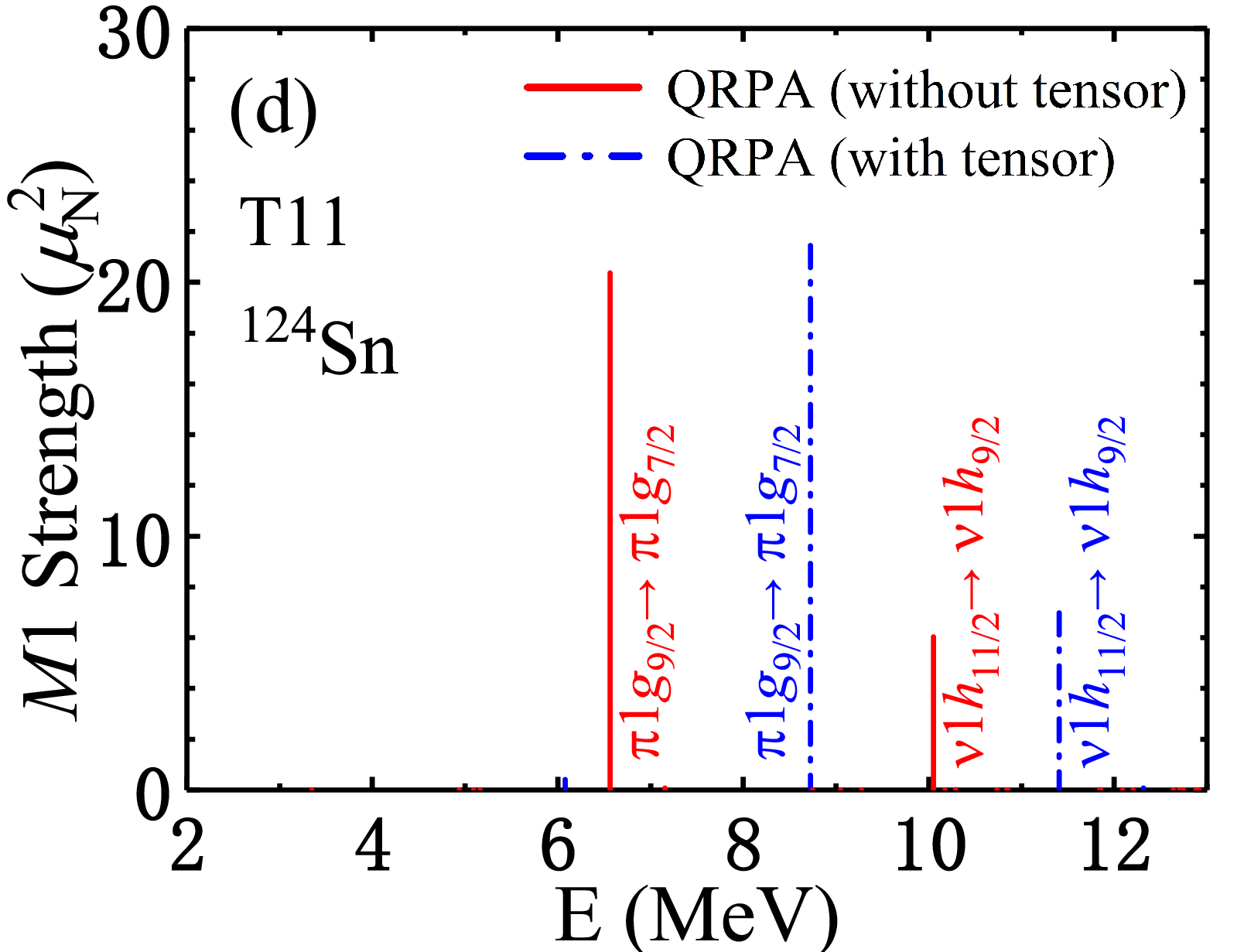}
	\caption{(color online) The $M$1 Hartree-Fock and QRPA strength distributions of $^{124}$Sn obtained using the	SLy5 and T11 parameter sets in the cases with and without tensor force.}  \label{f11}
\end{figure}

\begin{table}
\caption{The Hartree-Fock and QRPA energies, two-quasiparticle configurations which gives the main contribution to the excited state, and $X_\nu^2-Y_\nu^2$ (only for QRPA state). The results are calculated for $^{124}$Sn with SLy5 and T11 interactions  with and without tensor force.}\label{t3}
\begin{ruledtabular}
\begin{tabular}{cccccccccccccc}
force &     \multicolumn{6}{c}{without tensor}    &   &  \multicolumn{6}{c}{with tensor}   \\
     &     \multicolumn{2}{c}{Hartree-Fock} & & \multicolumn{3}{c}{QRPA}   &   &  \multicolumn{2}{c}{Hartree-Fock}  & & \multicolumn{3}{c}{QRPA}  \\
\hline
   SLy5  & E$_\nu$   & config. &    &  E$_\nu$  &  config.  &  X$_\nu$$^2$-Y$_\nu$$^2$  &  & E$_\nu$   & config. &    &  E$_\nu$  &  config.  &  X$_\nu$$^2$-Y$_\nu$$^2$     \\
\hline
         & 6.06      & $(1g_{\frac{9}{2}}1g_{\frac{7}{2}}^{-1})^\pi$  &    &  7.95  & $(1g_{\frac{9}{2}}1g_{\frac{7}{2}}^{-1})^\pi$  &  99.5  &  & 7.63   & $(1g_{\frac{9}{2}}1g_{\frac{7}{2}}^{-1})^\pi$&    &  8.85  &  $(1g_{\frac{9}{2}}1g_{\frac{7}{2}}^{-1})^\pi$  &  92.2     \\
         &10.63      & $(1h_{\frac{11}{2}}1h_{\frac{9}{2}}^{-1})^\nu$  &    &  11.35  & $(1h_{\frac{11}{2}}1h_{\frac{9}{2}}^{-1})^\nu$  &  97.9  &  & 11.02   & $(1h_{\frac{11}{2}}1h_{\frac{9}{2}}^{-1})^\nu$ &    &  11.61  &  $(1h_{\frac{11}{2}}1h_{\frac{9}{2}}^{-1})^\nu$  &  92.2     \\
  \hline
   T11  & E$_\nu$   & config. &    &  E$_\nu$  &  config.  &  X$_\nu$$^2$-Y$_\nu$$^2$  &  & E$_\nu$   & config. &    &  E$_\nu$  &  config.  &  X$_\nu$$^2$-Y$_\nu$$^2$     \\
\hline
         & 4.68      & $(1g_{\frac{9}{2}}1g_{\frac{7}{2}}^{-1})^\pi$  &    &  6.56  & $(1g_{\frac{9}{2}}1g_{\frac{7}{2}}^{-1})^\pi$  &  99.7  &  & 7.05   & $(1g_{\frac{9}{2}}1g_{\frac{7}{2}}^{-1})^\pi$ &    &  8.72  &  $(1g_{\frac{9}{2}}1g_{\frac{7}{2}}^{-1})^\pi$ &  96.8    \\
         & 9.37      & $(1h_{\frac{11}{2}}1h_{\frac{9}{2}}^{-1})^\nu$  &    &  10.05  & $(1h_{\frac{11}{2}}1h_{\frac{9}{2}}^{-1})^\nu$  &  99.3  &  & 10.77  & $(1h_{\frac{11}{2}}1h_{\frac{9}{2}}^{-1})^\nu$ &    &  11.41  &  $(1h_{\frac{11}{2}}1h_{\frac{9}{2}}^{-1})^\nu$  &  96.1     \\
 \end{tabular}
\end{ruledtabular}
\end{table}

\begin{figure}[htp]
 \includegraphics[width=0.4\textwidth]{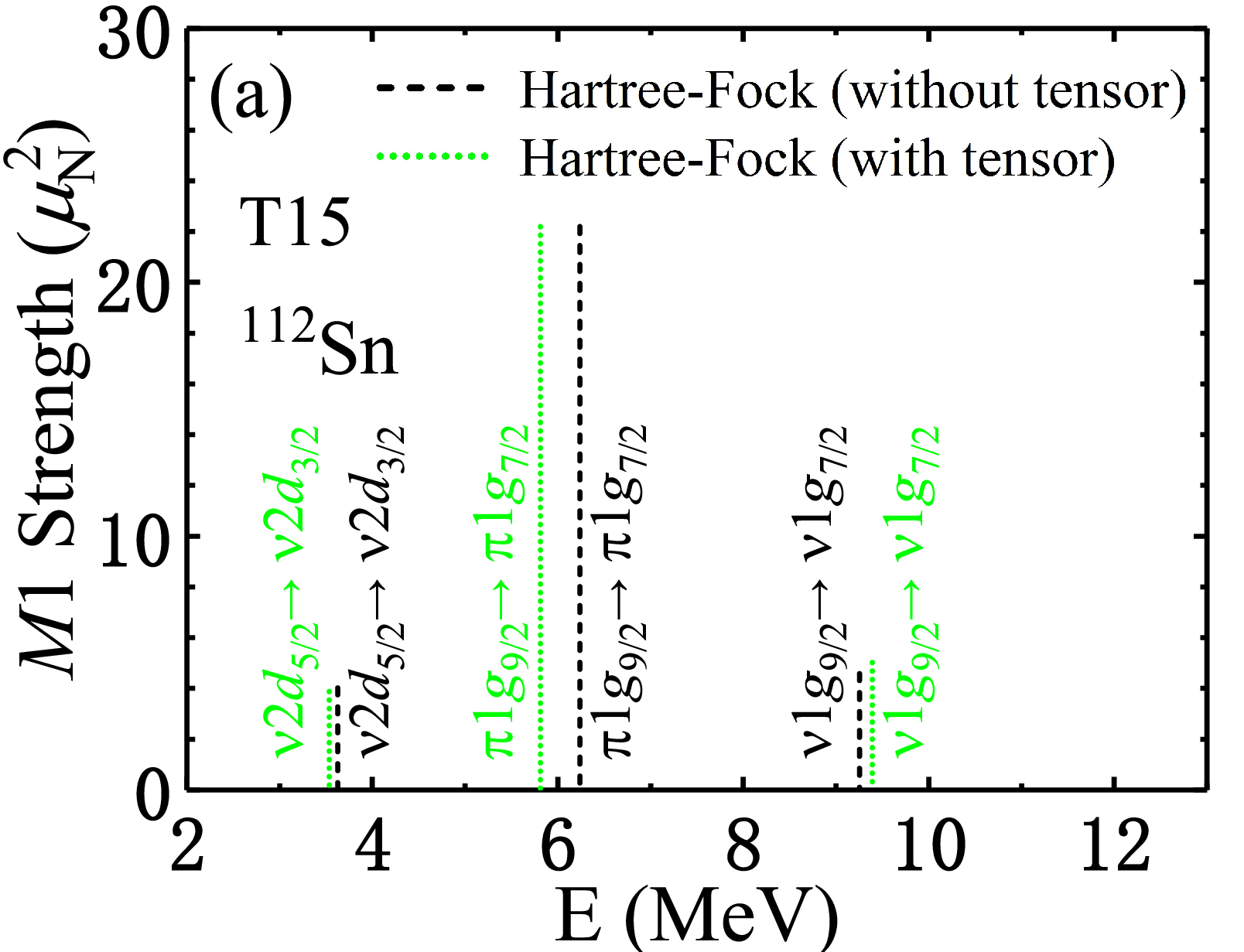}
 \includegraphics[width=0.4\textwidth]{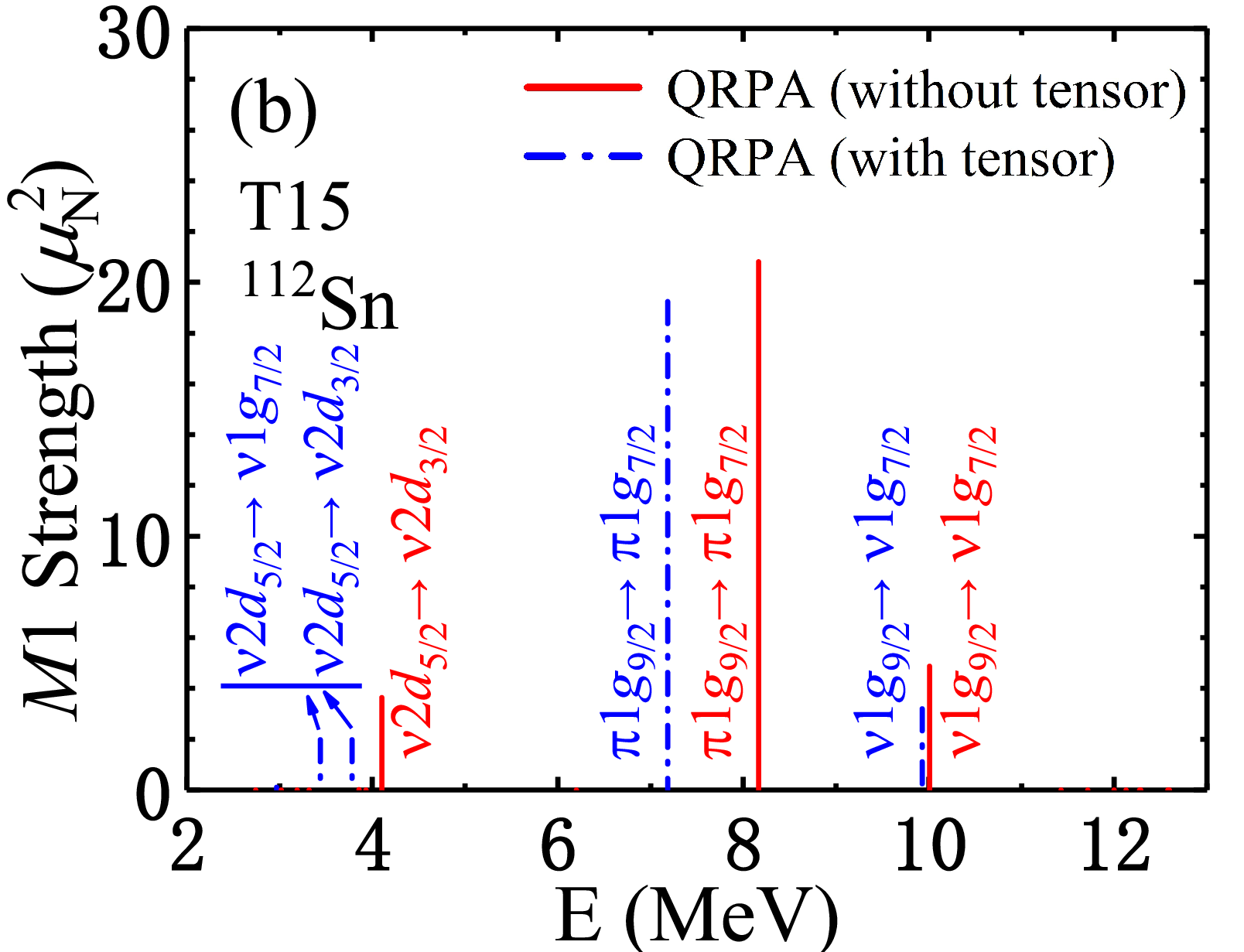}
 \includegraphics[width=0.4\textwidth]{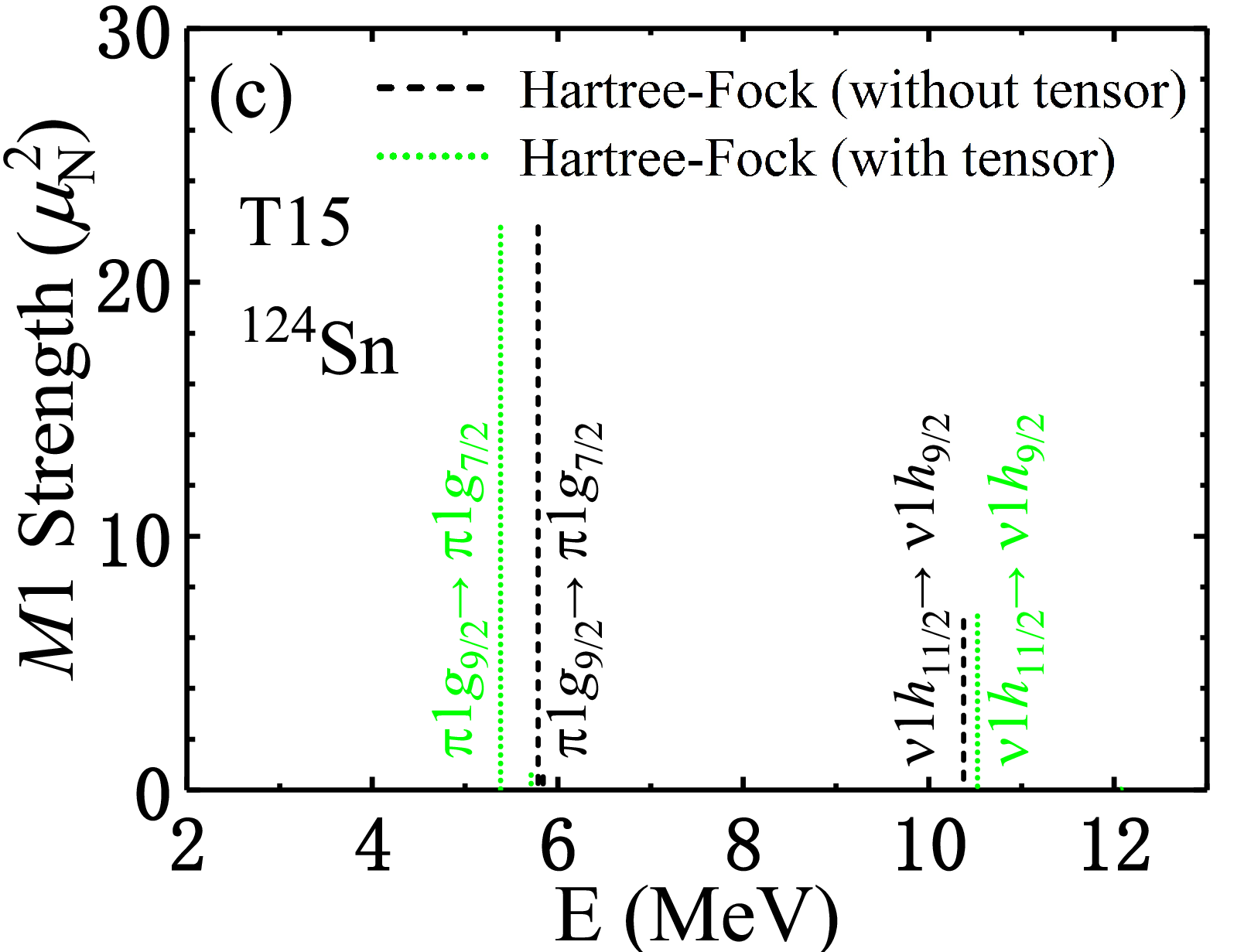}
 \includegraphics[width=0.4\textwidth]{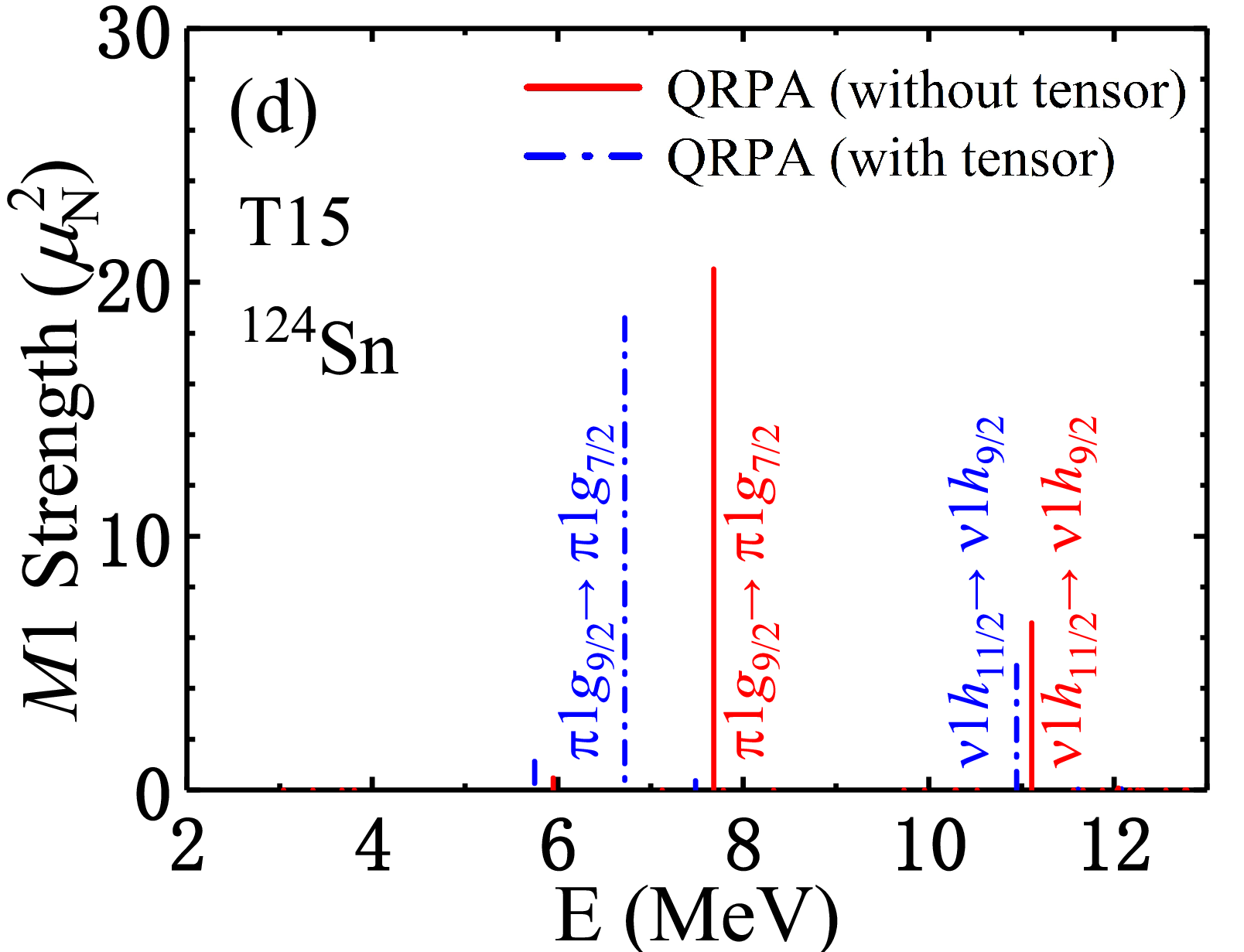}
\caption{(color online) The $M$1 Hartree-Fock and QRPA strength distributions of $^{112}$Sn and $^{124}$Sn obtained using the T15 parameter set in the cases with and without tensor force.
	}
\label{f12}
\end{figure}

\begin{table}
\caption{ The Hartree-Fock and QRPA energies, two-quasiparticle configurations which gives the main contribution to the excited state, and $X_\nu^2-Y_\nu^2$ (only for QRPA state). The results are calculated for $^{112}$Sn and $^{124}$Sn with T15 interaction  with and without tensor force.}\label{t4}
\begin{ruledtabular}
\begin{tabular}{cccccccccccccc}
     &     \multicolumn{6}{c}{without tensor}    &   &  \multicolumn{6}{c}{with tensor}   \\
     &     \multicolumn{2}{c}{Hartree-Fock} & & \multicolumn{3}{c}{QRPA}   &   &  \multicolumn{2}{c}{Hartree-Fock}  & & \multicolumn{3}{c}{QRPA}  \\
\hline
   $^{112}$Sn  & E$_\nu$   & config. &    &  E$_\nu$  &  config.  &  X$_\nu$$^2$-Y$_\nu$$^2$  &  & E$_\nu$   & config. &    &  E$_\nu$  &  config.  &  X$_\nu$$^2$-Y$_\nu$$^2$     \\
\hline
         & 3.63      & $(2d_{\frac{5}{2}}2d_{\frac{3}{2}}^{-1})^\nu$  &    &  4.10  & $(2d_{\frac{5}{2}}2d_{\frac{3}{2}}^{-1})^\nu$  &  99.3  &  & 3.54   & $(2d_{\frac{5}{2}}2d_{\frac{3}{2}}^{-1})^\nu$ &    &  3.44  &  $(2d_{\frac{5}{2}}1g_{\frac{7}{2}}^{-1})^\nu$  &  68.6     \\
         &        &    &    &     &     &  &     &      &   &  &  &  $(2d_{\frac{5}{2}}2d_{\frac{3}{2}}^{-1})^\nu$  &  29.7     \\
         & 6.24      & $(1g_{\frac{9}{2}}1g_{\frac{7}{2}}^{-1})^\pi$  &    &  8.17  & $(1g_{\frac{9}{2}}1g_{\frac{7}{2}}^{-1})^\pi$  &  99.7  &  & 5.81   & $(1g_{\frac{9}{2}}1g_{\frac{7}{2}}^{-1})^\pi$&    &  3.78  &  $(2d_{\frac{5}{2}}1g_{\frac{7}{2}}^{-1})^\nu$  &  30.3     \\
        &        &    &    &     &     &  &     &      &   &  &  &  $(2d_{\frac{5}{2}}2d_{\frac{3}{2}}^{-1})^\nu$  &  67.0     \\
         & 9.25      & $(1g_{\frac{9}{2}}1g_{\frac{7}{2}}^{-1})^\nu$  &    &  10.01  & $(1g_{\frac{9}{2}}1g_{\frac{7}{2}}^{-1})^\nu$  &  99.1  &  & 9.39   & $(1g_{\frac{9}{2}}1g_{\frac{7}{2}}^{-1})^\nu$&    &  7.19  &  $(1g_{\frac{9}{2}}1g_{\frac{7}{2}}^{-1})^\pi$  &  97.9     \\
        &        &   &    &     &    &     &  &    &   &    &  9.93  &  $(1g_{\frac{9}{2}}1g_{\frac{7}{2}}^{-1})^\nu$  &  98.5     \\
  \hline
 $^{124}$Sn  & E$_\nu$   & config. &    &  E$_\nu$  &  config.  &  X$_\nu$$^2$-Y$_\nu$$^2$  &  & E$_\nu$   & config. &    &  E$_\nu$  &  config.  &  X$_\nu$$^2$-Y$_\nu$$^2$     \\
\hline
         & 5.79      & $(1g_{\frac{9}{2}}1g_{\frac{7}{2}}^{-1})^\pi$  &    &  7.68  & $(1g_{\frac{9}{2}}1g_{\frac{7}{2}}^{-1})^\pi$  &  99.7  &  & 5.38   & $(1g_{\frac{9}{2}}1g_{\frac{7}{2}}^{-1})^\pi$ &    &  6.72  &  $(1g_{\frac{9}{2}}1g_{\frac{7}{2}}^{-1})^\pi$ &  96.5    \\
         & 10.38      & $(1h_{\frac{11}{2}}1h_{\frac{9}{2}}^{-1})^\nu$  &    &  11.11  & $(1h_{\frac{11}{2}}1h_{\frac{9}{2}}^{-1})^\nu$  &  98.5  &  & 10.53  & $(1h_{\frac{11}{2}}1h_{\frac{9}{2}}^{-1})^\nu$ &    &  10.95  &  $(1h_{\frac{11}{2}}1h_{\frac{9}{2}}^{-1})^\nu$  &  97.8     \\
 \end{tabular}
\end{ruledtabular}
\end{table}

\subsection{$M$1 of $^{112-120, 124}$Sn}

As discussed in the Introduction, the RPA or QRPA with Skyrme interactions has been used for many years in the description of $M$1 resonance in finite nuclei. Previously, we have systematically studied the effect of tensor
terms on the
magnetic dipole resonances in  $^{48}$Ca and $^{208}$Pb with various Skyrme interactions \cite{Caotens09,Caotens11}. Recently, the strength distributions of magnetic dipole resonances in even-even $^{112-120, 124}$Sn isotopes have been measured at RCNP \cite{M1exp-Sn120,M1exp-Sn}. This work extends our study to the magnetic dipole resonances in Sn isotopes using the QRPA approach with the SLy5 and T11 parameter sets. In Figs.~\ref{f8} and \ref{f9}, the $M$1 strength distributions of $^{112-120, 124}$Sn are shown, respectively. The results with and without tensor interaction are both given, and compared with the available experimental data~\cite{M1exp-Sn120,M1exp-Sn}.

From the figures, basically, one can find that the calculated response
functions with and without tensor force in $^{112-120}$Sn both
display three resonance peaks, namely, the low-lying $M$1 pygmy state,
the $M$1 main peak, and the one appearing at higher energy.
The three states are mainly formed by the neutron $\nu$2$d_{5/2}$$\rightarrow$$\nu$2$d_{3/2}$, proton $\pi$1$g_{9/2}$$\rightarrow$$\pi$1$g_{7/2}$,
and neutron $\nu$1$h_{11/2}$$\rightarrow$$\nu$1$h_{9/2}$ configurations,
respectively. For $^{112-116}$Sn, $M$1 main peaks with a shoulder are found. The state that forms the shoulder comes from the neutron $\nu$1$g_{9/2}$$\rightarrow$$\nu$1$g_{7/2}$ configuration, and its strength is reduced and disappears in $^{118}$Sn because the occupation probability of neutron state $\nu$1$g_{7/2}$ is becoming larger and the transition probability of neutron $\nu$1$g_{9/2}$$\rightarrow$$\nu$1$g_{7/2}$ configuration is becoming smaller with increasing mass number. The magnetic dipole strength distribution of $^{124}$Sn
displays the strong $M$1 peak arising from the
proton $\pi$1$g_{9/2}$$\rightarrow$$\pi$1$g_{7/2}$ configuration and the
higher energy state based on the neutron
$\nu$1$h_{11/2}$$\rightarrow$$\nu$1$h_{9/2}$ configuration,
while the pygmy state in the low energy region is not evident.

As for the experimental results shown in the figures, the current
researches~\cite{M1exp-Sn120,M1exp-Sn} on $^{112-120, 124}$Sn have only
provided the magnetic dipole strength distributions between
6.0 MeV and 12.0 MeV for all studied nuclei.
Our calculations reveal some pygmy strengths
emerging below 6.0 MeV in $^{112-120}$Sn, which mainly arise
from the neutron configuration $\nu$2$d_{5/2}$$\rightarrow$$\nu$2$d_{3/2}$.
It is seen that the strengths of these pygmy states become
weaker with increasing mass number.
This is because the occupation probabilities of neutron states $\nu$2$d_{3/2}$ in these nuclei are becoming larger,
and the transition probabilities between $\nu$2$d_{5/2}$ and $\nu$2$d_{3/2}$ are reduced along the Sn isotopes.
 As expected, the positions of the predicted pygmy peaks depend on the energy splittings of the two spin-orbit partners. 
 In the Skyrme~HF-BCS calculation, the spin$-$orbit potential~$U_{s.o.}$ has the dominant contributions from the spin$-$orbit strength~$W_{0}$ as well as the spin-orbit currents $J$
weighted by the tensor parameters~$\alpha_T$~and~$\beta_T$, as shown in Eqs. (7) and (10).
SLy5 and T11 interactions with tensor terms had been successfully applied to predict the spin$-$orbit splittings of finite nuclei. For example, in Refs. \cite{COLO2007227,Zou08}, it is shown that SLy5 interaction with tensor terms can fairly well explain the isospin dependence of energy differences $\varepsilon(\pi\,1h_{11/2})-\varepsilon(\pi\,1g_{7/2})$ along  Sn isotopes, and $\varepsilon(\nu\,1i_{13/2})-\varepsilon(\nu\,1h_{9/2})$ along  $N = 82$ isotones, as well as $\varepsilon(\pi\,2s_{1/2})-\varepsilon(\pi\,1d_{3/2})$ along Ca isotopes. 
Furthermore, the 1$f$ spin$-$orbit splittings in $^{40,48}$Ca calculated by T11  show reasonable agreement with the measurements as mentioned in Ref. \cite{Grasso13}. From these good features of the spin-orbit splittings, we expect that SLy5 and T11 EDFs with tensor terms can give reasonable predictions of the spin-orbit splittings of neutron states 2$d_{5/2}$
and 2$d_{3/2}$ in Sn isotopes. It will be quite interesting if these pygmy distributions
could be further confirmed by the experiments in the future.

\subsection{Effect of tensor force on $M$1 of $^{112}$Sn and $^{124}$Sn}

Although the experimental results have large error bars at high
energy, and in a few cases (the lighter
$^{112-114}$Sn isotopes) the main peaks do not emerge
clearly, still it is clear from
Figs.~\ref{f8} and \ref{f9} that the results calculated by
SLy5 and T11 with tensor can give a better description of
the experimental strength distributions of $^{112-120, 124}$Sn as
compared to the results  without the tensor force. To understand how the tensor force changes the strength distribution, we will take $^{112}$Sn and $^{124}$Sn as an example to show the mechanism. The effects of tensor force on the
Hartree-Fock and QRPA peaks have been discussed in Refs.
\cite{Caotens09,Caotens11}, and we will follow the same method in present analysis. The effect of tensor force on QRPA states can be estimated by the following formula where $\Delta E_{\textrm{QRPA}}$ represents the
difference between the QRPA results with and without tensor force
\begin{align}
\Delta E_{\textrm{QRPA}} \approx \Delta E_{\textrm{HF}} + \langle V_{\textrm{tensor}}\rangle.
\end{align}\label{erpa}
The first term in the right describes the change in the HF peak(s),
and the second term is the average of the effect from the
residual tensor interaction in QRPA calculation.  The calculated Hartree-Fock and QRPA $M$1 strength distributions of $^{112}$Sn
 and $^{124}$Sn obtained by using the SLy5 and T11 parameter sets with and without tensor force are shown in Figs.~\ref{f10} and ~\ref{f11}, respectively. The corresponding numerical data are shown in detail in Tables~\ref{t2} and ~\ref{t3}. For the Hartree-Fock strengths of $^{112}$Sn shown in Fig.~\ref{f10}(a),
 the results are obtained using the SLy5 interaction with and without tensor force. The unperturbed state associated with the configuration $\pi$1$g_{9/2}$$\rightarrow$$\pi$1$g_{7/2}$ is pushed upward from 6.57 MeV to 8.18 MeV when the tensor force is included in the calculation. Similarly, the tensor force moves the low-lying state related to $\nu$2$d_{5/2}$$\rightarrow$$\nu$2$d_{3/2}$ configuration from 3.70 MeV to 4.37 MeV.
 For the higher energy state arising from $\nu$1$g_{9/2}$$\rightarrow$$\nu$1$g_{7/2}$ configuration, its energy is  slightly shifted upward from 9.42 MeV to 9.74 MeV. This means that the spin-orbit splittings of partner levels are enlarged when the tensor force is involved, with this choice of parameters.
 In fact, it should be noticed by looking at Eq. (7) that the negative value
 of $U (\alpha_T)$ is essential to enlarge the spin-orbit splitting, and produces a better agreement with the experimental data as a net result.

The effect of tensor force on the QRPA strengths based on SLy5 interaction is shown in Fig.~\ref{f10} (b). Without including the tensor interaction, the main $M$1 resonance state coming from the proton configuration 1$g_{9/2}$$\rightarrow$1$g_{7/2}$ lies at 8.51 MeV, and the low-lying state formed from the
neutron configuration 2$d_{5/2}$$\rightarrow$2$d_{3/2}$ appears at 4.15 MeV. When the tensor force is taken into account, the $M$1 main resonance peak is pushed up to 9.31 MeV, being this peak mainly composed of the proton $\pi$1$g_{9/2}$$\rightarrow$$\pi$1$g_{7/2}$ configuration with an admixture of the neutron $\nu$1$g_{9/2}$$\rightarrow$$\nu$1$g_{7/2}$ configuration.
As for the low-lying QRPA states, there are two states located at energies 4.19 MeV and 4.84 MeV, which are formed
by the neutron configurations 2$d_{5/2}$$\rightarrow$1$g_{7/2}$ and 2$d_{5/2}$$\rightarrow$2$d_{3/2}$, where each configuration gives different contribution to the QRPA states as shown in Table~\ref{t2}.

From Table~\ref{t2}, one can extract that  $\Delta E_{\textrm{QRPA}}$ is 0.80 MeV and $\Delta E_{\textrm{HF}}$ equals to 1.61 MeV for the $M$1
main peak of $^{112}$Sn with the SLy5 interaction. Therefore, this leads to $\langle V_{\textrm{tensor}}\rangle = -0.81$ MeV.
Since the low-lying state is separated into two states when the tensor force is included, we use the average value as the QRPA result,
that is, 4.52 MeV. $\Delta E_{\textrm{QRPA}}$ is about 0.37 MeV and $\Delta E_{\textrm{HF}}$ is equal to 0.67 MeV for the low-lying state, so that the value of $\langle V_{\textrm{tensor}}\rangle$ is extracted to be $-0.30$ MeV. The extracted values of $\langle V_{\textrm{tensor}}\rangle$
 disclose that the tensor force is attractive. This is consistent with the conclusion in Ref.~\cite{Caotens09}.

For the results of $^{112}$Sn with the T11 interaction, both the Hartree-Fock and QRPA strengths are pushed upward when the tensor force is included in the calculations, as shown in  Fig~\ref{f10} (c) and (d). From Table~\ref{t2}, one can obtain that the unperturbed states associated with the  $\nu$2$d_{5/2}$$\rightarrow$$\nu$2$d_{3/2}$, $\pi$1$g_{9/2}$$\rightarrow$$\pi$1$g_{7/2}$ and $\nu$1$g_{9/2}$$\rightarrow$$\nu$1$g_{7/2}$ configurations move upward by 0.39, 2.75 and 1.03 MeV as an effect of the  tensor force. In the QRPA strengths, the energy changes of the low-lying, main $M$1, and higher energy states are 0.41, 2.48 and 1.26 MeV with the tensor force, respectively. Using Eq.~(13), we can extract that $\langle V_{\textrm{tensor}}\rangle$ = 0.02, -0.27 and 0.23 MeV, respectively. One can see that, in comparison with the results of SLy5 interaction, the residual tensor force of T11 interaction shows attraction for main peak but repulsion for the low-lying and high-lying states.

We now analyze  the results of $^{124}$Sn obtained using SLy5 and T11,
with and without tensor force. At variance with the case of $^{112}$Sn, there are only two Hartree-Fock or QRPA states shown
in Fig.~\ref{f11} and Table~\ref{t3}, one low-lying and one high-lying state, they are formed mainly from the proton $\pi$1$g_{9/2}$$\rightarrow$$\pi$1$g_{7/2}$ configuration and neutron $\nu$1$h_{11/2}$$\rightarrow$$\nu$1$h_{9/2}$ configuration, respectively. In the unperturbed strength obtained by SLy5 without the tensor force in Fig.~\ref{f11} (a), the low-lying unperturbed state lies at 6.06 MeV, while the high-lying state is at 10.63 MeV. As mentioned previously, when the tensor force is included, the spin-orbit splittings are enlarged, and the corresponding states are pushed upward to 7.63 and 11.02 MeV, respectively. For the QRPA states without the tensor force in Fig.~\ref{f11} (b), one can find that the low-lying state is located at 7.95 MeV while the high energy state is at 11.35 MeV, and they are pushed upward to 8.85 and 11.61 MeV by the tensor, respectively. According to Eq.~(13) and Table~\ref{t3}, the extracted values of $\langle V_{\textrm{tensor}}\rangle$ for low-lying and high-lying states are $-$0.67 and $-$0.13 MeV, respectively. These results  also reveal that the residual tensor force associated with the SLy5 set provides attractive contributions. In the case of the T11 interaction, the results obtained with and without tensor force are shown in Fig.~\ref{f11} (c and d) and Table~\ref{t3}. Because the tensor force enlarges the spin-orbit splittings also in the case of the T11 interaction, the calculated $\Delta E_{\textrm{HF}}$ of the unperturbed low-lying and high-lying states are 2.37 and 1.40 MeV, respectively. For the QRPA strengths, the values of $\Delta E_{\textrm{QRPA}}$ for the two states are about 2.16 and 1.36 MeV. So, the contributions of the residual tensor force in Eq.~(13) for the low-lying and high-lying states are extracted to be -0.21 and -0.04 MeV. Similar to the situation in $^{112}$Sn, the residual tensor force of T11 interaction for $^{124}$Sn shows weak attraction in the QRPA calculation.

In summary, SLy5 and T11 interactions with tensor force provide reasonable description of the experimental magnetic dipole data in $^{112-120, 124}$Sn in comparisons with other T$ij$ members and also the cases without tensor force. In other terms, we have checked that other forces, like those without tensor terms and the other T$ij$ sets, are less good when compared with experimental data. This is shown in Fig. 2 (cf. also the discussion).
By looking at
Table I, one can easily see that these results strongly suggest negative values of $\alpha$, whereas no clear constraint emerges for
$\beta$. We remind that $\alpha$ is associated with the tensor interaction between like-particle, while $\beta$ is associated with the tensor
interaction between protons and neutrons.

There are indeed other parameter sets with positive $\alpha_T$, like  the T15 interaction shown in Table I, which give an opposite effect on the $M$1 resonance. We take $^{112}$Sn and $^{124}$Sn as an example to explore the role of tensor force with the T15 interaction.  Figures ~\ref{f12} (a) and (b) show  the Hartree-Fock and QRPA strength distributions of $^{112}$Sn, while Fig.~\ref{f12} (c) and (d) do the same for the Hartree-Fock and QRPA strength distributions of $^{124}$Sn. As shown in Fig.~\ref{f12} (a), the $M$1 main peak in $^{112}$Sn coming from the proton configuration $\pi$1$g_{9/2}$$\rightarrow$$\pi$1$g_{7/2}$ has a clear downward shift, $\Delta E_{\textrm{HF}} = -0.43$ MeV, with the tensor interaction. This could be understood as follows: in the proton states in the $Z$ = 50 core, only the 1$g_{9/2}$ orbital gives positive contribution to the spin density $J_{p}>0$~\cite{COLO2007227}. On the other hand,  the neutron states $1g_{9/2}$, $2d_{5/2}$, $1g_{7/2}$, $2d_{3/2}$, $3s_{1/2}$ and $1h_{11/2}$ are partially occupied, and the neutron spin density $J_{n}$ is positive but smaller than $J_{p}$. According to Eq.~(7) and (10), and together with the values of $\alpha_T$ and $\beta_T$ of T15 (shown in Table~\ref{t1}), the tensor force provides a positive contribution to the proton spin$-$orbit potential $U_{\text {s.o.}}^{(p)}$, which makes  $U_{\text {s.o.}}^{(p)}$  weaker. As a result, the spin-orbit splittings of proton states are reduced, so one gets negative $\Delta E_{\textrm{HF}}$.  This leads to the downward shift of the main peak. In the QRPA case, based on Eq.~(13) and Table~\ref{t4}, we can extract that $\Delta E_{\textrm{QRPA}}$ is about -0.98 MeV, so the contribution of the residual tensor force  $\langle V_{\textrm{tensor}}\rangle$ is equal to -0.55 MeV.
A similar pattern is also seen in the case of $^{124}$Sn with the T15 interaction. The shift of the $M$1 main state in the Hartree-Fock response is -0.41 MeV, while $\Delta E_{\textrm{QRPA}}$ and $\langle V_{\textrm{tensor}}\rangle$ are -0.96 and -0.55 MeV, respectively.

\begin{table}[htb]
\caption{The total QRPA transition strengths $\sum\textit{B}^{\textrm{th.}}_{\textit{M}1}$ in $\mu^{2}_{N}$ for the Sn isotopes, calculated by using the Skyrme interactions SLy5 and T11 with and 	without tensor force. The values in the parentheses are obtained in the case of without tensor terms. The calculations are performed in the energy regions which are consistent with the experimental data in Table V of Ref.\cite{M1exp-Sn}. The experimental $\sum\textit{B}^{\textrm{exp.}}_{\textit{M}1}$ from Ref. \cite{M1exp-Sn} are also shown for comparison.  The quenching factors $g_{\textrm{eff}}/g_{\textrm{free}}$ for $M$1 resonances in each case are displayed. The calculated quenching factors  are also compared to the values
of the RQRPA model in Ref.\cite{PaarM121}.}\label{t5}
\begin{ruledtabular}
\begin{tabular}{clcccccc}
\specialrule{0em}{3pt}{3pt}
 & & $^{112}$Sn  & $^{114}$Sn & $^{116}$Sn & $^{118}$Sn & $^{120}$Sn & $^{124}$Sn\\
  \specialrule{0em}{3pt}{3pt}
\hline
\specialrule{0em}{3pt}{3pt}
~	&	$\text{\textrm{Exp.}}$	&	\text{14.7	 $\pm$	1.4	}	&	\text{19.6	 $\pm$	1.9	}	& \text{15.6	 $\pm$	1.3	}	&	\text{18.4	 $\pm$	2.4	} & \text{15.4 $\pm$ 1.4	}	&	\text{19.1	 $\pm$	1.7	} \\\specialrule{0em}{6pt}{6pt}.
\multirow{3}*{$\sum\textit{B}^{\textrm{th.}}_{\textit{M}1}$}	&	SLy5	&	27.76	(26.84)	&	24.84	(24.58)	&	23.07	(22.78)	&	22.60	(20.42)	&	25.35	(20.36)	&	25.85	(27.41)	\\\specialrule{0em}{2pt}{2pt}
~	&	T11 	&	28.86	(24.58)	&	26.11	(22.81)	&	23.94	(21.44)	&	22.37	(20.46)	&	25.88	(22.60)	&	21.88	(26.50)	\\\specialrule{0em}{2pt}{2pt}
\specialrule{0em}{3pt}{3pt}
\hline
\specialrule{0em}{3pt}{3pt}
\multirow{3}*{$g_{\textrm{eff}}/g_{\textrm{free}}$}	&	SLy5	&	0.73	(	0.74	)	&	0.89	(	0.89	)	&	0.82	(	0.83	)	&	0.90	(	0.95	)	&	0.78	(	0.87	)	&	0.86	(	0.83	)	\\\specialrule{0em}{2pt}{2pt}	
~	&	T11 	&	0.71	(	0.77	)	&	0.87	(	0.93	)	&	0.81	(	0.85	)	&	0.91	(	0.95	)	&	0.77	(	0.83	)	&	0.93	(	0.85	)	\\\specialrule{0em}{2pt}{2pt}
~	&	RQRPA 
	&	0.80	&	0.93	&	0.83	&	0.89	&	0.81	&	0.86	\\\specialrule{0em}{2pt}{2pt}	
\specialrule{0em}{3pt}{3pt}
\end{tabular}
\end{ruledtabular}
\end{table}

\subsection{Quenching factor}

Finally, we will discuss the quenching problem for $M$1 resonances.
Table~\ref{t5} shows the experimental total transition strengths \cite{M1exp-Sn},
the calculated total QRPA transition strengths and corresponding quenching factors for $^{112-120,124}$Sn isotopes. The results are calculated by using the Skyrme interactions SLy5 and T11 with and without tensor force. The values in the parentheses are obtained in the case of without tensor force. On the one hand, as shown in
Table~\ref{t5}, the calculated total QRPA transition strengths $\sum\textit{B}^{\textrm{th.}}_{\textit{M}1}$ with and without tensor both overestimate the experimental data. On the other hand, although the SLy5 and T11 interactions with tensor force can  describe well the strength distributions of $M$1 resonances for all nuclei, the summed transition probabilities obtained by SLy5 and T11 interactions deviate from the experimental data when the tensor terms are included.

The total $M$1 transition strengths predicted by QRPA are in general larger than experimental data when the free values of the $g$ factors are used. Therefore the free values are often modified
by taking implicity into account many-body effects, mesonic currents and $\Delta$-hole excitations, to obtain better description of the experimental data (details can be found in Ref. \cite{Metsch86}). The empirical  quenching factor is defined by
\be
q=\sqrt{\sum\textit{B}^{\textrm{exp.}}_{\textit{M}1}/{\sum\textit{B}^{\textrm{th.}}_{\textit{M}1}}},
\ee
where $\sum\textit{B}^{\textrm{exp.}}_{\textit{M}1}$ ($\sum\textit{B}^{\textrm{th.}}_{\textit{M}1}$) is the total experimental (theoretical) transition strength. The extracted quenching factors  are about 0.73-0.90 (0.74-0.95) for SLy5 with (without) the tensor terms. Similar values of quenching factors are obtained for T11 with (without)
tensor force, and are 0.71-0.93 (0.77-0.95). In Ref. \cite{PaarM121}, the B($M$1) strengths of $^{112-120,124}$Sn isotopes were obtained in the framework of relativistic QRPA (RQRPA), and compared with
the experimental data. The authors of that work also claimed that quenching factors were needed
to reproduce the data.  The calculated quenching factors from relativistic EDFs are also listed in Table~\ref{t5} and their values, around 0.80-0.93, are similar to those found in our work.

\section{Summary and Perspectives}\label{part4}
In this paper, we have investigated the magnetic dipole resonances of the even-even $^{112-120,124}$Sn isotopes, in the framework of the self-consistent Skyrme HF + BCS plus QRPA method. The Skyrme SLy5 and T11 interactions with and without tensor terms are used in the present calculations with a mixed type pairing interaction.

We have also checked other Skyrme sets and we have concluded that the SLy5 and T11 Skyrme interactions, with the tensor terms included, can give a better description of the  experimental $M$1 strength distributions of $^{112-120, 124}$Sn \cite{M1exp-Sn120,M1exp-Sn}, as compared with others. Taking $^{112}$Sn and $^{124}$Sn as examples, we have studied the role of tensor force in Hartree-Fock and QRPA response in detail. It is found that magnetic dipole resonances of $^{112-120,124}$Sn are  sensitive to the tensor parameter $\alpha_T$. A negative $\alpha_T$ leads to reproducing the experimental data.
On the other hand, a tensor force with positive $\alpha_T$, like that of the T15 interaction, gives an opposite contributions to the excitation energies of $M$1
resonances (compared to the case of SLy5 and T11), and the agreement with the experimental data is poorer.
This conclusion is not in conflict with that of previous works \cite{SAGAWA201476}, in which the tensor coupling constant $\beta$ is well constrained by the Gamow-Teller and spin-dipole states, while the coupling constant $\alpha$ has a large ambiguity.
In fact, the present study of $M$1 strength provides a complimentary constrain on the tensor coupling, the $\alpha$ value is rather well determined, but not the $\beta$ value.
Thus, we definitely need more observables to constrain the tensor terms.

The quenching problem is also discussed in the present work. In our calculations with the
tensor terms in the EDFs, we find that a quenching factor of about 0.71-0.93 is needed to reproduce
the total experimental transition probabilities for the nuclei we have studied. Without the tensor term,
we need quenching factors of more or less similar magnitude.

The calculated results show that low-lying magnetic dipole strength appears in the energy region below 6.0 MeV, it is mainly coming from the neutron configuration $\nu$2$d_{5/2}$$\rightarrow$$\nu$2$d_{3/2}$. However, no clear evidence of low energy $M$1 strength has been found so far in experiments.
It would be highly desirable to have further experimental investigations of the $M$1 strength. On the one hand, we would like to confirm or disprove our prediction regarding the low-lying $M$1 strength below 6.0 MeV. On the other hand, there is still some discrepancy between the theoretical results and the experimental results, while the experimental results have some non-negligible error bars at higher energy than 10 MeV. Additional efforts should be envisioned in the future, both on the experimental side and theoretical side. Eventually, it may reveal necessary to further improve the Skyrme energy density functional in the spin-isospin channel.

In Ref. \cite{PaarM121}, the evolution of magnetic dipole strength of Sn isotopes had been studied in the
RQRPA model. The quasi-particle configurations of $M$1 states are essentially
the same as  those of our calculations, they are mainly the proton configuration $\pi$1$g_{9/2}$$\rightarrow$$\pi$1$g_{7/2}$ and the neutron configurations $\nu$2$d_{5/2}$$\rightarrow$$\nu$2$d_{3/2}$, $\nu$1$g_{9/2}$$\rightarrow$$\nu$1$g_{7/2}$, and $\nu$1$h_{11/2}$$\rightarrow$$\nu$1$h_{9/2}$.  The importance of neutron configurations depends on
whether the spin-orbit partners are fully occupied or not. The energy dependence of the RQRPA strength
is similar to ours, but the energy location of the main $M$1 peak is different because of the different nuclear EDFs adopted. The RQRPA also predicted low-energy $M$1 strengths below 6.0 MeV in $^{112,116}$Sn (see Fig.~2 of Ref. \cite{PaarM121}), but the strengths are relatively small, which may be due to  different occupation probabilities of the involved orbitals in the pygmy states of those nuclei in RQRPA calculations.
 In Table II of Ref. \cite{PaarM121}, the total RQRPA transition strengths for $M$1 excitations in $^{112-120,124}$Sn were compared with the experimental data from inelastic proton scattering in Ref. \cite{M1exp-Sn}.  It has
been shown that the calculated values are larger than the experimental data, and in order to reproduce the experimental data, quenching factors of about 0.80-0.93 are needed in RQRPA.
These quenching factors are  similar to the extracted ones in our case, as shown in Table~\ref{t5}.

We should clarify the role of the correlations of beyond mean field.
The HF+RPA model has been a very successful model to describe collective states such as low-lying collective states
and giant resonances, not only in spherical nuclei but also in deformed nuclei.  The width of Gamow-Teller (GT) resonances, and the missing GT
strength, cannot be accounted for by the standard  mean-field models
and the description of these features is much improved
by models beyond mean-field like second RPA or particle-vibration
coupling models  \cite{Niu16,Gamba20,Yang22}.
The shift of the excitation energies
induced by models beyond mean-field is not completely negligible
but less important at the level we discuss in this work. Moreover,
the importance of the tensor force was recognized in the splittings
of spin-dipole (SD) excitations, already at
the RPA level \cite{Baitens10}.
Therefore, as is done in the present study,
the effect of tensor force on the excitation energy of $M$1 state
can be discussed at the QRPA level in a solid manner.

\section{ACKNOWLEDGEMENTS}

This work is partly supported by the National
Natural Science Foundation of China under Grant Nos. 12275025, 11975096, 12135004, 11961141004, 11635003, the Fundamental Research Funds for the Central Universities under Grant No. 2020NTST06, and the Japanese Grant-in-Aid for Scientific Research
(C) under Grant No. 19K03858.

\end{document}